\documentclass[lettersize,journal]{IEEEtran}
\usepackage{graphicx}
\usepackage{graphics}
\usepackage{amsmath,amsfonts,amsthm}
\usepackage{algorithmic}
\usepackage{algorithm}
\usepackage{array}
\usepackage[caption=false,font=normalsize,labelfont=sf,textfont=sf]{subfig}
\usepackage{textcomp}
\usepackage{stfloats}
\usepackage{url}
\usepackage{verbatim}
\usepackage{color}
\usepackage{xcolor}
\usepackage{cite}
\usepackage{makecell}
\usepackage{multirow}
\usepackage{hyperref}
\usepackage{framed}
\usepackage{tcolorbox}

\ifodd 1
\newcommand{\com}[1]{\textbf{\color{red} (COMMENT: #1)}} 
\newcommand{\comg}[1]{\textbf{\color{green} (COMMENT: #1)}}
\newcommand{\response}[1]{\textbf{\color{magenta} (RESPONSE: #1)}} 
\else

\newcommand{\com}[1]{}
\newcommand{\comg}[1]{}
\newcommand{\response}[1]{}
\fi

\begin{document}

\title{Edge Graph Intelligence: Reciprocally Empowering Edge Networks with Graph Intelligence}

\author{Liekang Zeng,
Shengyuan Ye, 
Xu Chen,
Xiaoxi Zhang,
Ju Ren, \\
Jian Tang,~\IEEEmembership{Fellow,~IEEE,}
Yang Yang,~\IEEEmembership{Fellow,~IEEE,}
and Xuemin (Sherman) Shen,~\IEEEmembership{Fellow,~IEEE}
\thanks{Liekang Zeng was with Sun Yat-sen University and The Hong Kong University of Science and Technology (Guangzhou). 
Shengyuan Ye, Xu Chen, and Xiaoxi Zhang are with Sun Yat-sen University and the Key Laboratory of Machine Intelligence and Advanced Computing, Ministry of Education.
Ju Ren is with Tsinghua University.
Jian Tang is with Midea Group.
Yang Yang is with the IoT Thrust and the Research Center for Digital World with Intelligent Things (DOIT) at the Hong Kong University of Science and Technology (Guangzhou), and also with Peng Cheng Laboratory, and also with Terminus Group.
Xuemin (Sherman) Shen is with The University of Waterloo.
}
}



\maketitle

\begin{abstract}
Recent years have witnessed a thriving growth of computing facilities connected at the network edge, cultivating edge networks as a fundamental infrastructure for supporting miscellaneous intelligent services.
Meanwhile, Artificial Intelligence (AI) frontiers have extrapolated to the graph domain and promoted Graph Intelligence (GI).
Given the inherent relation between graphs and networks, the interdiscipline of graph learning and edge networks, i.e., Edge GI or EGI, has revealed a novel interplay between them -- GI aids in optimizing edge networks, while edge networks facilitate GI model deployment.
Driven by this delicate closed-loop, EGI is recognized as a promising solution to fully unleash the potential of edge computing power and is garnering growing attention.
Nevertheless, research on EGI remains nascent, and there is a soaring demand within both the communications and AI communities for a dedicated venue to share recent advancements.
To this end, this paper promotes the concept of EGI, explores its scope and core principles, and conducts a comprehensive survey concerning recent research efforts on this emerging field.
Specifically, this paper introduces and discusses:
1) fundamentals of edge computing and graph learning,
2) emerging techniques centering on the closed loop between graph intelligence and edge networks,
and 3) open challenges and research opportunities of future EGI.
By bridging the gap across communication, networking, and graph learning areas, we believe that this survey can garner increased attention, foster meaningful discussions, and inspire further research ideas in EGI.
\end{abstract}

\begin{IEEEkeywords}
Edge computing, edge intelligence, graph learning, artificial intelligence, wireless communication.
\end{IEEEkeywords}

\newtcolorbox{mybox}{
  colback=gray!20,
  colframe=black,
  boxrule=1pt,
  fonttitle=\bfseries,
  boxsep=0pt,
  toptitle=1mm,
  bottomtitle=1mm,
  left=5mm,
  right=5mm,
  top=1mm,
  bottom=1mm
}

\section{Introduction}

\IEEEPARstart{E}{dge} networks are swiftly proliferating.
By assembling progressively spreading computing facilities at the network edge, edge networks have hosted ever-increasing amounts of data, storage, and computing resources.
They have become a fundamental infrastructure supporting miscellaneous applications like smart industrial manufacturing \cite{qiu2020edge, cao2021survey}, streaming video analytics \cite{xu2023edge, siriwardhana2021survey}, and Internet of robotics and vehicles \cite{mcenroe2022survey, ji2020survey}, etc.
As a complementary symmetry of the centralized core network, edge networks locate at the end of the Internet and encompass users in their physical vicinity, allowing for user-centric services with reduced response latency, improved resource efficiency, and enhanced privacy and security.
Benefited from these unique architectural superiorities, edge networks have been a vital experimentation arena for advanced communication techniques.
They are practically favorable for emerging intelligence services with delay-sensitive, resource-demanding, and privacy-preserved requirements, and have been widely recognized as a promising prospect for bridging the last mile between Artificial Intelligence (AI) and human beings \cite{shi2016edge, zhou2019edge}.

Meanwhile, AI is also rapidly booming.
To fully unleash the potential of big data in diverse forms, recent AI advances have extrapolated representation learning over massive data from Euclidean structure to graph topology, pushing Deep Learning (DL) frontiers to a new stream of models named Graph Neural Network (GNN) \cite{wu2020comprehensive, zhou2020graph}.
Different from traditional DNN (e.g., CNN, RNN) that typically applies 1D/2D convolutions, GNN introduces graph embedding techniques to digest information from graph relations \cite{abadal2021computing, besta2022parallel}.
Specifically, it applies neighbor aggregation on an input graph iteratively and captures hierarchical patterns through neural network operators from subgraphs of varying sizes.
This enables GNNs to abstract and learn the properties of specific vertices, links, or the entire graph, and thus generalize to unobserved graphs.
Leveraging such powerful expressiveness, learning with GNN, i.e., Graph Learning (GL), has exhibited superior graph analysis performance and empowers various graph-related tasks from node classification and link prediction to graph isomorphism and categorization \cite{zhang2020deep, ward2022practical}.

\begin{figure}[t]
  \centering
  \includegraphics[width=\linewidth]{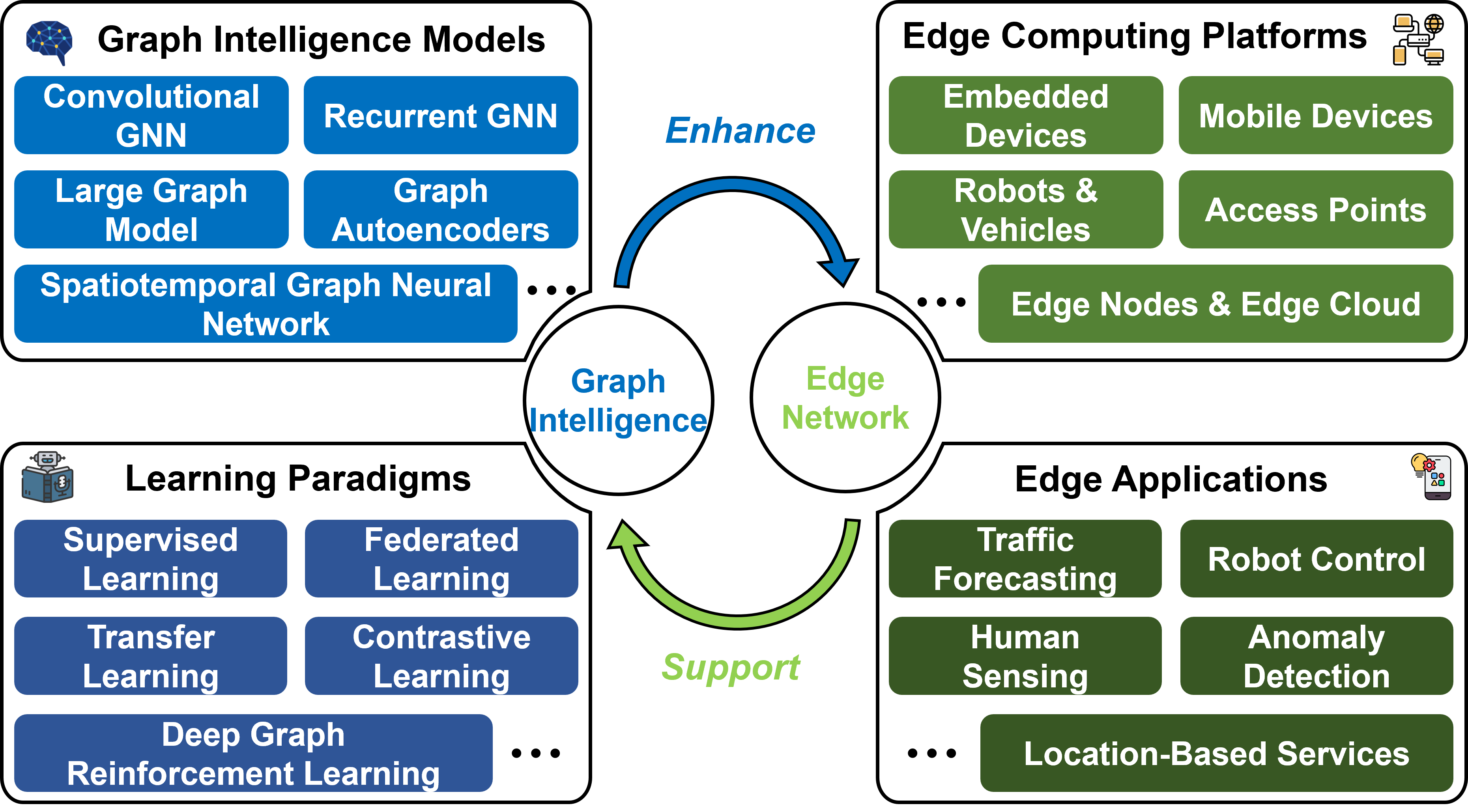}
  \caption{Illustration of the interplay between GI and edge networks, where GI can be applied as a data-driven tool to optimize edge networks, and conversely, edge networks perform as digital infrastructure to support GI deployment.
  }
  \label{fig:interplay}
\end{figure}

Given the remarkable success of Graph Intelligence (GI) and edge networks in their respective fields, the inherent connection between graphs and networks impels them to a confluence.
As illustrated in Fig. \ref{fig:interplay}, GI provides a vast zoo of empirical learning models (e.g., convolutional and recurrent GNNs, graph autoencoders) as well as various learning paradigms like Transfer Learning (TL) and Reinforcement Learning (RL), allowing advanced learning ability from graph data.
Symmetrically, edge networks generally comprise of a rich set of platforms including mobile devices, robots, vehicles, and edge nodes, which host miscellaneous graph-based applications such as traffic forecasting and network resource management.
Their bidirectional interaction, where GI enhances and optimizes edge networks and edge networks support and enable GI computation, draws a closed loop with mutual empowerment and nurtures a reciprocal interplay of their integration, namely ``\textbf{Edge Graph Intelligence}" or ``\textbf{EGI}" for brevity.

While the term EGI is fresh to come, research and practices have begun early.
Since the development of GCN in 2015 \cite{kipf2016semi}, GI has increasingly gained popularity in the AI community and ignited a wave of building GNNs over various real-world graphs.
Meanwhile, edge networks and edge computing are also rapidly evolving and actively embracing AI since 2019, giving rise to the concept of edge AI or edge intelligence \cite{zhou2019edge, li2019edge, wang2019inedge}.
Currently, the interplay of EGI has attracted growing attention from both the industry and academia and propelled a plethora of innovative optimizations, techniques, and applications at the network edge, e.g., traffic flow forecasting \cite{li2017diffusion, wang2020traffic}, location-based recommendation \cite{zhong2020hybrid, chang2020learning}, and vehicle trajectory prediction \cite{jeon2020scale, zhou2021ast}.
As a substantial extension of edge AI, EGI sheds light on its fundamental questions - how deeply edge networks and AI techniques can be fused and how much potential their fusion can shine - and demonstrates its powerful capability through plentiful realistic applications.

\begin{figure}[t]
  \centering
  \includegraphics[width=\linewidth]{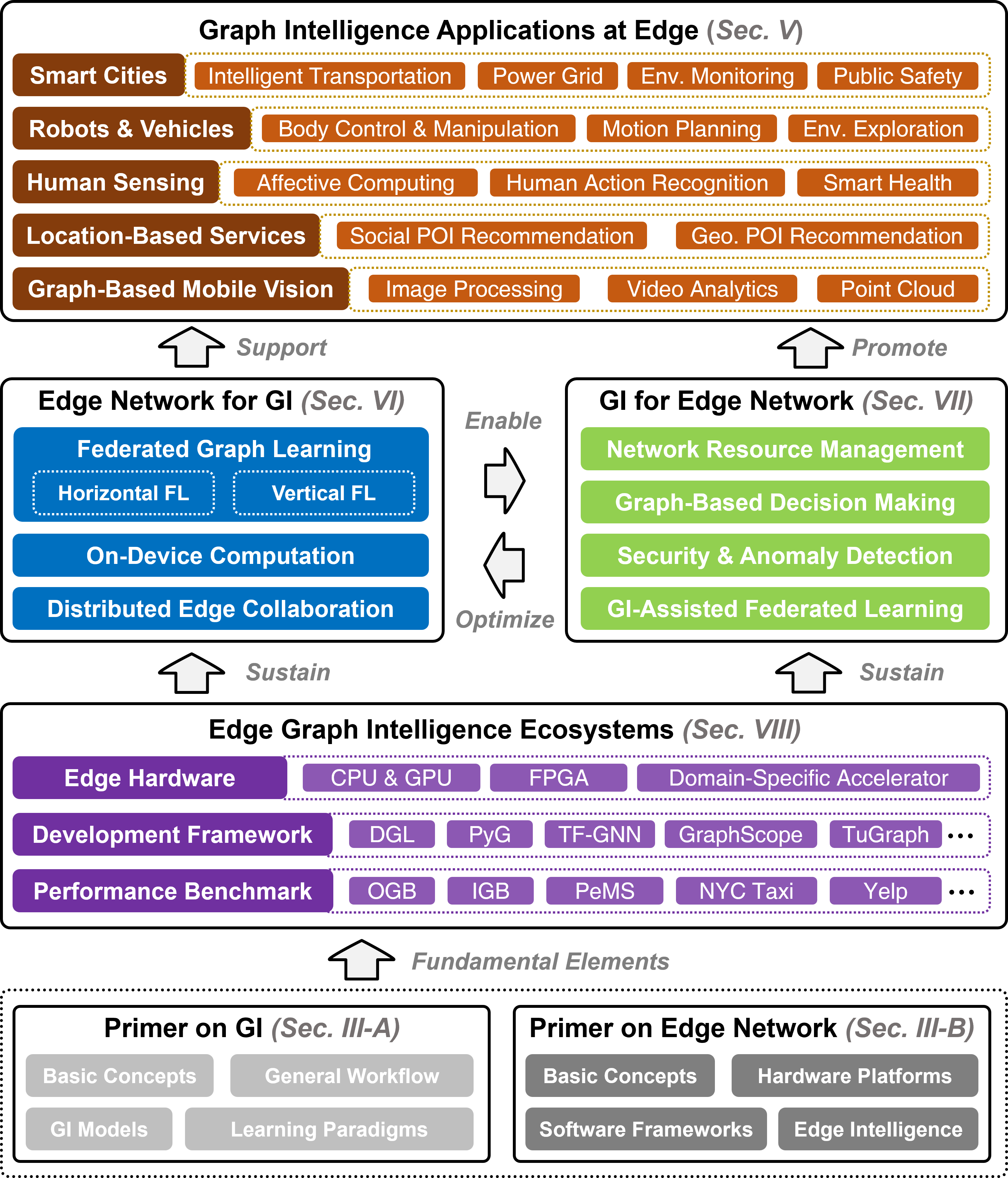}
  \caption{
  Outline and conceptual relationships of EGI aspects discussed in this survey: Based on fundamental elements of GI (Sec. \ref{sec:fundamental-gnn}) and edge networks (Sec. \ref{sec:fundamental-edge}), EGI ecosystems (Sec. \ref{sec:edge-infra-gnn}) sustain all stakeholders in the closed loop of edge networks and graph intelligence.
  Edge Network for GI (Sec. \ref{sec:edge_computation_gnn}) reviews techniques for supporting edge computation of GI models and GI for Edge Network (Sec. \ref{sec:gnn-optimization-edge}) discusses GI-based optimizations on edge networks.
  Both of them serve as support for a rich set of EGI applications (Sec. \ref{sec:gnn-application-edge}). 
  }
  \label{fig:outline}
\end{figure}

In this paper, we discuss in-depth how GI and edge networks are reciprocal to each other, and conduct a comprehensive and concrete survey of the recent research efforts on EGI.
In particular, centering around the inherently interconnected nature of graphs and networks, this paper reveals the bilateral interplay, for the first time, between GI and edge networks, and provides a concise rating in accordance with their mutually beneficial interactions.
In light of the rating, our survey identifies the four primary enablers essential for EGI, as illustrated in Fig. \ref{fig:outline}:

\begin{itemize}
    \item Edge applications of GI Models (Sec. \ref{sec:gnn-application-edge}): Typical application scenarios and use cases for applying GI in edge networks;
    \item Edge Networks for GI (Sec. \ref{sec:edge_computation_gnn}): Paradigms of GI model computation, including model training and inference, for GI over edge networks;
    \item GI for Edge Networks (Sec. \ref{sec:gnn-optimization-edge}): Practical GI-based methods for optimizing edge networks concerning their specific functionalities;
    \item EGI ecosystems (Sec. \ref{sec:edge-infra-gnn}): full-stack infrastructural support for high-performance EGI computation in terms of hardware, software, and benchmarks.
\end{itemize}

\begin{table*}[t]
\centering
\caption{List of main abbreviations.}
\label{tab:abbr}
\begin{tabular}{clclcl}
\hline
\textbf{Abbr.} & \multicolumn{1}{c}{\textbf{Definition}} & \textbf{Abbr.} & \multicolumn{1}{c}{\textbf{Definition}} & \textbf{Abbr.} & \multicolumn{1}{c}{\textbf{Definition}} \\ \hline
AI             & Artificial Intelligence                 & FPGA           & Field Programmable Gate Array           & LLM            & Large Language Model                    \\
AR             & Augmented Realtiy                       & GAE            & Graph Autoencoder                       & LSH            & Locality-Sensitive Hash                 \\
CL             & Contrastive Learning                    & GAN            & Generative Adversarial Network          & LSTM           & Long Short Term Memory networks         \\
CDN            & Content Delivery Network                & GAT            & Graph Attention Network                 & ML             & Machine Learning                        \\
CNN            & Convolutional Neural Network            & GCN            & Graph Convolutional Network             & MLP            & Multilayer Perceptron                   \\
CPS            & Cyber-Physical Systems                  & GFL            & GI-Assisted Federated Learning          & MOT            & Multi Object Tracking                   \\
DGRL           & Deep Graph Reinforcement Learning       & GFM            & Graph Foundation Model                  & NAS            & Neural architecture search              \\
D2D            & Device-to-Device                        & GI             & Graph Intelligence                      & NPU            & Neural Processing Unit                  \\
DAG            & Directed Acyclic Graph                  & GL             & Graph Learning                          & QoS            & Quality of Service                      \\
DL             & Deep Learning                           & GNN            & Graph Neural Networks                   & POI            & Point-of-Interest                       \\
DNN            & Deep Neural Network                     & HAR            & Human Action Recognition                & RecGNN         & Recurrent Graph Neural Network          \\
DP             & Differential Privacy                    & HE             & Homomorphic Encryption                  & RL             & Reinforcement Learning                  \\
DQN            & Deep Q-Network                          & HLS            & High-Level Synthesis                    & RNN            & Recurrent Neural Network                \\
DRL            & Deep Reinforcement Learning             & IID            & Independent and IDentically             & SDN            & Software-Defined Networks               \\
DSA            & Domain-Specific Accelerators            & IoT            & Internet of Things                      & SLO            & Service Level Objective                 \\
DT             & Digital Twin                            & IoV            & Internet of Vehicles                    & SpMM           & Sparse-dense Matrix Multiplication      \\
EGI            & Edge Graph Intelligence                 & ITS            & Intelligent Transportation Systems      & STGNN          & Spatio-Temporal Graph Neural Network    \\
EHR            & Electronic Health Record                & KD             & Knowledge Distillation                  & TL             & Transfer Learning                       \\
EI             & Edge Intelligence                       & KG             & Knowledge Graph                         & VAE            & Variational Autoencoder                 \\
FER            & Facial Expression Recognition           & LAN            & Local-Area Network                      & VNF            & Virtual Network Function                \\
FGL            & Federated Graph Learning                & LBS            & Location-Based Service                  & VR             & Virtual Reality                         \\
FL             & Federated Learning                      & LGM            & Large Graph Model                       & WAN            & Wide-Area Network                       \\ \hline
\end{tabular}
\end{table*}

In general, these key enablers can be well accommodated in the closed loop, i.e., ``edge for GI" and ``GI for edge" as described in Fig. \ref{fig:interplay}, 
In the ``edge for GI" course, edge networks provide physical platforms and software stacks to graph intelligence, serving as infrastructure to support GI models training and inference processes.
More specifically, GI models' intensive training workload can be resolved by means of pools of edge resources (e.g., federated edge learning), and edge inference techniques are developed for deploying and accelerating GI models under resource constraints and SLO requirements. 
Alternatively, in the ``GI for edge" course, GI models with these inference solutions can thereafter be efficiently executed upon edge platforms, which enables miscellaneous graph-based applications and optimizes various aspects of edge networks.
Besides reviewing these key enablers, our survey provides fundamental and friendly premiers of GI and edge networks that assume no prior knowledge of GI or edge computing.
We also discuss various open challenges and research directions toward future EGI, encouraging both AI and communications communities to advance EGI for a broader range of people.

In summary, this paper makes the following key contributions.

\begin{itemize}
    \item A brief tutorial on the primer of  EGI key components that explains basic and necessary concepts about GI and edge networks.
    \item We submit an in-depth discussion on EGI advantages as well as an EGI rating taxonomy that can be used as a criterion to identify the advancement of EGI systems. 
    \item We conduct a comprehensive survey on various aspects of EGI, including edge applications of GI, edge networks for GI, GI for edge networks, and EGI ecosystems.
    \item We discuss open challenges and research directions of EGI, acted as a reference for future EGI innovations.
\end{itemize}

The rest of this paper is organized as follows: 
First, Sec. \ref{sec:fundamental-gnn} and Sec. \ref{sec:fundamental-edge} briefly review the primers of graph intelligence and edge computing networks, respectively.
Next, the subsequent sections introduce research efforts with respect to the four enablers: GI applications at Edge (Sec. \ref{sec:gnn-application-edge}), edge networks for GI (Sec. \ref{sec:edge_computation_gnn}), GI for edge networks (Sec. \ref{sec:gnn-optimization-edge}), and EGI ecosystems (Sec. \ref{sec:edge-infra-gnn}).
Finally, Sec. \ref{sec:challenge} discusses open challenges and future research opportunities of EGI and Sec. \ref{sec:conclusion} concludes.
Table \ref{tab:abbr} lists the main abbreviations used in this survey.

\section{Related Survey}
\label{sec:related_work}

While EGI intersects edge networks and GI, its investigation is still narrowly limited to unilateral dimensions.

On the GI side, a volume of literature has reviewed the general landscape of GL, providing insights on model taxonomy \cite{zhou2018graph}, capability boundaries \cite{zhang2020deep}, and practical tutorials \cite{ward2022practical}.
As flourishing attention is attracted, publications relevant to GI topics explode.
Following this trend, some researchers begin to deliver reviews concentrated on specific types of models, such as GAE \cite{park2019symmetric, khodayar2019convolutional}, STGNN \cite{jin2023spatio, li2023graph} and DGRL \cite{munikoti2023challenges, nie2023reinforcement}.
Surveys on certain scenarios are also released.
For instance, some researchers review the use of GI in traffic domain \cite{jiang2022graph, bui2022spatial, rahmani2023graph}, discussing how traffic networks can be constructed as graphs and how traffic patterns can be resolved with GI; several surveys target power grids, where the electrical network are naturally graphs \cite{liao2021review, abdelmalak2022survey, huang2022applications}.
Nevertheless, these investigations either center on graph learning landscapes with limited discussion about their roles in edge networks \cite{zhang2020deep, wu2020comprehensive, zhou2020graph}, or focused particularly on applying GI techniques on some specific edge scenarios, yet ignoring the big picture of general edge networks spectrum.
Some recent literature \cite{he2021overview, dong2023graph, moorthy2024survey} also reviews the progress of GI in the context of IoT and wireless networks, aiming at exploiting GI for optimizing communication channels in IoT services. 
These works, however, mainly focus on GI applications in their discussed scopes and lack a systematic taxonomy on how GI can be computed in hierarchical edge networks, which is one of the fundamental pillars in EGI's ecosystem.


On the edge AI side, surveys on the broader edge computing topics have been carried out for years, pertaining to visions and challenges \cite{shi2016edge}, systems and tools \cite{liu2019survey}, communication and IoT perspectives \cite{mao2017survey, kong2022edge, yu2017survey}, etc. 
With the popularization of AI in edge networks, many researchers turned their attention to the potential and future directions of edge AI.
For instance, Zhou et al. \cite{zhou2019edge} discuss the motivation and advantages of edge intelligence along with a grading mechanism for edge intelligence by assessing the training and inference on edge platforms.
Wang et al. \cite{wang2020convergence} submit a comprehensive taxonomy of edge AI and analyze a set of open challenges toward future edge AI development.
Although these surveys have extensively investigated edge intelligence systems, a majority of them \cite{zhou2019edge, wang2020convergence, murshed2021machine, liu2022bringing, hua2023edge} center on general AI computation or are dedicated to traditional DL workloads such as CNN or RNN.
GI models, which possess distinct capabilities and unique computing characteristics, are much less understood in the edge AI context.

In summary, while the AI and edge computing communities have pushed their investigation to their respective frontiers, a thorough review of EGI, the combination of both lines, is still absent and desires actions.
\section{Primer on Edge Graph Intelligence}

Before diving into various aspects of EGI, we briefly review the basic concepts and relevant techniques of GI and edge networks, respectively.

\subsection{Graph Intelligence}
\label{sec:fundamental-gnn}

As one of the key flywheel actuating the loop within EGI, graph representation learning is devoted to the algorithm side and contributes enhanced ability in graph data processing.
Before diving into EGI, this section introduces graph representation learning with respect to its basic concepts, general workflow, and representative models and learning paradigms. 
For a more comprehensive treatment of GI, concentrated reviews \cite{zhou2018graph, wu2020comprehensive, zhang2020deep, ward2022practical} on GL are highly recommended.
Table \ref{tab:notation-gl} lists the main notations used in this section.

\subsubsection{Basic Concepts}

\textbf{Graphs.} 
Graphs are a way to organize data, and with graphs, one can succinctly characterize relationships across scattered data points.
The input of GI models, i.e., GNNs, are graphs, which typically contain two types of data.
One is the adjacency matrix or adjacency list, which interprets the graph topology, and the other is the feature vectors that describe vertices and edges’ actual properties.
Formally, an input graph is denoted as $\mathcal{G} = \langle \mathcal{V}, \mathcal{E} \rangle$, with vertices and links collected in $\mathcal{V}$ and $\mathcal{E}$, respectively.

\textbf{Vertices and Links.}
Vertices or nodes in a graph can be items, objects, or entities, and are not necessarily homogeneous when constructing a graph.
For instance, a location-based knowledge graph can represent its vertices as human users, IoT devices, scenic spots, and any other entities of various types within a specific district.
Links are another essential component in graphs that characterizes the relationships between these items, objects, or entities.
Note that to avoid misunderstanding, we exclusively use ``links" to indicate the connection between vertices in an input graph while leaving ``edge" for edge networks.
A link can be defined with respect to the two (not necessarily unique) vertices associated with it.
For $\mathcal{V} \in \mathcal{G}$ and $\mathcal{E} \in \mathcal{G}$, we denote their size, i.e., the number of vertices and links, as $V$ and $E$, respectively, and use $v$ and $e$ to index arbitrary vertex and link in them.

\textbf{Neighbors.}
Neighbors are ego-networks centering on specific vertices within a graph.
For a vertex $v$, its neighbors cover the vertices directly connected to $v$ and their adjoining links.
Note that a vertex's neighbors can be iteratively expanded by considering the neighbors of its neighbors.
Formally, given $\mathcal{N}^{(k)}_v$ as vertex $v$'s $k$-hop neighbors, we have $\mathcal{N}^{(k+1)}_v = \{\mathcal{N}^{(1)}_u | \forall u \in \mathcal{N}^{(k)}_v \}$, where $\mathcal{N}^{(1)}_v$ indicates $v$'s one-hop direct neighbors.

\textbf{Representation Vectors, Features, and Embeddings.}
Representation vectors are the numerical vectors associated with vertices and links, and are also referred to as encodings, representations, latent vectors, or high-level feature vectors depending on the context.
In this section, we respectively denoted representation vectors by $h^{(l)}_{v}$ and $h^{(l)}_{e}$ at the $l$-th GNN layer.
Upon input to the model, the initial representation vectors $h^{(0)}_{v}$ and $h^{(0)}_{e}$ are exactly the features attached to vertices and links, which quantify physical properties in specific applications.
Extending the above knowledge graph example, the features of a vertex may include the users' age and food preferences, and for a scenic spot, it can be location and popularity.
After processing through model layers, the exported representation vectors are embeddings, a form of compressed feature representations of vertices, links, neighbors, or graphs.
Embeddings can be viewed as the mappings of original data in latent spaces, which effectively reserve the semantics implicated in the input graph while can be used by downstream models for specific tasks (e.g., vertex classification and link prediction).

\begin{table}[t]
    \centering
    \caption{List of main notations in graph representation learning.}
    \label{tab:notation-gl}
    \begin{tabular}{cp{6.5cm}}
        \hline
        \textbf{Notation} & \multicolumn{1}{c}{\textbf{Definition}} \\ \hline
         $\mathcal{G}$ & The graph input to the GNN model.\\
         $\mathcal{V}$, $V$, $v$ & The input graph $\mathcal{G}$ includes a set $\mathcal{V}$ of vertices, where its size is $V$ and $v$ is an arbitrary vertex in $\mathcal{V}$. \\
         $\mathcal{E}$, $E$, $e$ & The input graph $\mathcal{G}$' includes a set $\mathcal{E}$ of links, where its size is $E$ and $e$ is an arbitrary link in $\mathcal{E}$. \\
         $\mathcal{N}^{(k)}_v$ & The set of vertex $v$'s $k$-hop neighbors in $\mathcal{G}$. \\ 
         $\mathcal{L}$, $L$, $l$ & The GNN model's layers set $\mathcal{L}$ is of size $L$, where $l$ is an arbitrary layer in $\mathcal{L}$. \\
         $h^{(l)}_{v}$, $h^{(l)}_{e}$ & The representation vector of vertex $v$ and link $e$ of layer $l$, respectively. When $l=0$ and $l=L$, they represent the cases of input and output representation vectors of vertex $v$ (link $e$), respectively. \\
         $\varphi^{(l)}_{V}$, $\varphi^{(l)}_{E}$ & Aggregation function of vertices and links of layer $l$. \\
         $\phi^{(l)}_{V}$, $\phi^{(l)}_{E}$ & Update function of vertices and links of layer $l$. \\
         $\theta^{(l)}$, $\vartheta^{(l)}$ & Sample function and pooling function of layer $l$. \\
         $\psi$ & Readout function. \\
         
         $y$ & Global output vector. \\
         \hline
    \end{tabular}
\end{table}

\textbf{Model Output.}
\label{sec:model-output}
The final output of GI models depends on the way of processing embeddings, i.e., the readout function.
In general, the outcome of GI models can be grouped into three types:
1) Vertex-level output, where the result is vertex-wise predictions (e.g., classes, scores) for some dedicated vertices.
2) Link-level output, where the result is link-wise predictions for some dedicated links.
3) Graph-level output, where the results are the prediction of the whole graph (e.g., the operational status of a power grid).

\begin{figure*}[t]
  \centering
  \includegraphics[width=0.95\linewidth]{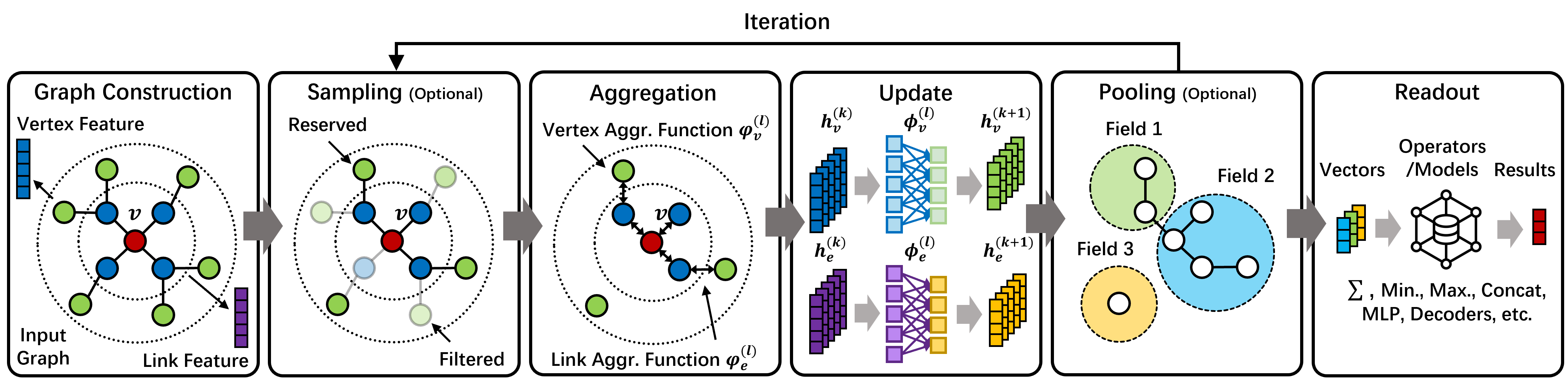}
  \caption{General workflow of GI models. Given an input graph with feature vectors, a GI model iteratively performs sampling, aggregation, update, and pooling through consecutive model layers. The obtained embeddings will be finally converted to results in an expected form through a readout function.
  }
  \label{fig:gnn-workflow}
\end{figure*}

\subsubsection{General Workflow}

A GI model is an algorithm that essentially leverages graph topology to abstract and learn the relationships between vertices and links.
It takes an attributed graph as input, and output embeddings or predictions in an application-admitted format.
Among versatile GI models, the GNN series is the state-of-the-art genre and is prevalent in various types of edge applications, thus we illustrate the general workflow of GI model based on GNN models.
Fig. \ref{fig:gnn-workflow} depicts the general workflow of them.

\textbf{Preprocessing.}
The first step serves as an initialization to prepare data in a format aligned with the targeted GI model's requirement.
This can be, for example, reorganizing the adjacency matrix in a dense format or the compressed sparse row format and dropout some irrelevant elements from the feature vectors.
Since the GI model is often known ahead of runtime, graph preprocessing is usually done offline.

\textbf{Sampling.}
With the preprocessed input graph, the GI model dives into iterations.
The number of iterations required is exactly the number of layers the GI model possesses.
Within each iteration, it first applies sampling on the input graph $\mathcal{G}$ to reduce the computational complexity of subsequent steps.
Assuming the sampling function as $\theta^{(l)}$, this step can be formally written in:
\begin{align}
    \mathcal{G}' = \theta^{(l)}(\mathcal{G}).
\end{align}
The result of sampling is a sampled graph $\mathcal{G}'$, where $\mathcal{G}' = \langle \mathcal{V}', \mathcal{E}' \rangle$.
Note that this is an optional step and inference processes typically deactivate this step for prediction accuracy.

\textbf{Aggregation.}
Upon the sampled graph $\mathcal{G}'$, the GI model performs neighbor aggregation where each vertex/link pulls feature vectors from its neighbors.
Taking vertices aggregation as example, let $\varphi^{(l)}_{V}$ be the aggregation function of vertices, we have
\begin{align}
    a^{(l)}_v = \varphi^{(l)}_{V}(h^{(l)}_{v}, h^{(l)}_{u}), \ \forall v \in \mathcal{V}', \ \forall u\in\mathcal{N}(v),  \label{eq:aggregation}
\end{align}
where $\{h^{(l)}_{u}|u\in\mathcal{N}(v) \}$ collects $v$'s neighboring vertices' representation vectors and $a^{(l)}_v$ is the aggregation result.

\textbf{Update.}
The GI model then passes the aggregation $h^{(l)}_{v}$ through a neural network operator $\phi^{(l)}_{V}$ to update $v$'s representation vector:
\begin{align}
    h^{(l+1)}_{v} = \phi^{(l)}_{V}(h^{(l)}_{v}), \ \forall v \in \mathcal{G}', \label{eq:update}
\end{align}
This operator is usually learnable  Multi-Layer Perceptron (MLP) and non-linear activations like the sigmoid function.
Note that Eq. \eqref{eq:aggregation} and Eq. \eqref{eq:update} are merely described for vertices for simplicity.
Repeating the same procedure to all vertices and links yields the complete representation vectors for the whole graph.

\textbf{Pooling.}
Pooling is also an optional step that aims at reducing the original graph to a smaller graph for lower computational complexity.
It directly operates the sampled graph with updated representation vectors, pooling fields of graphs:
\begin{align}
    \bar{\mathcal{G}} = \vartheta^{(l)}(\mathcal{G}').
\end{align}

\textbf{Readout.}
The above sampling, aggregation, update, and pooling steps will iterate until all model layers are processed, and thereafter generate embeddings of all vertices and links.
To attain the desired results demanded by applications, these embeddings are obliged to the final readout step, which applies a pre-defined operator or model to transform embeddings to a global output, as explained in Sec. \ref{sec:model-output}.
Given a readout function $\psi$, the global output vector can be obtained by:
\begin{align}
    y = \psi(h^{(L)}_{v}, h^{(L)}_{e}|v,e\in\mathcal{G}).
\end{align}

In summary, executing a GI model can be regarded as processing a collection of operators and neural networks iteratively over a graph, where each iteration, i.e., each model layer, comprises weights that specify the computation of vertices' and links' feature vectors.
These weights are learnable via model training, i.e., through the means of backpropagation algorithms such as gradient-based optimization.
Specifically, during model training, a GI model with $L$ layers undergoes a forward pass: first transforms input graph through model layers to graph embeddings, and next converts the obtained embeddings into desired results.
With the exported result and known labels, the model computes the pre-defined loss function, where the gradient is then backpropagated across the layers, updating the shared weights.
This process is carried out iteratively with multiple samples, which are often in batches until an expected accuracy is attained.
For model inference, it directly goes through a forward pass and generates the predictions.


\subsubsection{Graph Learning Models}

There are multifarious GI model variants developed for multifarious applications with multifarious ability requirements.
For brevity, here we enumerate several representatives that are commonly adopted in edge scenarios.

\textbf{Recurrent Graph Neural Network (RecGNN)} is a pioneering architecture that builds the conceptual foundation in the field of graph representation learning \cite{wu2020comprehensive}.
They are primely proposed to learn node representations using recurrent neural architectures, integrating a recurrent hidden state and graph signal processing (GSP) to exploit the spatial structural information inherent in graph processes.
As a genre of GI models dating back to the ``pre-deep-learning" era, RecGNNs have inspired numerous subsequent research such as convolutional variants.

\textbf{Convolutional Graph Neural Network (ConvGNN)} extend convolution operations from grid data to graph data, aggregating the features of vertices and links with their neighbors to generate representations \cite{zhou2020graph, ward2022practical}.
Contrary to RecGNNs, successive graph convolutional layers are stacked in ConvGNNs to extract hierarchical patterns from subgraphs \cite{kipf2016semi}.
Among GNNs, ConvGNNs serve as a foundational component in constructing various advanced GI models \cite{hamilton2017inductive}.

\textbf{Graph Attention Network (GAT)} inherit spatial ConvGNNs by incorporating the attention mechanism into the aggregation functions, which effectively improves the capacity as well as the expressiveness of GI models \cite{velivckovic2017graph}.
The rationale behind this combination is to differentiate the contribution of vertices' neighbors in a learning manner \cite{sun2023attention}.
Based on this, many types of attention mechanisms are derived such as self-attention, gating attention, and semantic-level attention.

\textbf{Graph Autoencoder (GAE)} are unsupervised frameworks that encode vertices, links, or graphs into a latent vector space and reconstruct graph data by decoding their encoded information \cite{zhang2020deep}.
GAEs are particularly useful for graph generation tasks, where a GAE model employs graph convolutional layers to compute embeddings for vertices and rebuild the graph adjacency matrix via decoders.
Variational GAE goes beyond traditional GAE by transforming vertex embeddings as distribution, where each node's embedding is represented by a mean and variance, allowing the model to capture uncertainties by sampling from these distributions \cite{kipf2016variational}.

\textbf{Spatio-Temporal Graph Neural Network (STGNN)} analyze dynamic graphs from both the spatial and temporal dimensions \cite{jin2023spatio, sahili2023spatio}.
Each vertex and link in these graphs attaches a feature vector that describes their behaviors within a time window, and STGNNs aim to learn their patterns and predict their changes in the incoming time slots.
To extract information from spatial-temporal dependencies, many sequential decision-making approaches are applied, e.g., Long Short-Term Memory (LSTM).

\textbf{Graph Transformer and Graph Foundation Model.}
Graph transformers explore embracing transformer architecture to the graph domain in pursuit of improved graph modeling ability \cite{yun2019graph,hu2020heterogeneous, rampavsek2022recipe}.
Typically, graph transformers incorporate GNN with transformers in dual ways:
1) design tailored positional embedding modules and graph-specific attention matrices, and
2) exploit GNNs as an auxiliary module by combining GNNs into transformer architectures \cite{min2022transformer}.
Following this line, scaling small transformers to huge foundation models leads to Graph Foundation Models (GFMs) \cite{liu2023towards, liu2023towards} and Large Graph Models (LGMs) \cite{zhang2023large}. 
Motivated by the success of Large Language Models (LLMs), many researchers believe that utilizing pretraining techniques to resolve massive graph data can breed a comprehensive GI model, which possesses advanced abilities such as in-context graph understanding and versatile graph reasoning \cite{xia2024opengraph}.
Nevertheless, GFMs and LGMs are still in a very early stage of their development and demand for further exploration.

\subsubsection{Learning Paradigms}

Given a rich zoo of GI models, distinct learning paradigms empower them with distinct abilities for edge applications.
We provide several example learning paradigms as follows.

\textbf{Supervised Learning} is one of the most fundamental training paradigms in ML, which requires labeled data to guide the optimization of models \cite{chong2020graph, song2022graph}.
For GNNs, supervised learning enables to capture both local and global graph structures and the latent information of vertices and links.
It iteratively optimizes model parameters by minimizing the difference between the model's predictions and the labels, where the performance of the model is evaluated on a separate test set of labeled objects.
Supervised GL has found success in various domains such as social network analysis, intelligent transportation, and recommendation systems.
However, in many real-world scenarios where labels are unavailable or expensive, supervised learning is constrained and other semi-supervised or unsupervised learning paradigms are introduced.

\textbf{Transfer Learning (TL)} for GI models builds upon the assumption that the learned representations and knowledge from a source graph can be effectively transferred and applied to a targeted graph \cite{zhu2021transfer, han2021adaptive}.
Based on this, it exploits knowledge distillation methods to extract the knowledge from one graph data to improve the learning performance on another different but related graph data.
By transferring knowledge across graphs, the target graph benefits from the learned representations, enabling improved prediction performance for the targeted graph, even with limited labeled data.

\textbf{Contrastive Learning (CL)} for GI models is a self-supervised learning paradigm that learns meaningful representations by contrasting positive and negative samples, where positive samples are pairs of vertices, links, or subgraphs that are similar or related, and negative ones are the contrary \cite{you2020graph, zhu2021graph, zhu2021empirical}.
With these samples, contrastive learning maximizes the similarity between positive samples and minimizes the similarity between negative samples.
Various techniques have been applied to enhance contrastive learning on graphs, such as graph augmentation and sample generation with random walks.

\textbf{Variational Learning} extends the idea of variational inference to GNN, introducing a probabilistic framework to learn node or graph-level representations. This technique is mainly applied in Variational GAEs \cite{kipf2016variational}, where each node's embedding is treated as a latent variable sampled from a learned distribution.
The variational framework enables the model to capture uncertainties in graph data by learning distributions over node embeddings rather than fixed values.
This allows for uncertainty quantification by modeling distributions over node embeddings, improving robustness to noisy data, and preventing overfitting through regularization. This leads to more expressive representations and enables generative tasks like predicting missing links or generating new graph structures \cite{Kingma2014}.

\textbf{Federated Graph Learning (FGL).}
FGL applies Federated Learning (FL) on graphs, which allows multiple clients to collaboratively train a GI model without sharing their local data \cite{chen2024fedgl, chen2021fedgraph, wang2022federatedscope}.
In particular, each client trains the GI model using its local graph data and shares only the model updates, i.e., gradients or weights, with a central coordinator, and the coordinator aggregates the updates and sends a merged model update back to each client.
Graph data are typically distributed across clients, with respect to structural segments, i.e., subgraphs, or feature segments.
More details on FGL are discussed in Sec. \ref{sec:fgl}.

\textbf{Deep Graph Reinforcement Learning (DGRL).}
DGRL combines GI models with Deep Reinforcement Learning (DRL) techniques for interacting with graph environments \cite{munikoti2023challenges, nie2023reinforcement}.
Typically, in DGRL, the GI model is responsible for processing the graph data, extracting features, and capturing the relationships between vertices and links. The DRL component then uses the embeddings computed by the GI model to learn a policy and make decisions, often by taking actions like vertex selection, link insertion/removal, or graph modification.
Benefiting from the superior capability of representing graphs, DGRL has become a powerful tool for graph-based sequential decision-making tasks.

\subsection{Edge Networks and Edge computing}
\label{sec:fundamental-edge}

\subsubsection{Basic Concepts}

\begin{figure}[t]
  \centering
  \includegraphics[width=0.95\linewidth]{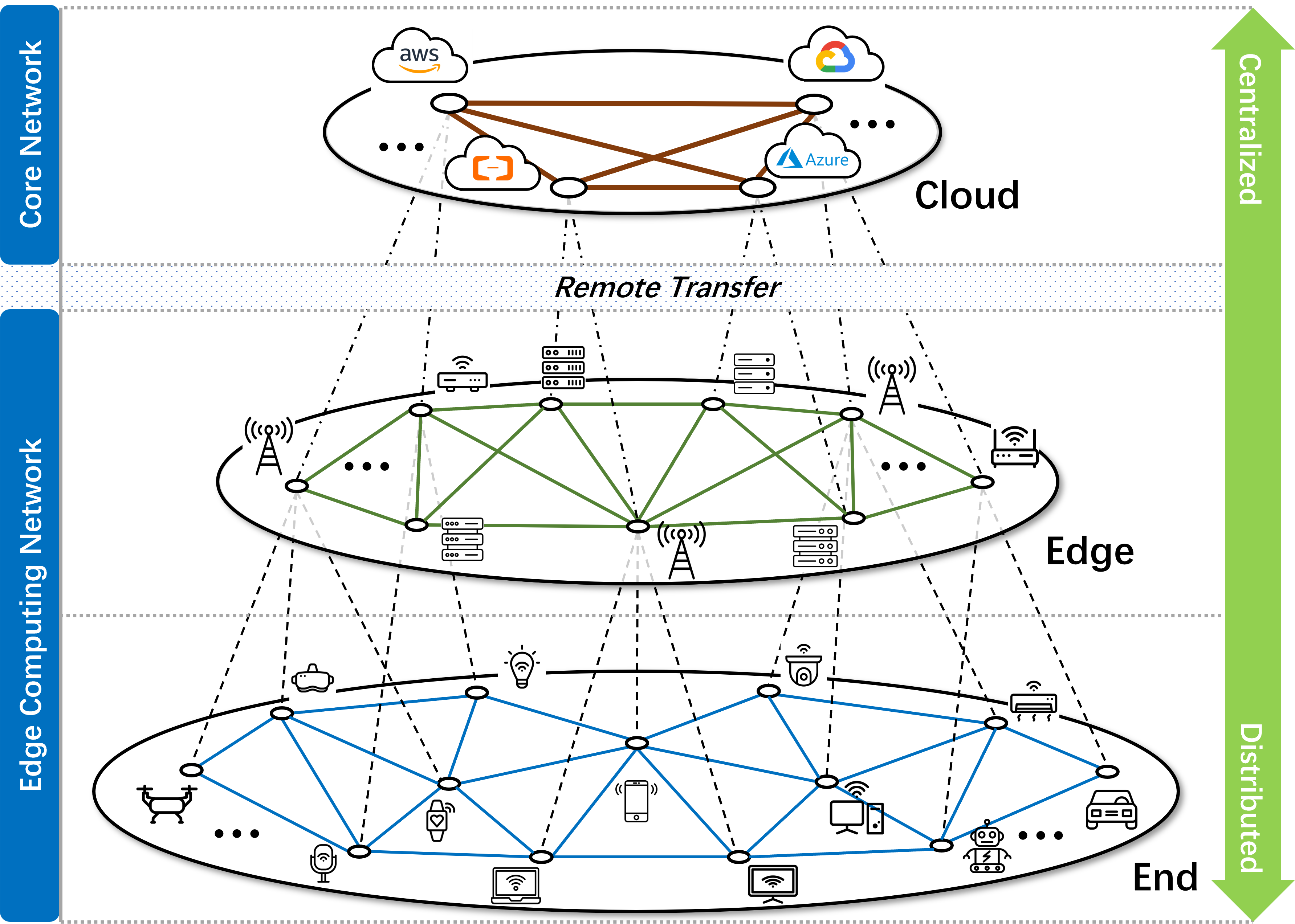}
  \caption{Architecture overview of cloud-edge-end hierarchy, where distributed edge devices within edge networks serve as infrastructure for graph-intelligent applications and their networked data can be analyzed by graph representation learning models.
  }
  \label{fig:edge-network}
\end{figure}

Edge networks have emerged as a pivotal architecture, facilitating the transition from centralized cloud-based core networks to the much more decentralized edge \cite{wang2017survey, pooyandeh2021edge, yang2019multi}. 
Fig. \ref{fig:edge-network} illustrates the edge network as a cohesive architecture that integrates the \textit{Edge} and \textit{End} layers, encompassing devices naturally situated closer to end users.
Besides traditional network-style topology, edge networks can exhibit versatile organizations such as peer-to-peer connection (e.g., client-server mode) and star-like interactions (e.g., one-to-many subscription).

Specifically, in contrast to the centralized core networks, several salient features distinguish edge networks: (1) \textit{Distributed}: Edge networks are characterized by their geographically dispersed resources, such as edge servers and mobile devices, in stark contrast to the centralized nature of cloud-based data centers. (2) \textit{Heterogeneity}: Edge networks encompass a diverse array of edge devices and servers, each varying in computational capacity, network bandwidth, and hardware architecture, embodying an extremely heterogeneous computing environment. (3) \textit{Resource Constraints}: Unlike cloud-based data centers equipped with high-performance accelerators and dedicated ultra-speed links, devices within edge networks often operate under resource limitations, facing constraints in aspects like computational throughput, device-to-device communication bandwidth, and memory capacity. (4) \textit{Dynamic Resources}: The proximity of edge network devices to end users inherently results in greater dynamism in resource availability. Their mobility across various networking domains and multitasking with multiple applications simultaneously contribute to frequent fluctuations in both network and computational resources.

Emerging from the advanced architecture of edge networks, edge computing stands as a new computing paradigm in contrast to cloud computing \cite{zhou2019edge, wang2020convergence}. Edge computing represents a shift in computational paradigm that brings data processing exponentially closer to the point of data collection and consumption, enabling low-latency and high-bandwidth communication essential for real-time applications \cite{zhang2024large, wang2024socialized}. 
With lower reliance on the core network, edge computing not only alleviates both the stress of centralized cloud and backhaul bandwidth usage but also enhances privacy preservation through localized data processing.


\subsubsection{Components of Edge Networks}
\label{sec:spectrum_edge_network}

\begin{figure}[t]
  \centering
  \includegraphics[width=0.95\linewidth]{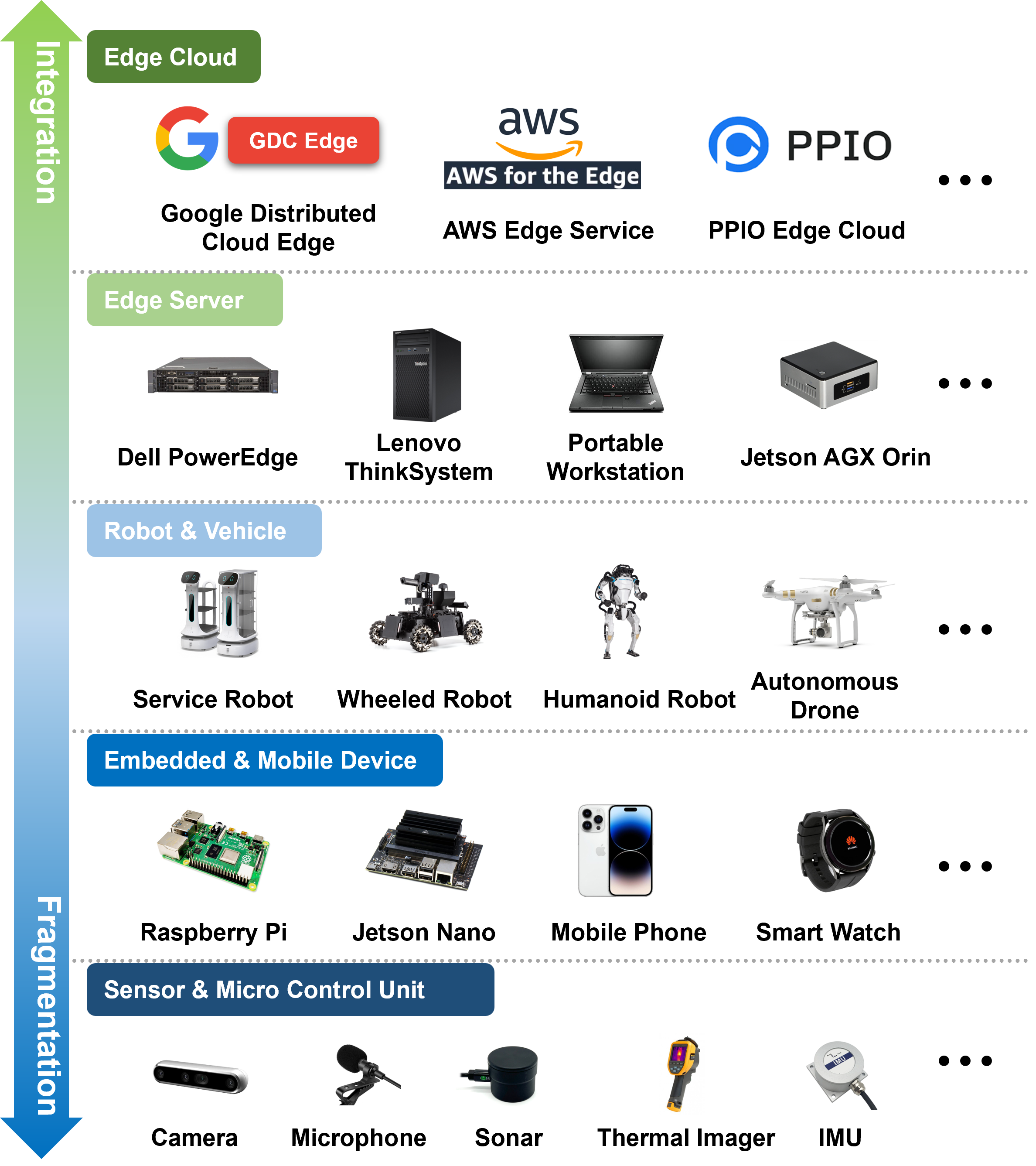}
  \caption{
  The spectrum of edge networks can be classified into five categories: sensors and micro control units, embedded and mobile devices, robotics and Vehicles, edge servers, and edge cloud.
  } 
  \label{fig:edge-range}
\end{figure}

Locating at the periphery of the Internet, edge networks cover a wide spectrum of diverse and heterogeneous platforms.
From fragmentation to integration, we enumerate example components in edge networks in five levels as in Fig. \ref{fig:edge-range}.

\textbf{Sensors} are devices that capture real-time data from the physical environment, such as image, temperature, motion, and gas composition. 
This type of device is the most scattered IoT and is thus located at the bottom level in Fig. \ref{fig:edge-range}, where several example sensors as shown.

\textbf{Embedded and Mobile Devices} are advanced systems that integrate multiple sensors and MCUs into a single, often portable, framework. Unlike standalone sensors or MCUs that primarily capture or process data, these devices offer comprehensive functionality, including data processing, analysis, and user interaction.
Embedded devices are specialized computing units designed for specific functions within larger systems.

\textbf{Robots and Vehicles} usually excel in functionality, autonomy, and complexity. While embedded devices are integral to these systems, robotics and vehicles incorporate them into more complex frameworks, designed for higher autonomy and minimal human intervention. 

\textbf{Edge Servers} act as micro data centers to deliver services in a way similar to using cloud servers but with a key difference: they are situated closer to the data source. This proximity not only ensures adherence to data localization laws but also significantly reduces data transfer latency.

\textbf{Edge Cloud} services extend cloud computing's convenience to the network's edge, with providers like AWS and Google offering solutions such as AWS Edge Service \cite{aws_edge} and Google Distributed Cloud Edge \cite{google_cloud_edge}. Edge clouds are usually hosted by micro-data centers comprising edge servers that store, analyze, and process data faster than is possible using a connection to a cloud data center.

\subsubsection{Software Frameworks for Edge Networks}

Software frameworks for edge computing across edge networks are often cross-platform, cross-protocol, language-agnostic, and resource-efficient. EdgeX Foundry \cite{edgexFoundry} is an open-source edge platform that lets users create IoT gateway functionality from edge devices, which acts as a dual transformation engine sending and receiving data to and from cloud and edge applications. Eclipse Kura \cite{eclipseKura} has taken a similar approach to EdgeX Foundry to provide a platform for developing IoT gateways. Apache Edgent \cite{apacheEdgent} is an open-source stream processing framework for edge computing, enabling developers to process sensor-collected streaming data in real time on edge devices. RedisEdge \cite{redisEdge} is a purpose-built, multi-model database for the demanding conditions at the IoT edge, which can ingest millions of writes per second with less than 1ms latency and a very small footprint (less than 5MB). TensorFlow Lite \cite{tf-lite} and MNN \cite{jiang2020mnn} is an open-source lightweight framework that enables on-device machine learning by helping developers run their AI models on embedded and mobile devices to achieve edge intelligence.

Some software frameworks offer a simulation environment to model and simulate edge network infrastructures and services, providing timely, repeatable, and controllable methods for evaluating the performance of new edge applications and policies prior to actual development. EdgeCloudSim \cite{EdgeCloudSim} provides a simulation environment based on CloudSim \cite{CloudSim} but adds considerable functionality so that it can be efficiently used for edge network scenarios. IFogSim2 \cite{iFogSim} is a toolkit for modeling and simulation of resource management techniques in IoT, edge, and fog computing environments. YAFS \cite{YAFS} is a Python-based simulator tool for architectures like Fog Computing ecosystems, facilitating analyses related to resource placement, deployment costs, and network design.

\subsubsection{Edge Intelligence}
AI stands at the forefront of modern technological advancements, enabling computer systems to perform tasks that typically require human intelligence. This encompasses a wide range of activities including learning, decision making, and problem solving. In recent years, deep learning (DL), a branch of machine learning, has become synonymous with AI's progress. Characterized by its use of multi-layered neural networks, DL enables the processing of complex data, driving innovations in areas such as image recognition, natural language processing, and automated decision-making. The marriage of edge computing and AI catalyzes the birth of \textit{edge intelligence} \cite{zhou2019edge, wang2020convergence}.

In examining the advancements in the field of AI, particularly in the realms of DL model training and inference, it becomes clear that edge intelligence holds distinct advantages over traditional cloud intelligence: (1) \textit{Model Training}: 
In the current AI landscape, there has been a significant shift in data sources from centralized cloud data centers to increasingly ubiquitous edge devices, such as mobile and IoT devices.
These methods require the collection of massive amounts of data in the cloud, exerting immense pressure on core networks due to high data traffic. Additionally, this centralized approach strains data center resources, both in terms of storage and data processing capabilities. Edge intelligence offers a strategic solution to this challenge. By facilitating the processing and training of deep learning models directly at the network edge, closer to where the data is generated \cite{ye2024asteroid, cai2023efficient, xu2022mandheling}. (2) \textit{Model Inference}: Upon completion of model training, these models are deployed to provide inference services to end users. Routing user requests to cloud-based systems for inference poses significant privacy concerns, as it could lead to the leakage of sensitive information, such as personal identifiers and location data. Moreover, relying on cloud-based inference services, which depend on the stability of core networks, may introduce significant propagation delays. This aspect is particularly critical for applications requiring real-time response, such as autonomous driving and medical-surgical robots. Edge intelligence, by processing these inference requests locally, not only enhances data privacy but also ensures low-latency, efficient service delivery, crucial for such time-sensitive applications \cite{ye2024galaxy, jia2022codl}.
EGI can be regarded as a particular genre of edge intelligence concerning GI and is yet in its seed time for exploration and exploitation.
In the following sections, we will discuss EGI in detail from different aspects.

\section{Edge Graph Intelligence as a New Edge AI Paradigm}
\label{sec:egi-insight}

Stemed from edge AI, EGI dives into the fusion of edge networks and GI techniques, serving as a brand-new avenue for the evolution of AI deployment.
In what follows, we discuss the benefits of EGI and provide a rating taxonomy of EGI in six levels.

\subsection{Reciprocal Benefits of Edge Graph Intelligence}

EGI provides reciprocal benefits in the following aspects.

\begin{itemize}
\item \textbf{Edge networks for GI.} 
With the rapid proliferation of mobile and IoT devices, data generated at edge networks have skyrocketed in both quantity and modality (e.g., physical signals, digital audio, and visual content).
As predicted by IDC \cite{rydning2018digitization}, the billions of IoT devices in edge networks are expected to generate over 90 Zettabytes of data in 2025.
This naturally provides rich data nurseries and abundant application scenarios for modifying, training, and fine-tuning GI models with real-world data, thus boosting GI models toward higher-degree intelligence.
Besides, edge networks as user-nearby infrastructures can effectively enhance GI model computing performance, e.g., edge networks assisted with cloud as the back-end can largely reduce the computing latency, thus boosting the GI performance.
\item \textbf{GI for edge networks.} 
Given the rich relational data collected at edge networks, GI enables modern graph analysis for understanding, diagnosing, and optimizing edge networks, which steers the enhancement of network performance such as robustness and Quality of Service (QoS).
Applying GI to edge networks thus unlocks their extended capability in securing edge networks from anomalies, developing new graph-based applications, and intelligently serving graph-related tasks.
\item \textbf{Reciprocal technology advancement.}
The combination of GI and edge computing not only carries out mutual empowerment to improve each other (as mentioned above) but also enhances the advancement of both research and technology.
For instance, EGI catalyzes a number of new edge applications (Sec. \ref{sec:gnn-application-edge}) and incubates an ecosystem including GNN-oriented hardware accelerators and programming frameworks (Sec. \ref{sec:edge-infra-gnn}), which actually expedites edge AI popularization.
Additionally, and perhaps more notably, new technologies and system models for edge AI are being nourished and flourishing, extending beyond the traditional scopes and standards of these two.
\end{itemize}

\begin{figure*}[t]
  \centering
  \includegraphics[width=0.85\linewidth]{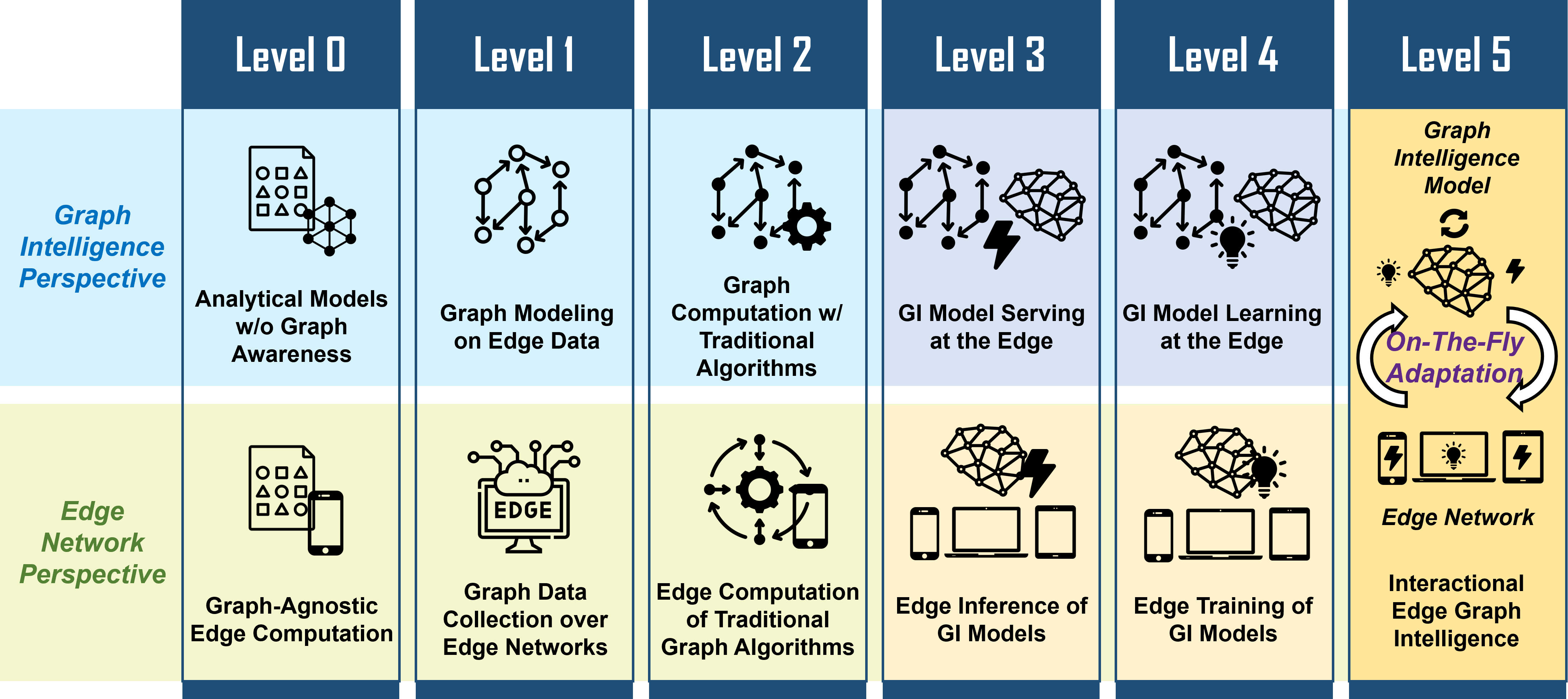}
  \caption{Rating of EGI from the perspectives of both graph intelligence and edge network. In particular, EGI in Level 5 coalesces both lines and renders interactional EGI that can dynamically adapt the GI model for optimal model performance and the edge network for optimal runtime performance on the fly.
  }
  \label{fig:egi-grade}
\end{figure*}

\subsection{Rating of Edge Graph Intelligence}

Given the reciprocal benefits of EGI, we advocate that EGI should not be restricted to merely applying GI on edge data or running GI on edge platforms.
Instead, GI and edge networks are blending a confluence and EGI should be treated as a whole to reflect the inherent interplay between GI and edge networks. 
This indicates that their bilateral empowerment desires a comprehensive exploration such that the degree of EGI can be identified and measured.
Specifically, according to the fusion of GI and edge networks, we may rate EGI into six levels from their respective perspectives, as shown in Fig. \ref{fig:egi-grade}.

\begin{itemize}
    \item \textbf{Level 0}: Given the graph data implicated in the edge network infrastructure, analytical models in Level 0 are unaware of the graph structures. The edge computing systems also process data in a graph-agnostic way.
    In other words, neither the model side nor the infrastructure side explicitly tackles ``graphs", and thus they are categorized to the initial level.
    For instance, visual data mining with traditional ML models and conventional neural networks like CNN \cite{krizhevsky2012imagenet, he2016deep} and RNN \cite{greff2016lstm, yu2019review} can be categorized in Level 0 since their data are not formulated in graphs and their data processing pipeline has no specific relations with graphs. 
    \item \textbf{Level 1}: Data collected from edge networks are modeled in graphs but are not exploited in analytics.
    Systems at Level 1 push one step further over Level 0 by endowing graph semantic to edge data through graph modeling with collected edge data, but process data with general computing methods (instead of graph-related methods).
    As an example, while wireless sensory data can be generally abstracted in graphs \cite{he2021overview, dong2023graph}, Level-1 systems treat them in a vertex-wise manner, e.g., separately analyzing individual sensors' data arrivals, instead of processing as a whole graph.
    \item \textbf{Level 2}: Edge data in graph formats are processed with traditional graph computing algorithms such as PageRank \cite{brin1998anatomy} and single-source shortest path algorithms \cite{dijkstra1959note}.
    In this respect, edge data are modeled in graphs and processed with traditional graph algorithms rather than ML methods. 
    Systems at Level 2 outperform Level 1 by enabling graph-oriented computing capability though they do not introduce AI techniques in graph processing.
    \item \textbf{Level 3}: Edge networks serve GI model inference with graph data, where the models may be trained on the cloud.
    Compared with lower levels, systems at Level 3 not only allow graph formulation and computing of edge data but also initiate AI to edge networks and embrace GI models such as GCN \cite{kipf2016semi} and GraphSAGE \cite{hamilton2017inductive}.
    In other words, edge systems that compute GNN inference such as GNN-based traffic forecasting systems \cite{li2017diffusion,yu2017spatio} and DGRL decision systems \cite{chen2021multitask,swaminathan2021graphnet} are located at Level 3.
    \item \textbf{Level 4}: Edge networks perform GI model training with graph data.
    Example includes SUGAR \cite{xue2023sugar} and HGNAS \cite{zhou2023hardware}.
    The key difference between Level 4 and Level 3 lies in the ability to learn edge-native GI models, e.g., fine-tuning model parameters with edge data.
    Otherwise stated, systems at Level 4 can customize their GI models in-situ, thereby enabling tailored ability for edge AI services. 
    \item \textbf{Level 5}: Interactional EGI, where GI and edge networks can dynamically adapt their configurations during the runtime for optimal EGI performance.
    Systems at Level 5 outperform all other levels because they can adjust GI and edge networks on the fly, whereas lower levels are all static settings.
    Both perspectives of GI and edge networks reach a convergence since they are in complete harmony.
\end{itemize}

The EGI rating can be mainly divided into three intervals.
The first interval covers from Level 0 to Level 2, where EGI is less related to AI and even processes non-graph data.
The second interval comprises Level 3 and Level 4, where EGI incorporates GI models by either inference or training on edge networks.
The third interval is exactly Level 5, which stands at the highest level because its GI and edge networks have profoundly blended as integration and can adapt to diverse scenarios on the fly. 
As EGI systems are located at higher levels, their fusions of GI and edge networks go deeper.
As a result, the intelligence resources of GI and infrastructural resources of edge networks are progressively exploited for better EGI performance.
Nonetheless, this may also come at the cost of additional development effort and system overhead.
This conflict implies that there is no ``silver bullet" in all cases.
Instead, the panacea of EGI in practice should align with user demand, anticipating a joint consideration of specific application scenarios as well as available resources budgets.
In this survey, given the focus on advanced GNN in edge environments, the reviewed systems mostly lie in Level 3 and beyond. 
\section{Edge Applications of Graph Intelligence}
\label{sec:gnn-application-edge}

\begin{figure*}[t]
  \centering
  \includegraphics[width=0.95\linewidth]{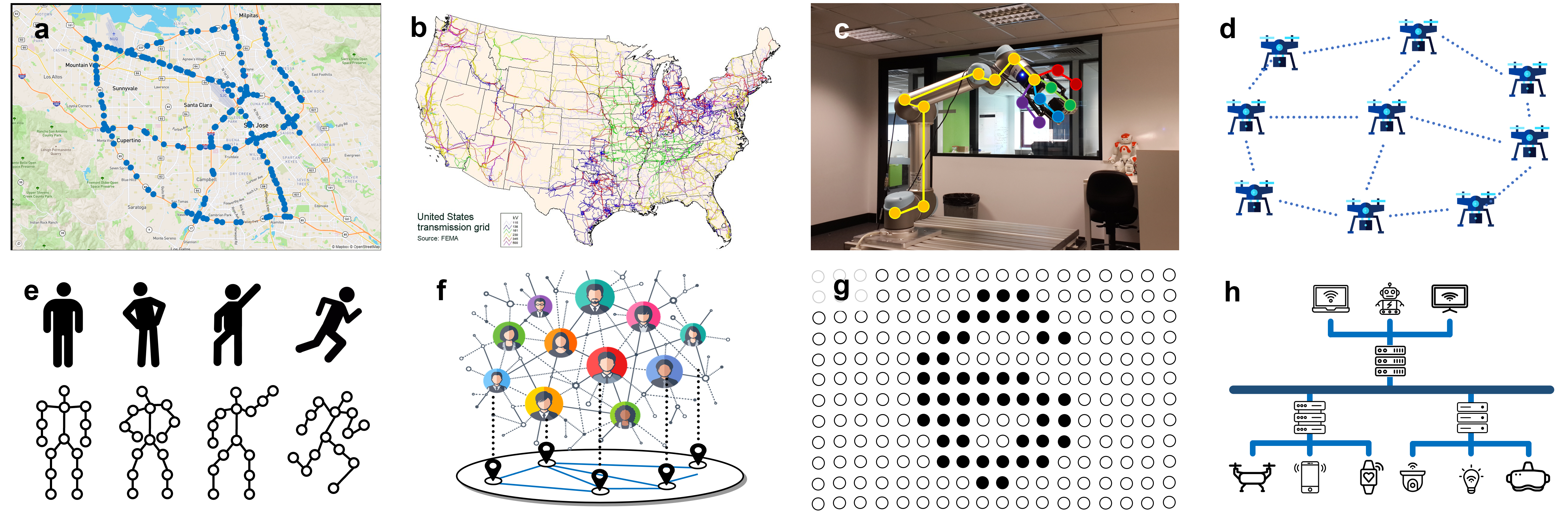}
  \caption{Edge networks host a broad spectrum of platforms including wireless sensors, robots, vehicles, gateways, and cameras. These facilities spawn miscellaneous smart GI applications with their data organized in graphs. (a) A traffic sensory network is a graph with sensors as vertices and links aligned with the road net. (b) A power grid is naturally a graph that connects electrical facilities (image adapted from \cite{US_power_grid}). (c) A robotic arm can be abstracted in a graph with joints as vertices (image adapted from \cite{hosseinzadeh2019structure}). (d) A drone swarm comprises vehicles and their communication relationships shape a graph. (e) Human skeletons can be represented in graphs. (f) Location-based social networks are constructed upon users' fellowship. (g) A graph representation of an 18$\times$12-pixels image of the digit ``6'', where pixels are vertices and their direct adjacency are links. (h) An edge network hierarchy is a tree-like graph. 
  }
  \label{fig:graph-example}
\end{figure*}

Graphs are highly abstract data structures and can be used to represent a spectrum of data generated in edge networks.
These graph data drive miscellaneous applications at the last mile of the Internet, as exemplified in Fig. \ref{fig:graph-example}, and catalyze graph learning as a promising principle for edge intelligence.
Following the insights discussed in Sec. \ref{sec:egi-insight}, we first provide a panorama of representative EGI applications before showing how edge networks and GI interact with each other in subsequent sections.
For clarity, we categorize them into five groups, i.e., smart cities, robots and vehicles, human sensing and analysis, location-based recommendation, and graph-based mobile vision.
Note that edge applications of GI models mostly only compute GI model inference at the edge and these related EGI systems therefore lie at Level 3 in the rating taxonomy.

\subsection{Smart Cities}

Managing large cities is one of the key problems for modernization, especially under the stresses of continuously booming urbanization and the rapidly growing population.
To enable resource-efficient management, AI has been widely recognized as an advantageous way to assist, promoting the concept of ``smart cities'' \cite{anthopoulos2015understanding, shelton2015actually}.
Given the ubiquity of wireless sensor networks and graph data in cities, GI shines among AI techniques and enables numerous smart applications.
Note that many use cases may not be fully deployed on edge networks but involve the cloud as a backend, our review focuses on the processing of graph data collected at the edge and demonstrates how these applications run.
In the following, we discuss representatives concerning transportation, energy, environment, and public safety.

\subsubsection{Intelligent Transportation Systems}
The transportation system spans geographical areas and acts like the blood vessels of a city.
To keep it smooth and unimpeded, GI models are applied for forecasting its states, predicting the supply and demand of users, as well as coordinating its resources.

Traffic state forecasting involves forecasting traffic amount, traffic flow, and travel time between two locations, etc. 
Typically, it takes data from roadside sensors and organizes them in a graph along with the structure of the road net, as visualized in Fig. \ref{fig:graph-example}(a), where the feature vector of each vertex is the collected data of each sensor.
GI models can thereafter be utilized to learn the semantics of traffic networks and export predictions.
For example,
Li et al. \cite{li2017diffusion} introduce the Diffusion Convolutional Recurrent neural network to model traffic flow as a diffusion process on a directed graph, capturing both spatial dependencies through bidirectional random walks and temporal dynamics using an encoder-decoder architecture.   
Recognizing the critical role of spatial and temporal dependencies in traffic flow, Yu et al. \cite{yu2017spatio} present Spatio-Temporal Graph Convolutional Networks, a novel approach for time series prediction in traffic analysis. Instead of applying regular convolutional and recurrent units, STGCN utilizes a fully convolutional structure on graphs, enhancing training efficiency and reducing parameter count.

Supply and demand predictions are a type of trip prediction task though they are often viewed as an ensemble, namely the origin-destination demand prediction.
This can be relevant in many ride-hailing services (e.g., Uber \cite{uber}, DiDi \cite{didi}, and Lyft \cite{lyft}), where accurately predicting users' demand is crucial for dispatch orders concerning efficiency and profit.
Building on the importance of capturing dynamic dependencies in traffic flow, Lu et al. \cite{lu2021dual} developed a dual attentive GNN that can effectively predict the distribution of metro passenger flow considering the spatial and temporal influences.
Further expanding on the application of attention mechanisms in GNNs, Makhdomi et al. \cite{makhdomi2023gnn} introduce a GAT method for passenger origin-destination flow prediction that leverages diverse linear and non-linear dependencies among requests originating from distinct locations, capturing both the repetition pattern and contextual data of each location.
Their superior performance demonstrates the capability to capture inherent connections between regions or origin-destination pairs, and some of them \cite{wang2021gallat, lu2021dual} have been deployed in the wild.

Traffic resource management is to detect or schedule dedicated components in transportation systems.
As an example, Wang et al. \cite{wang2022contrastive} and Yu et al. \cite{yu2021deep} both introduce GI-based approaches aimed at addressing critical issues in intelligent transportation systems. These frameworks capture complex dependencies within their respective applications, showcasing the versatility of GNNs in this field.
However, their focuses differ significantly. Wang et al. \cite{wang2022contrastive} specifically design their GI model for traffic anomaly analysis in IoT-based intelligent transportation systems, aiming to identify and analyze traffic anomalies, while Yu et al. \cite{yu2021deep} focus on the operational problem of relocating and repositioning idle vehicles, a crucial aspect of traffic resource management.

\subsubsection{Power Grid}

As basic facilities for smart cities, power grids deliver electricity from power plants to end users through transmission and distribution lines, substations, and equipment.
To serve as many people in the city as possible, modern power grids have evolved to a colossal scale, becoming an extremely complex system that is vulnerable to attacks and system failure.
Given the graph nature of power grids, GI models are employed to ensure their reliability and stability, i.e., for failure detection and energy harvesting prediction.

Failure detection aims at localizing potential or yet-occurring failures within the power grid.
Liao et al. \cite{liao2020fault} utilize the adjacency matrix to represent the similarity between unknown and labeled samples, and propose graph convolutional layers for identifying complex nonlinear relationships between dissolved gas and fault type. 
Building on the use of GCNs in power system diagnostics, Liu et al. \cite{liu2021searching} further present a technique that employs GCNs for swiftly identifying key failure points in power systems. This method simplifies and speeds up the process of detecting critical cascading failures, making it highly effective for complex power networks.
Extending GI to the realm of cybersecurity in power grids, Boyaci et al. \cite{boyaci2022cyberattack} introduce a deep learning model utilizing Chebyshev GCNs for detecting cyberattacks, specifically false data injection attacks, in large-scale AC power grids.

Energy harvesting prediction is for power grids with vulnerable energy input, e.g., solar energy and wind energy.
Accurately predicting their future behavior is of great importance for power systems with high penetration of RESs for stable operation and economic consideration.
Karimi et al. \cite{karimi2021spatiotemporal} focus on improving photovoltaic power forecasting by leveraging spatial and temporal coherence among power plants.
They propose an STGNN to harness the relationship between power plants, observing that plants in a region experience similar environmental conditions and can thus inform each other's power forecasts.
Khodayar et al. \cite{khodayar2019convolutional} introduce a convolutional GAE for capturing continuous probability densities on graph nodes.
By integrating spectral graph convolutions and variational Bayesian inference, convolutional GAE generates samples from these densities. This approach is applied to probabilistic solar irradiance prediction, using an undirected graph model of solar radiation measurement sites for future irradiance estimation.
In their subsequent research, Khodayar et al. \cite{khodayar2020spatiotemporal} further tackle the challenge of forecasting behind-the-meter load and rooftop photovoltaic generation in power systems using a spatio-temporal GAE and graph dictionary learning optimization.

\subsubsection{Environment Monitoring}

Environmental science assumes a crucial role in comprehending the dynamics of natural systems and their complex interactions with human activities, particularly for smart cities.
DL, as a powerful tool, has emerged to support environmental science in smart cities, enabling the analysis of vast amounts of data to model and predict environmental phenomena such as air quality concentration.
In this context, GI techniques have found extensive applications in the study of wireless sensor networks for environmental monitoring in the realm of smart cities, with respect to environmental assessment and prediction as well as meteorological monitoring and forecasting.

Environmental assessment and prediction aims at extracting environmental information from sensory data collected across smart cities.
Chen et al. \cite{chen2023group} introduce the group-aware GNN (GAGNN) for nationwide air quality forecasting, in order to understand latent dependencies among geographically distant cities.
GAGNN constructs a hierarchical model with a city graph and city group graph, combined with a differentiable grouping network to discover and encode these complex inter-city relationships.
In addition to the use of GCNs, both Gao et al. \cite{gao2021graph} and Mao et al. \cite{mao2021hybrid} incorporate LSTM to integrate spatio-temporal information for more accurate predictions.
For example, Mao et al. \cite{mao2021hybrid} develop a graph convolutional temporal sliding LSTM model to predict various air pollutants. They use GCNs to model spatial dependencies and a temporal sliding LSTM strategy to capture dynamic changes over time, allowing for a broader application beyond just PM2.5 prediction.

Meteorological monitoring and forecasting leverage GI models to analyze meteorological data for understanding current weather conditions and predicting future weather patterns.
Remote automatic weather stations can collect geo-distributed meteorological data for temperature prediction.
The targeted sensory data are usually in a spatio-temporal form, covering atmospheric parameters such as temperature, humidity, pressure, wind speed and direction, precipitation, and other relevant variables.
As an example, Lin et al. \cite{lin2022conditional} present the Conditional Local Convolution Recurrent Network for spatio-temporal forecasting in weather prediction, overcoming challenges of high nonlinearity and complex spatial patterns. It introduces a GCN that captures local spatial patterns through conditional local convolutions and incorporates these into an RNN architecture to model temporal dynamics.
Addressing the challenge of accurately forecasting frost events, which significantly impact agriculture, Lira et al. \cite{lira2021frost} describe a GNN with spatio-temporal attention architecture. The model, adept at handling the localized nature of frost influenced by various environmental factors, maps weather station data onto a graph structure, optimizing an adjacency matrix during training to capture complex environmental interactions.
Similarly focusing on the spatial relationships among weather stations, Stanczyk et al. \cite{stanczyk2021deep} introduce GI models for wind speed prediction, focusing on handling data from multiple weather stations.

\subsubsection{Public Safety}
Ensuring residents' physical safety and security is of primary importance in building smart cities.
Towards that, GI techniques are employed to analyze graph data collected from city-wide edge networks and to find patterns related to residents' health.
We next discuss them in three respects, namely natural disaster prevention, crime warning, and public health.

Natural disaster prevention utilizes GI to aid people in mitigating the impact of unexpected extreme hazards, including earthquakes, wildfires, floods, and hurricanes.
Bilal et al. \cite{li2023graph} presents a batch normalization GCN for early earthquake detection, which integrates a CNN with a GNN.
The model employs batch normalization to reduce training epochs and stabilize activation value distributions, enhancing training efficiency and prediction accuracy for key earthquake parameters like magnitude, depth, and location. This approach of integrating CNNs with GCNs for efficient feature extraction and spatial relationship modeling is not limited to seismology. 
Similarly, Jin et al. \cite{jin2020ufsp} tackle the challenge of predicting urban fire dynamics, which is crucial for urban safety and emergency response. They propose a model that combines CNNs with GCNs, where CNNs extract pixel-level latent representations from fire situation awareness images and GCNs manage the spatial relationships between different urban areas. By integrating the strengths of CNNs and GCNs, \cite{jin2020ufsp} effectively illustrates how this combined approach can address complex spatio-temporal phenomena, highlighting its versatility and potential across different domains.

Urban crime is another critical factor imperiling residents' properties and lives.
To raise the attention of people potentially in criminal danger, researchers develop GI models to predict crime frequency in dedicated regions.
Xia et al. \cite{xia2022spatial} present the Spatial-Temporal Sequential Hypergraph network (ST-SHN), aimed at improving crime prediction by addressing the dynamic nature of criminal patterns in both spatial and temporal domains, and the evolving dependencies between different crime types. ST-SHN employs a graph-structured message-passing architecture combined with hypergraph learning to manage the spatial-temporal dynamics and global context of crime.
In parallel, Sun et al. \cite{sun2022spatial} present AGI-STAN, a framework that improves crime prediction by discarding reliance on domain-specific knowledge and predefined graphs. It uses adaptive graph learning to autonomously discern interdependencies among urban communities and integrates a time-aware self-attention mechanism to model temporally varying crime incidents.
Toward the same problem, these two methods take different optimization paths: ST-SHN focuses on global context through hypergraph learning, while AGI-STAN discovers community interdependencies and processes temporal information adaptively.

Epidemic outbreaks, notably the recent novel coronavirus, pose significant threats to global public health systems. Accurate epidemic prediction is crucial, as it plays a vital role in safeguarding public health by informing and guiding proactive strategies to mitigate the impact of such health crises. In this context, GI has emerged as a powerful tool for modeling the complex interactions inherent in epidemic data. Both Kapoor et al. \cite{kapoor2020examining} and Keicher et al. \cite{keicher2023multimodal} leverage the expressiveness power of GNNs to address different facets of COVID-19 prediction, underscoring the versatility and efficacy of graph-based modeling in this domain. While both studies employ GNNs to harness the complexity of COVID-19 data, they differ in their specific applications and data integration strategies. Kapoor et al. focus on macro-level predictions of case numbers across regions, emphasizing the epidemiological aspect of the disease spread. In contrast, Keicher et al. concentrate on micro-level patient outcomes within a clinical setting, emphasizing the individual variability in disease progression and response to treatment.

\subsection{Intelligent Robots and Vehicles}

Robots and vehicles are typical facilities deployed in edge networks and have undertaken multifaceted services for human beings \cite{huang2022edge}.
Graphs are also widely used to represent their behaviors, thereby promoting GI as a promising tool to empower more autonomous machine intelligence.
Regarding the objects that are represented in graphs, we discuss GI for robots and vehicles in three respects, i.e., body control and manipulation, motion prediction and planning, and environment exploration.

\subsubsection{Body Control and Manipulation}
Controlling the body of a robot is an initial step to steer its work, which requires an accurate modeling of it.
However, body modeling is non-trivial, since it usually interacts with objects, environments, and human beings.
For instance, when directing a robotic arm to grasp a compliant object, e.g., a soft bread, it very likely suffers a non-linear deformation, depending on the force and torques made by the arm.
To address that, the robots are formulated in graphs with the spatial and kinetic relationships between keypoints, as depicted in Fig. \ref{fig:graph-example}(c), and GI models are applied to abstract them.
Almeida et al. \cite{almeida2021sensorimotor}  address the complex task of modeling soft robots using a GNN to simulate a non-rigid kinematic chain, like a robotic soft hand.

With a model of the robot, we may envisage to command it to firmly grasp or manipulate some objects.
This requires, beyond body modeling, to characterize the interaction between robots, objects, and environments.
In this context, Li et al. \cite{li2018learning} make a significant contribution by introducing dynamic particle interaction networks (DPINets), a differentiable, particle-based simulator. This innovative approach leverages dynamic interaction graphs to adeptly handle the manipulation of a variety of materials, including fluids, deformables, and rigid bodies.
Building upon the foundation laid by the DPINets' focus on material interaction, Xie et al. \cite{xie2020deep} present a deep imitation learning framework for bimanual robotic manipulation in continuous state-action spaces, focusing on the challenge of generalizing manipulation skills to objects in varying locations.
Further extending the application of graph-based learning in robotics, Liang et al. \cite{liang2021learning} introduce a self-supervised learning method, target-oriented Deep Q-Network (DQN), which employs visual affordance graphs for a complex object grasping task, guiding robot actions using environmental cues in both simulated and real-world scenarios. These three works collectively exemplify the power of graph-based modeling in advancing robotic manipulation capabilities

\subsubsection{Motion Prediction and Planning}
\begin{figure}[t]
  \centering
  \includegraphics[width=0.85\linewidth]{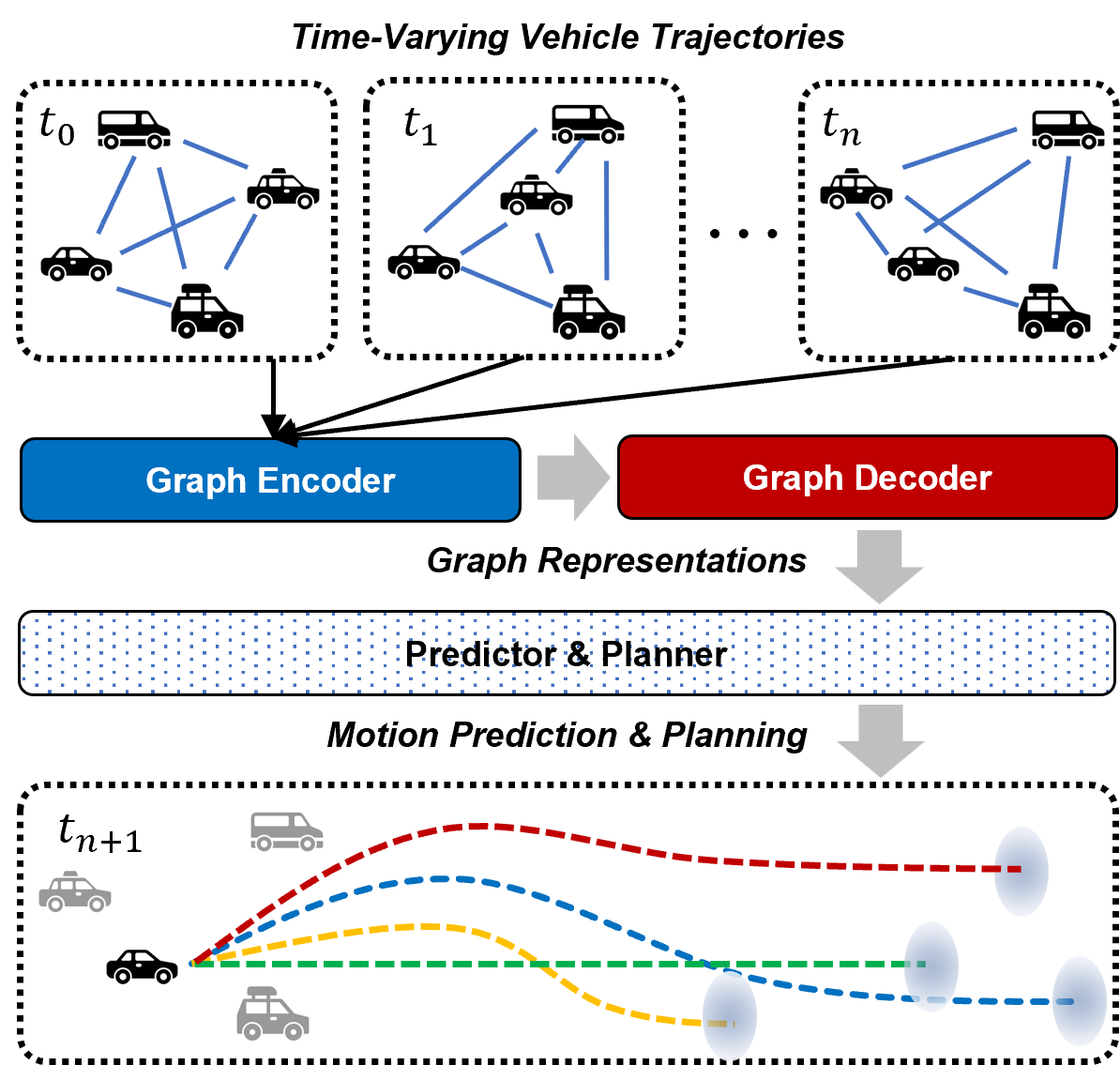}
  \caption{
  An illustration of vehicle trajectory planning via GI models, where the historical trajectories of vehicles are modeled as a time-varying graph.
  }
  \label{fig:vehicle-planing}
\end{figure}

In many circumstances, vehicles are not launched individually but collaboratively, which requires global planning of vehicles to attain mutual goals \cite{chen2022adadrone}.
For these cases, GI is also profitable with vehicles modeled in graphs.
For instance, Fig. \ref{fig:graph-example}(d) shows a drone swarm committed to collision-freely flying from a starting point to a targeted location, where they can be viewed as a graph with links indicating their communication paths.
Another illustration is shown in Fig. \ref{fig:vehicle-planing}, where the movement of vehicles is modeled in a time-varying graph and fed into a GAE for learning graph representations.
Through a predictor and a planner, the EGI system can export the prediction and planning of the vehicle's next motion.
Li et al. \cite{li2020graph} tackle the challenge of decentralized multi-robot path planning by introducing a model that synthesizes local communication and decision-making policies for robots in constrained workspaces. The model combines a CNN for extracting features from local observations and a GNN for efficient communication among robots.
A similar structure is exploited by \cite{li2023dynamic}, which presents the dynamic motion planning model based on graph neural networks and historical information to enhance path-finding in decentralized multi-robot systems.

\subsubsection{Environment Exploration}

Contrary to motion planning where destinations are known, multi-robot exploration tasks demand exploring an unknown environment a key aspect of automation applications like cleaning, searching, and rescue.
For these tasks, scheduling robot swarms to efficiently cover the underexplored field while avoiding conflicts is vital \cite{zeng2023a3d, moller2021survey}.

To achieve that, researchers define the spatial relationships between robots and regions of the environment and intend to apply GI to regulate this complex process.
Luo et al. \cite{luo2019multi} were among the pioneers in this domain, introducing a graph-based DRL approach for autonomous exploration in unknown territories. By translating the exploration task into a graph domain through a hierarchical map segmentation, they paved the way for utilizing a GCN to effectively assign exploration targets to agents. This seminal work highlighted the potential of GNNs in enhancing multi-agent exploration efficiency. Building upon this, Gosrich et al. \cite{gosrich2022coverage} advanced the application of GNNs by developing a decentralized control policy for multi-robot sensor coverage. Their innovative approach allowed robots with limited sensing capabilities to efficiently detect events in regions of varying importance.
In parallel, Zhang et al. \cite{zhang2022h2gnn} have taken the application of GNNs a step further with the introduction of Hierarchical-Hops Graph Neural Networks (H2GNN) for multi-robot coarse-to-fine exploration. Their approach stands out by enabling selective integration of key environmental information through a multi-head attention mechanism, which discerns the importance of information from different hops.

Together, these studies provide a comprehensive view of how GNNs can be leveraged to address a spectrum of challenges in multi-agent robotics, setting the stage for future advancements in the field.

\subsection{Human Sensing and Analysis}

Human sensing is one of the core applications in many edge services, which use edge devices to collect and process data related to human activities and behaviors.
Various platforms are involved, including sensors, smartphones, wearables, cameras, etc., to monitor various aspects of human behavior such as movement, gestures, vital signs, facial expressions, and environmental interactions.
GI models are employed in this context for those data that can be formulated in graphs.
In general, these human sensing applications are mainly recognized in three types: affective computing with facial graphs, action recognition with human skeletons, and smart health with human-centric sensor networks.

\subsubsection{Affective Computing}
Facial expressions contribute to more than 55\% of the information perceived by individuals during verbal communication \cite{mehrabian1971silent, mehrabian1974approach}, which plays a crucial role in conveying significant information related to emotional states and reactions within human interactions.
To enable user-centric services deployed at the edge, the community has developed various intelligent affective computing methods to understand users' facial affects.
Notably, some researchers establish graph representations for faces, where facial landmarks are marked as vertices and are connected in facial shapes, and introduce GI models to reason users' states. 

In a harmonious convergence, Xie et al. \cite{xie2020assisted}, Liu et al. \cite{liu2021sg}, and Zhao et al. \cite{zhao2021geometry} have each harnessed GNNs to decode the complex language of human emotions manifested through facial movements. Specifically, Xie et al. \cite{xie2020assisted} delve into the micro-level of facial expressions, integrating action units with emotion category labels within a graph framework, thereby enhancing the precision of recognition tasks that are critical in various real-world applications.
Building upon this graph-centric perspective, Liu et al. \cite{liu2021sg} propose a semantic graph-based dual-stream network that addresses a key limitation of conventional CNN methods by incorporating semantic information into the recognition process. Their model not only learns from physical appearance but also understands the deeper meaning behind facial movements.
Similarly, Zhao et al. \cite{zhao2021geometry} contribute to the field by developing a geometry-aware FER framework, which combines the structural analysis capabilities of a GNN with the feature extraction prowess of a CNN. This dual-pronged strategy allows for a detailed examination of both the geometric and appearance-based aspects of facial expressions, leading to a robust recognition system that is sensitive to the minute details of human emotions.

\subsubsection{Action Recognition}
\label{sec:action_recognition}
The rapid advancement of Human Action Recognition (HAR) and tracking techniques has emerged as a crucial catalyst for diverse edge applications, e.g., for security surveillance and robot imitation.
The rationale behind using GI models for action recognition is to model human bodies into skeleton-based graphs, as exemplified in Fig. \ref{fig:graph-example}(e).
Specifically, body joints are regarded as vertices with each attached a feature vector indicating its 3D/2D coordinates and confidence scores, while they are connected along with the human skeletons. 

In the realm of graph-based models for skeleton dynamics, a spectrum of approaches has been explored, categorized by their focus on spatial or spatio-temporal characteristics.
Wang et al. \cite{wang2020graph} present Graph-PCNN, a framework that refines keypoint localization through a two-stage process, enhancing the accuracy of human pose estimation. This model-agnostic framework underscores the importance of accurate initial localization, which is further improved through a graph pose refinement module.
Building upon the spatial foundation, Peng et al. \cite{peng2020learning} take a leap into the spatio-temporal domain by employing NAS to evolve an adaptive GCN for HAR. Their approach integrates dynamic graph modules and multiple-hop connections, which not only bolsters the network's capacity to capture the nuances of motion in skeleton data but also aligns with the spatial-temporal emphasis of Graph-PCNN, albeit in a different context. Similarly, Jin et al. \cite{jin2020differentiable} address the complexity of multi-person pose estimation by proposing a hierarchical graph grouping method. This work, like Graph-PCNN, leverages the power of graph structures but diverges by reformulating the human part grouping into a graph clustering task, offering a fresh perspective on the integration of graph theory in pose estimation.

\subsubsection{Smart Health}
Smart health, also known as digital health or eHealth, has been a widely-concerned field in edge networks.
GI techniques are widely adopted in the healthcare domain and their applications are mainly about Electronic Health Records (EHRs) analysis and health condition analysis.
In this paper, we particularly focus on the latter since it relates to edge networks.

Health condition analysis integrates GI with sensing techniques to enable healthcare monitoring and intervention.
The synergy of these technologies is exemplified in Dong et al.'s work \cite{dong2021influenza}, where they introduce a GI model that harnesses mobile sensing data to detect early signs of influenza-like symptoms by encapsulating the dynamics of state transitions and internal dependencies within human behaviors through graph representation.
Building upon this foundation, the concept of semi-supervised learning is elegantly integrated in the work \cite{dong2021semi}, where they propose a Graph Instance Transformer (GIT). GIT not only combines multi-instance learning with contrastive self-supervised learning but also demonstrates the adaptability of GNNs in predicting early signs of mental health disorders.
Similarly, Jia et al. \cite{jia2020graphsleepnet} contribute to the field by introducing GraphSleepNet, a GI framework specifically tailored for automatic sleep stage classification. This framework adeptly tackles the challenge of utilizing brain spatial features and the transition information among sleep stages, showcasing the versatility of GNNs in addressing complex health-related problems.

\subsection{Location-Based Services}

Social relationships in edge networks usually appear in a location-based manner, i.e., with social entities distributed geographically.
Provided with the nature graph structure of social networks as shown in Fig. \ref{fig:graph-example}(f), GI models bring advanced learning ability for mining location-based social information and exploring Point-of-Interest (POI), raising significant attention from the community.
In the context of location-based social networks, we discuss their applications for social POI recommendation and geographical POI recommendation, respectively.

\subsubsection{Social POI Recommendation}
Social POI recommendation utilizes users' activities on social media platforms to provide recommendations for friends they may know or products they may like to purchase.
In a pioneering approach, Kefalas et al. \cite{kefalas2018recommendations} enhance the traditional recommender system by introducing a hybrid tripartite graph that intertwines user, location, and session networks. This structure is analyzed using an advanced random walk with a restart algorithm, which captures the fluid nature of user preferences over time and space.
Building upon this foundation, Salamat et al. \cite{salamat2021heterographrec} introduce HeteroGraphRec, a sophisticated recommender system that considers social networks a heterogeneous graph. Continuing this thread of innovation, Wang et al. \cite{wang2022spatiotemporal} tackle the specific challenge of session-based recommendations by proposing an STGCN model. This model predicts a user’s immediate interests by considering the sequence of their recent anonymous behaviors, thereby providing a more nuanced and responsive recommendation strategy.

\subsubsection{Geographical POI Recommendation}
The objective of geographical POI recommendation is to exploit users' behaviors to predict their potentially interested locations.
Both Luo et al. \cite{luo2021stan} and Yang et al. \cite{yang2022stam} have introduced models that ingeniously incorporate spatio-temporal dynamics to predict user preferences for future locations.
A common thread between these works is their reliance on attention mechanisms to dissect the complex interplay of spatial and temporal factors.
Luo et al.'s STAN model \cite{luo2021stan} introduces self-attention to directly model the spatial and temporal aspects of user check-ins, excelling at capturing the significance of non-adjacent and non-consecutive interactions.
In a parallel yet distinctive approach, Yang et al.'s STAM model enhances GNNs by incorporating scaled dot-product and multi-head attention mechanisms, which enriches the learning of neighbor embeddings.
The transition from STAN to STAM illustrates the field's maturation, moving from specific spatio-temporal interactions to a more integrated and comprehensive understanding of user preferences in recommendation systems.

\begin{figure}[t!]
  \centering
  \includegraphics[width=0.95\linewidth]{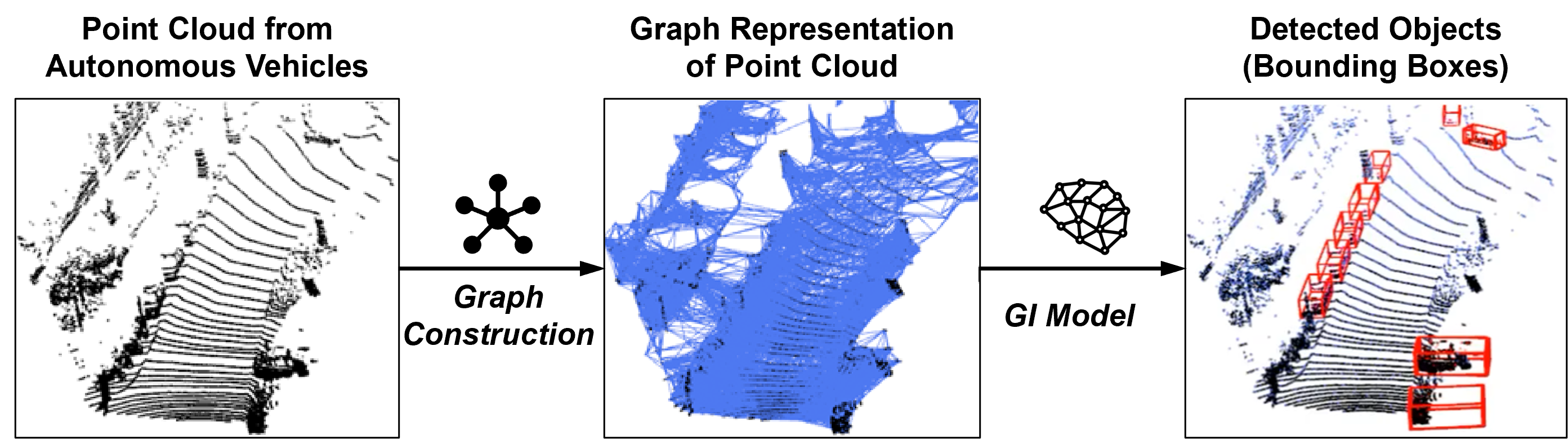}
  \caption{Example point cloud processing with GI models. The point cloud data captured by autonomous vehicles can be reorganized in a graph format and processed with GI models for high-precision object detection. Image adapted from \cite{shi2020point}.
  }
  \label{fig:point-cloud}
\end{figure}

\subsection{Graph-Based Mobile Vision}
Computer vision tasks accept inputs in the forms of images, videos, and point clouds, where many of them are represented by graphs.
For instance, Fig. \ref{fig:graph-example}(g) shows an image of the digit ``6" in 18$\times$12 pixels, whereby viewing pixels as vertices and their direct adjacency as links, it can be recognized as a graph.
Another example is the skeleton-based graphs in many images of human actions, as discussed in Sec. \ref{sec:action_recognition} and depicted in Fig. \ref{fig:graph-example}(e).
GI models are applied to these graphs to assist various traditional computer vision tasks like image classification, object detection, point cloud analysis, etc.

\subsubsection{Image Processing}
In the evolving landscape of image processing, Graph Inference (GI) models have emerged as a powerful paradigm, particularly for tasks that require an understanding of the relationships and context within visual data.

At the forefront of this advancement is the work by Liu et al. \cite{liu2018structure}, who introduced the Structure Inference Network (SIN). This pioneering approach in object detection treats objects as nodes and their interactions as links within a graph, providing a more holistic and contextual understanding of the visual scene. Expanding on this foundation, Chen et al. \cite{chen2019multi} delve into the domain of multi-label image recognition. They propose a GCN model that constructs a directed graph of object labels, capturing the intricate dependencies and co-occurrences among them. This model not only builds upon the graph-based reasoning initiated by Liu et al \cite{liu2018structure}. but also extends the concept to the challenges of recognizing multiple objects within an image. 
Following this thread, Lu et al. \cite{lu2019graph} first apply the GCN to image semantic segmentation with their Graph-FCN, an approach that redefines the task as a graph node classification problem, successfully integrating local location information often overlooked by standard deep learning methods.

\subsubsection{Video Analytics}
Video analytics with GI models are mainly for HAR and Multi-Object Tracking (MOT).
While HAR has been thoroughly explored in Section \ref{sec:action_recognition}, this section delves into the latter, focusing on the innovative approaches that have emerged.

Braso et al. \cite{braso2020learning} set a message-passing network that not only integrates learning within the MOT paradigm but also crucially extends it to the data association phase (which is traditionally a bottleneck).
Simultaneously, Liu et al. \cite{liu2020gsm} introduce the graph similarity model for MOT, which integrates a unique graph representation of object features and relationships, coupled with a graph matching module, to enhance robustness against occlusion and similar appearances in tracking scenarios.
Both of them have independently demonstrated the potential of GI models to transform MOT by focusing on different yet complementary aspects of the problem, i.e., global reasoning and robust feature representation.

\subsubsection{Point Cloud Processing}
Different from images in 2D representations, point clouds are organized in 3D structure and provide richer semantics in describing scenes and objects.
For instance, many autonomous vehicles use LiDAR to scan the environment and export point cloud data for perception, as in Fig. \ref{fig:point-cloud}.
Yet the 3D data are still accommodated in a graph representation, and thereby GI models are leveraged for efficient processing.

Wang et al. \cite{wang2019dynamic} develop EdgeConv, which enhances the representation power of point clouds by recovering topological information. This method uses dynamic graph construction to capture local geometric features, proving its suitability for high-level tasks in computer graphics and vision.
Building on this idea of dynamic feature extraction, Shi et al. \cite{shi2020point} propose Point-GNN, a GI model for object detection from LiDAR point cloud, which incorporates an auto-registration mechanism to reduce translation variance and a box merging and scoring operator for accurate detection over multiple vertices.

Extending these advancements, Lin et al. \cite{lin2020convolution} develop 3D-GCN to address the challenges of processing and summarizing information in unstructured point clouds for 3D vision applications. 
Their approach is particularly effective when dealing with data variations such as shift and scale changes, aligning with the principles seen in EdgeConv.
Furthermore, Zhang et al. \cite{zhang2021linked} present the linked dynamic GNN, which tackles the challenge of processing sparse, unstructured point clouds by linking hierarchical features from dynamic graphs. This method optimizes the network's architecture for more effective point cloud classification and segmentation, showing how linking hierarchical features can avoid the vanishing gradient problem and improve overall performance.

\begin{mybox}
\textbf{Lessons Learned (Sec. \ref{sec:gnn-application-edge})}

EGI has catalyzed a broad range of applications beyond traditional edge AI scenarios. These applications typically follow a general working procedure: the system collects data from components of edge networks and formulates them in graphs. The graph data are ingested into a GI model for graph embedding, along with a subsequent model, e.g., FCN and LSTM, for downstream tasks. The key to building a well-performed EGI application is carefully modeling the graph data and properly selecting the GI model. 
\end{mybox}
\section{Edge Networks for Graph Intelligence}
\label{sec:edge_computation_gnn}

Given the applications introduced in Sec. \ref{sec:gnn-application-edge}, this section discusses how edge networks serve GI execution in their multi-tier hierarchy.
In the following, we survey various computing systems in different architectures according to their reliance on the cloud.
Note that in each architecture, we review both model training and model inference systems, and their EGI ratings thus diverge.

\subsection{Federated Graph Learning}
\label{sec:fgl}

Edge-cloud synergy extends cloud resources as remote computing assists in optimizing the performance and efficiency of user-centric computing and data processing.
Specifically, it enables certain computing tasks reserved at the edge for immediate processing, while employing the cloud for offloading more resource-intensive tasks with long-term storage and complex data analysis.
This hybrid computing paradigm combines the differentiated capabilities of edge and cloud computing, thereby admitting flexible service computing for various applications \cite{ye2022eco}.

For edge computation of GI models, typical workloads in the form of edge-cloud synergy are FL of GI models, i.e., Federated Graph Learning (FGL).
FGL is the Level-4 EGI system in the rating taxonomy but is distinct from GI-assisted FL (GFL) as discussed in Sec. \ref{sec:GFL}.
Although they both combine GL and FL, the targeted model to be federatively trained in FGL is exactly GI models but GFL allows the targeted model to be any model not only GI models.
Readers particularly interested in the combination of GI and FL may find dedicated reviews \cite{fu2022federated, liu2024federated} for more information.
With respect to the segmentation of the data space, FGL can be divided into horizontal ones (graph topology space) and vertical ones (feature data space).

\begin{figure}[t]
  \centering
  \includegraphics[width=0.9\linewidth]{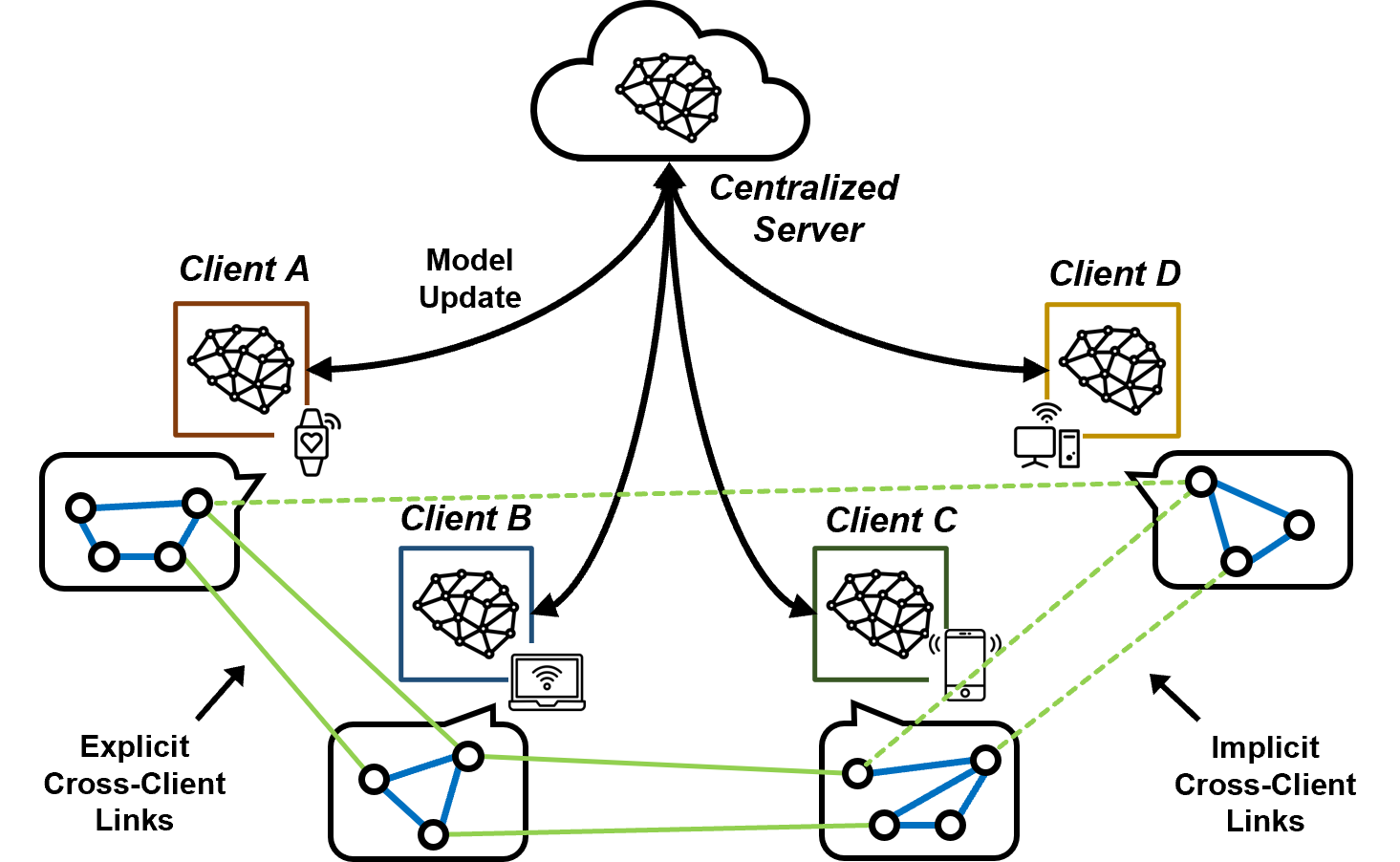}
  \caption{Federated Graph Learning (FGL) leverages FL mechanisms to collaboratively train GI models with distributed clients. Specifically, horizontal FGL considers data splitting in the graph topology dimension, where the relationship across client data may be explicit or implicit to the centralized server.
  }
  \label{fig:horizontal-fgl}
\end{figure}

\subsubsection{Horizontal Federated Graph Learning}

Horizontal Federated GNNs pertain to the scenario in which multiple clients participate in a federated learning setting, sharing identical node feature spaces while employing distinct node IDs \cite{chen2021fedgraph}.
Each client possesses a minimum of one graph or a collection of graphs, and vertices are distributed across diverse clients with interconnections.
Concerning the explicity of links between subgraphs on clients, horizontal FGL works can be categorized into two classes, namely with explicit cross-client links and with implicit cross-client links.

The first setting considers FGL with explicit links across subgraphs on FL clients, as in Fig. \ref{fig:horizontal-fgl}.
The conventional approach in this setting entails training local GI models within individual clients to initially acquire the local representations or node embeddings of the respective graphs.
Subsequently, a Federated Learning (FL) algorithm is applied to facilitate the aggregation process, and the FL server, which is usually a cloud server, accumulates the model parameters or gradients received from the clients and performs FL aggregation.
The resultant updated parameters are then returned to the clients for subsequent rounds of training.
The introduction of FGL within this framework is contingent upon the specific research problems they address.
For Non-IID graph data issues, many approaches intend to enhance the adaptation performance of local models or acquire a robust global model through model adjustment like model interpolation \cite{zheng2021asfgnn,zhang2023fedego} and meta learning \cite{wang2022graphfl}.

The second setting assumes that certain links connecting vertices across different clients are implicit and absent, illustrated in Fig. \ref{fig:horizontal-fgl}.
For the case where the implicit links connect vertices with dissimilar IDs, a prevalent strategy involves rectifying the local graphs by reconstructing the missing links.
This ensures the local graph's completeness, thereby facilitating high-quality graph representation.
Furthermore, the inclusion of links connecting clients aids in mitigating the challenges associated with non-IID data.
FASTGNN \cite{zhang2021fastgnn} introduces a straightforward link generator in the FL server, which is responsible for reconstructing absent links between clients by introducing Gaussian randomly generated links.
These reconstructed edges are broadcast to all clients during FL training, prompting them to update their local graphs.
In a similar vein, the link generator in FedGL \cite{chen2024fedgl} produces a global pseudo graph with vertex embeddings provided by clients.
Subsequently, this global pseudo-graph is disseminated to all clients, enabling them to amend their local graphs to facilitate GI model training.
Based on the simple yet effective idea of link generation, follow-up works further develop more complex techniques for better performance \cite{zhang2021subgraph,peng2022fedni}, consider privacy issues with security means \cite{qiu2022privacy,wu2022federated,wu2021fedgnn}, and explore to improve communication efficiency \cite{du2022federated,yao2022fedgcn}.

For the case where links are implicit in different clients, Knowledge Graph (KG) techniques are involved to complete information between these vertices.
Once the local graphs have been rectified, the application of an FL algorithm facilitates the training of GI models in a similar manner as observed in scenarios with explicit links.
FKGE \cite{peng2021differentially} adapts a GAN-based module \cite{jordon2018pate} to facilitate the translation of aligned entity and relation embeddings across paired KGs, where the refined embeddings will be broadcast as long as the paired KGs are improved.
FedE \cite{chen2021fede} establishes an entity table on the server that catalogs unique entities from multiple clients.
This table employs the FedAvg aggregation to process the aligned entity embeddings, and once processed, the updated embeddings are distributed back to the clients.
Upon FedE, FedR \cite{zhang2022efficient} and FedEC \cite{chen2022federated} further improve the capability of privacy preservation and non-IID data resilience.

\begin{figure}[t]
  \centering
  \includegraphics[width=0.85\linewidth]{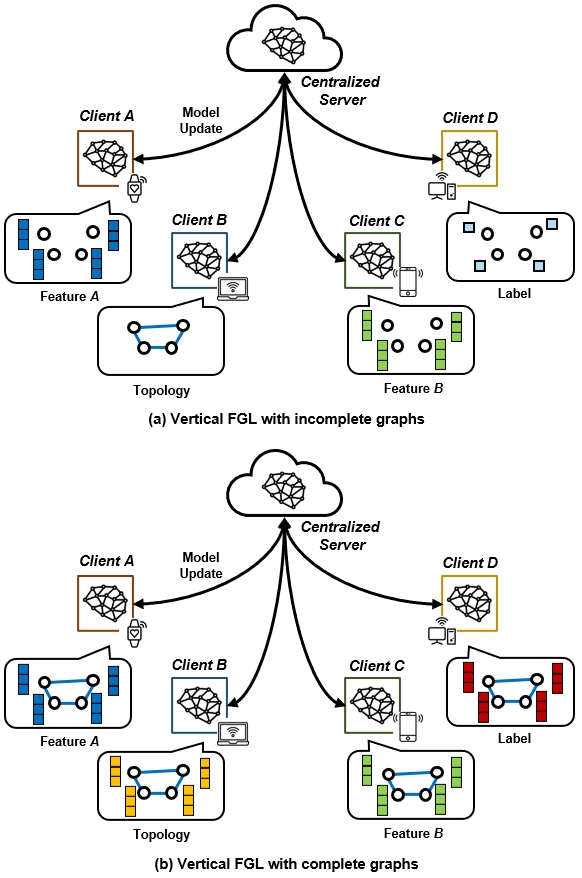}
  \caption{Vertical FGL considers data split in the feature space with (a) incomplete graphs or (b) complete graphs.
  }
  \label{fig:vertical-fgl}
\end{figure}

\subsubsection{Vertical Federated Graph Learning}

Vertical FGL assumes that clients possess nodes characterized by fully overlapping node IDs, albeit with distinct feature spaces.
Through the utilization of FL, clients collaborate in training a global Graph Neural Network (GNN) model using features sourced from multiple clients.
Within vertical FGL, there exist two distinct cases.

In the first case, graph topology, vertex features, and vertex labels are distributed across different clients.
Consequently, clients may not be able to obtain the full graph data with respect to both vertex features and graph topology, as shown in Fig. \ref{fig:vertical-fgl}(a).
To orchestrate these clients in such semi-blind situations, SGCNN \cite{mei2019sgnn} computes a similarity matrix based on the dynamic time-warping algorithm to sketch the graph's topology without knowing the underlying structure.
One-hot encoding is employed to describe vertices' features under privacy requirements, which will be transmitted to the server to train the global GI model for GI-based tasks.
FedSGC \cite{cheung2021fedsgc} considers a two-client decentralized setting with vertices' features and graph topology being respectively possessed.
In this scenario, FedSGC proposes an additively Homomorphic Encryption (HE) method to ensure privacy preservation during FGL.

The second case entails the ownership of solely disparate vertex feature spaces by clients, while the graph topology remains accessible to all clients (Fig. \ref{fig:vertical-fgl}(b)).
For the case with only two clients and a central server, FedVGCN employs HE to encode intermediate data transferred across clients, where the server generates encryption key pairs and performs global FL aggregation.
FMLST \cite{li2022federated} assumes a shared global spatio-temporal pattern across clients and applies an MLP to fuse local patterns and global patterns for personalizing client models.
Graph-Fraudster \cite{chen2022graph} investigates adversarial attacks on this vertical FGL scenario and shows differential privacy mechanisms and top-k mechanisms as two viable defense strategies against such attacks.

\subsection{Distributed Edge Collaboration}
\label{sec:distributed_edge_collaboration}

\begin{figure}[t]
  \centering
  \includegraphics[width=0.95\linewidth]{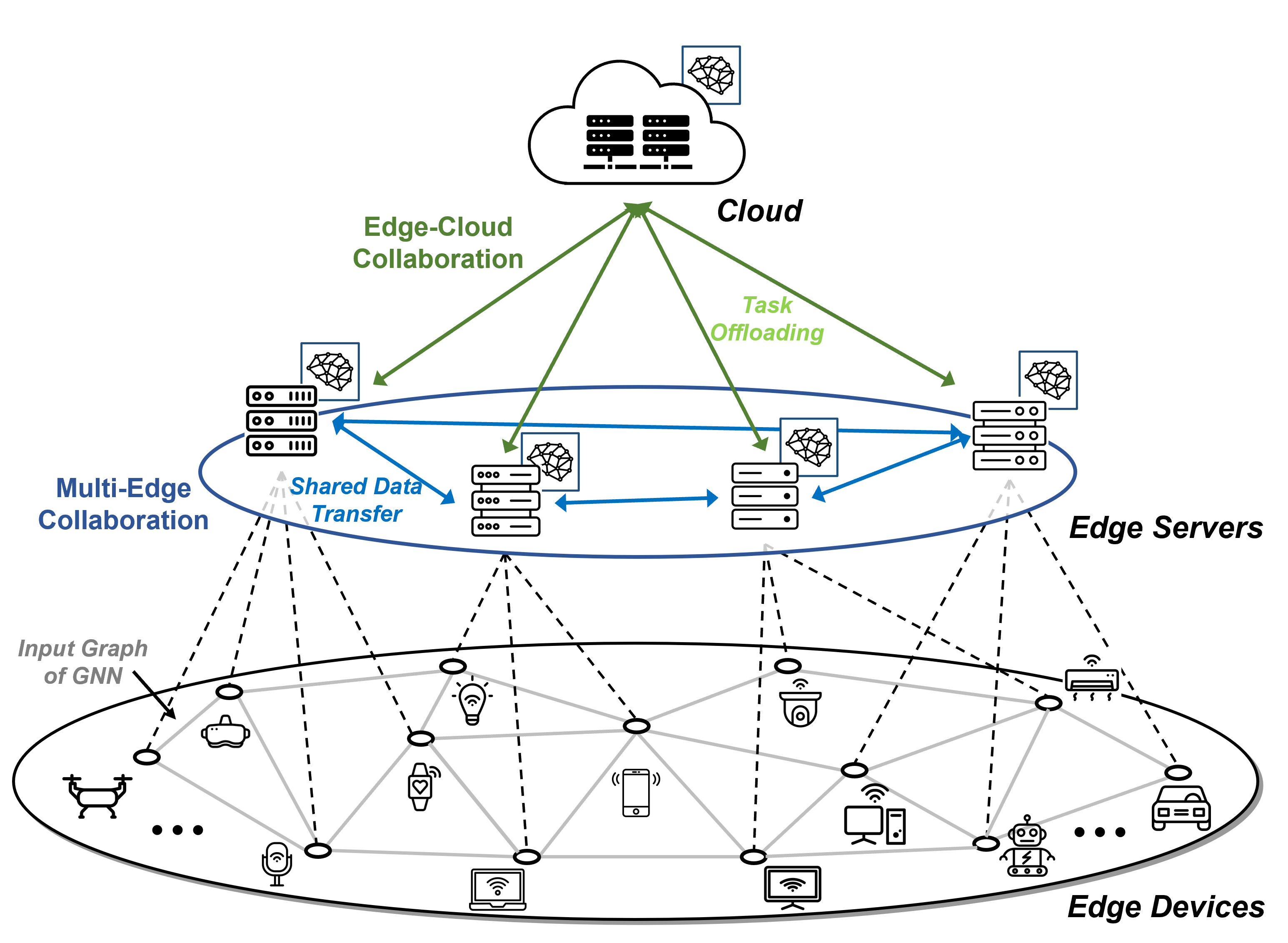}
  \caption{The graph data collected from edge networks can be processed via distributed edge collaboration, i.e., edge-cloud collaboration or multi-edge collaboration. The former bridges the cloud and the participants from edge networks through GI model computation offloading, while the latter usually assembles multiple edge platforms as a whole to run GI services in a cooperative manner.
  }
  \label{fig:edge-collaboration}
\end{figure}

Besides completely offloading to the cloud or executing in-situ, distributed edge collaboration, as a balanced compromise between them, introduces distributed training or inference of GI models within an edge network or between the edge-cloud continuum as in Fig. \ref{fig:edge-collaboration}.
This compromise offers several advantages, such as reduced remote transfer without cloud communication, (conditional) privacy preservation, and resource-augmented scalability \cite{ren2023survey, zhou2024digital, hu2023adaptive, qu2023stochastic}.
According to the dependency of cloud servers, distributed edge collaboration typically exhibits in two forms, i.e., edge-cloud collaboration and multi-edge collaboration.

\subsubsection{Edge-Cloud Collaboration}

Edge-cloud collaboration differs from traditional integrated cloud offloading by splitting computation workload between edge platforms and cloud servers and thus can fortify the duty of the edge side and alleviate the reliance on the remote cloud \cite{li2019edge, zeng2019boomerang, ye2025fast}.
This allows edge facilities to be a key processor that blocks sensitive raw data from the cloud and eliminates redundant data transmission in the backhaul.
Following this technical path, Branchy-GNN \cite{shao2021branchy} concentrates on GI-based point cloud processing and applies edge-cloud collaboration with a learning-based source-channel coding scheme for bandwidth-efficient intermediate data transfer.
It also borrows the idea of joint split learning and early exiting to reduce the computational cost on edge devices \cite{li2019edge, laskaridis2020spinn}. 
Pyramid proposes a hierarchical architecture for edge-cloud synergetic GNNs that trains a local, small model on edge servers with its collected data and a global model on a cloud server to capture inter-region spatio-temporal relationships.
During the runtime, Pyramid \cite{he2022pyramid} allows GI-based predictions with either the local model or the global model in an on-demand fashion and achieves lower inference latency against baselines.
Branchy-GNN is a Level-3 EGI system since it only tackles GNN inference at the edge, while Pyramid rates Level-4 because of its joint training and inference designs.

\subsubsection{Multi-Edge Collaboration}

Multi-edge collaboration \cite{zeng2024implementation, yang2022online, zhang2022ents} provisions all data flow within the edge networks, where edge devices individually perform local computation on the local data they possess and collectively train or infer GI models by sharing information and merging the results.
On the one hand, compared to the cloud-assisted solutions, multi-edge collaboration fully reserves data without leaking them to the remote datacenter, thus avoiding users' privacy concerns from cloud providers \cite{xu2023mobile, ye2024galaxy, li2024distributed}.
On the other hand, collaborating multiple edge devices augments the available resources for targeted GI tasks beyond a single edge device, which therefore enables superior performance against on-device computation \cite{ye2024asteroid,zeng2020coedge}.
Besides, privacy requirements are also guaranteed if the participating edge devices are mutually trusted, which can be relevant for many edge scenarios, e.g., in a smart home with edge devices owned by the same user and in a manufacturing pipeline with cameras and sensors possessed by the same company \cite{ye2022eco, dong2021collaborative}.

Under this paradigm, Fograph \cite{zeng2022fograph, zeng2023serving} proposes a distributed GNN inference system that leverages multiple edge servers in proximity to IoT data sources for real-time GI model serving.
To address resource heterogeneity and dynamics among edge servers, Fograph designs a load-aware inference execution planner as well as an agile workload scheduler for maximum parallelization.
It also introduces a degree-aware compression mechanism to minimize data transmission overhead.
Following this principle, GLAD \cite{zeng2022gnn} studies the cost optimization problem for distributed GNN serving in multi-tier edge networks.
By formulating cost factors in the whole lifecycle of edge-enabled GI services, it separately considers static graphs and dynamics graphs and derives a holistic graph layout scheduling solution with theoretical performance guarantees.
Both Fograph and GLAD are Level-4 systems for their functionality of GNN inference at the edge.

\subsection{On-Device Computation}

On-device computation for GI models refers to performing GI model computations directly on edge devices rather than relying on cloud or remote servers.
This entails executing the training or inference tasks locally on edge devices such as smartphones, laptops, or tablets, without requiring a constant internet connection or external server support.
This paradigm is particularly useful for privacy-sensitive applications because it ensures users' data to be fully preserved in situ.
Nonetheless, given the intensive workload of computing GI models, on-device edge computation of them is non-trivial, and researchers have developed several routines to tackle that by designing resource-friendly procedures and models or compression for compact GI models.

\subsubsection{Resource-Efficient Procedure}

Without modifying GI models, resource-efficient procedure design indicates the endeavors that develop GI processing flows in minimal resource usage, e.g., memory footprint.
For instance, centering on the out-of-memory issues where GI models' sizes are too large to fit into edge devices' memory, Zhou et al. \cite{zhou2021brief} (Level-3 rating) design a feature decomposition approach that decomposes the dimension of feature vectors and performs aggregation respectively, reducing up to 5$\times$ memory footprint against generic GI frameworks.
Building on the theme of memory efficiency, SUGAR \cite{xue2023sugar} (Level-4 rating) targets resource-efficient training on resource-constrained edge devices through a subgraph-level training scheme that first partitions the initial graph into a set of disjoint subgraphs and then performs local training for individual subgraphs.
Complementing these approaches, RAIN \cite{liu2024efficient}  (Level-3 rating) explores the opportunity of conservatively processing similar inference batches based on local sensitive hash and reusing repeated vertices' data among successive batches to reduce redundant data loading.

There are also some works that utilize channel pruning methods to reduce input feature dimensions for GI model inference acceleration.
Zhou et al. \cite{zhou2021accelerating} set the stage by employing a LASSO regression-based pruning technique, which identifies the most influential feature channels, thus ensuring that the pruning process is both effective and minimally disruptive to the model's accuracy.
Yik et al. \cite{yik2022input} extend this line of inquiry by conducting a meticulous analysis of the trade-offs inherent in channel pruning. They develop algorithms that not only preserve the model's predictive prowess but also enhance computational efficiency, aligning seamlessly with the work \cite{zhou2021accelerating} by further honing in on the most valuable features.

\subsubsection{Resource-Efficient Model}

Designing resource-efficient GI models usually involves simplifying operators in GI semantics to improve computational efficiency in both GI model training and inference.
Recent advances in this routine have explored eliminating redundant operators like linear aggregation and non-linear activation.
For instance, SGC \cite{wu2019simplifying} removes the non-linear ReLU operators in conventional GI models to reduce the model complexity and reserves only the final Softmax function for probabilistic output generation, which significantly accelerates the model training yet maintains comparable accuracy in many generic tasks like text and graph classification.
Other examples include LightGCN \cite{he2020lightgcn} and UltraGCN \cite{mao2021ultragcn} for location-based social recommendation tasks, where the former drops the feature transformation and non-linear activation and the latter modifies the message-pass process into a simplified approximate embedding method.
As a result, both of them effectively decrease the computational load against baselines and attain remarkable speedup.

To automate the resource-efficient model design for edge platforms, some researchers \cite{zhou2023hardware, zhang2022pasca, odema2023magnas} introduce NAS techniques to the GI domain and present hardware-aware NAS frameworks tailored to GNNs, aiming to strike a balance between model accuracy and efficiency corresponding to the characteristics of targeted devices.
Zhou et al. \cite{zhou2023hardware} present HGNAS, a hardware-aware graph neural architecture search (GNAS) framework that optimizes GNN designs for edge devices by balancing accuracy and efficiency using a hardware performance predictor. Similarly, PaSca \cite{zhang2022pasca} constructs scalable GNNs through a novel scalable GNAS paradigm, enabling systematic exploration of a vast design space to optimize GNNs for both accuracy and efficiency via multi-objective optimization. Building upon these concepts, MaGNAS \cite{odema2023magnas} is a mapping-aware GNAS framework that efficiently processes vision-based GNN workloads on heterogeneous multi-processor platforms by identifying optimal GNN architectures and mappings for maximized resource efficiency and performance trade-offs. Together, these works highlight the growing trend towards hardware-aware and scalable GNAS frameworks that address both the accuracy and efficiency requirements crucial for deploying GNNs on edge devices.
This line of work represents Level-4 EGI systems since they enable both edge training and inference.

\subsubsection{Model Compression}
Compressing a complete model into a smaller one that appropriately fits the capability of the targeted edge device has been a popular deployment means for many edge intelligence services.
The smaller model, which usually with fewer parameters, allows much less computing cost and thus significant acceleration for on-device computation.
To obtain such a lightweight model for EGI, researchers typically employ three ways, i.e., model quantization, knowledge distillation, and neural architecture search.

Model quantization intends to quantize GI model parameters from an origin, larger bitwidth to a targeted, smaller bitwidth, and thus reduce the total size of the model.
In this context, SGQuant \cite{feng2020sgquant} proposes a multi-granularity method that can quantize feature vectors at the component level, topology level, and layer level to meet diverse data precision demands.
It also designs an automatic bitwidth-selecting algorithm to attain proper configuration among different quantization granularities.
Building upon the quantization paradigm, Degree-Quant \cite{tailor2020degree} pioneers a quantization-aware training scheme specifically crafted for GI models. This scheme is designed to facilitate low-precision inference on devices with constrained resources, without compromising the model's predictive capabilities.
VQ-GNN \cite{ding2021vq} takes a quantum leap in this domain by introducing a unified framework that applies vector quantization (VQ) for dimensionality reduction and learns a small number of quantized reference vectors of global node representations, which viably avoids the “neighbor explosion” problem and enables faster mini-batch training and inference of GI models.
Following this quantization principle, some researchers go one step further with GI model binarization, i.e., binarizing parameters in GI models for aggressive execution speedup \cite{wang2021bi}.

\begin{figure}[t]
  \centering
  \includegraphics[width=0.95\linewidth]{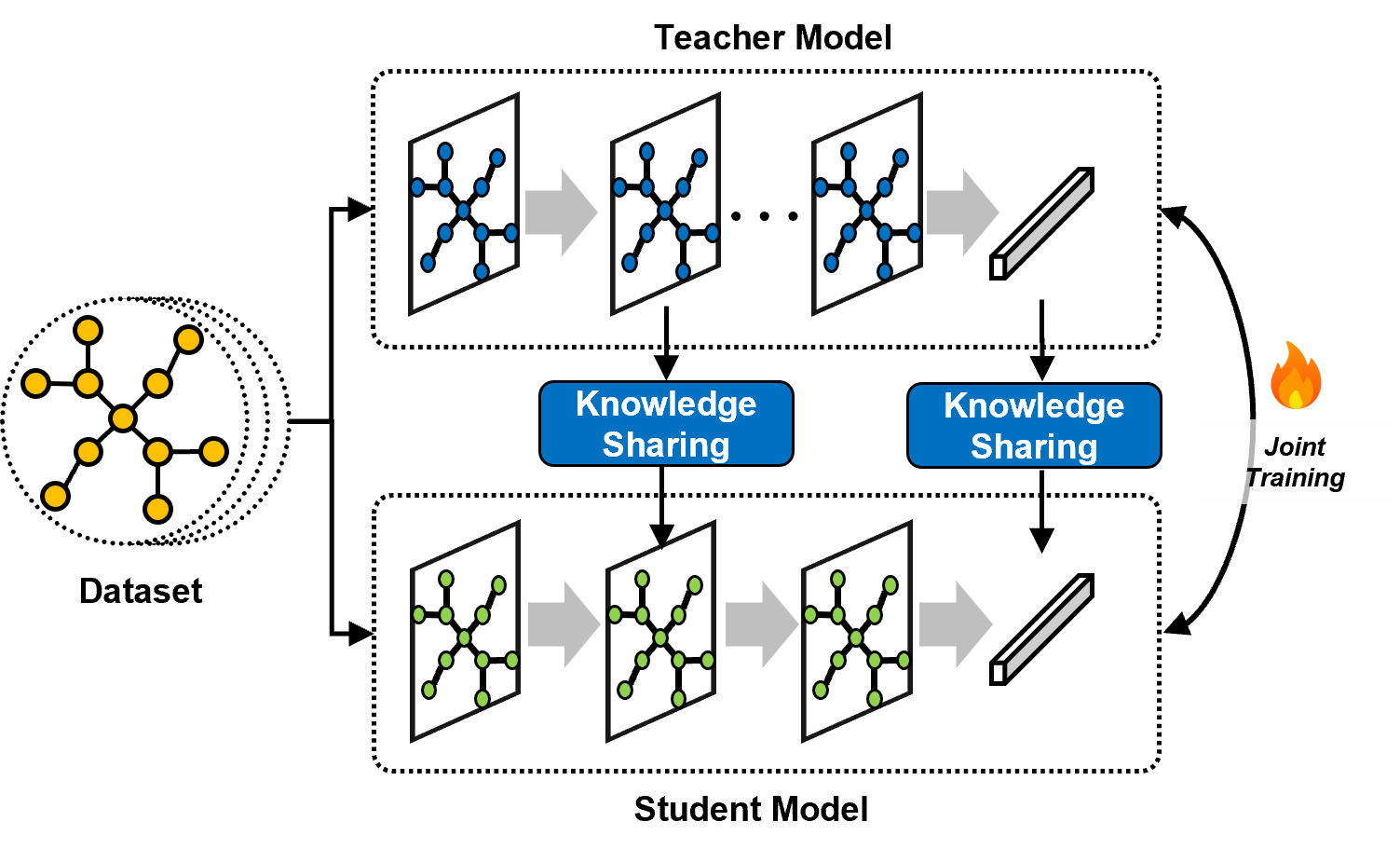}
  \caption{
  In general, KD for GNN follows a similar procedure of KD for traditional DL models, where we share the knowledge of the teacher model with the student model to improve its accuracy.
  }
  \label{fig:graph-kd}
\end{figure}

Knowledge Distillation (KD) aims at extracting knowledge from a complex teacher model and transferring it into a compact student model while maintaining comparable performance \cite{hinton2015distilling,liu2023graph}, as depicted in Fig. \ref{fig:graph-kd}
While KD is originally applied in the computer vision domain \cite{cho2019efficacy,chen2020knowledge,gou2021knowledge}, Yang et al. \cite{yang2020distilling} first introduce KD for GI models along with a local structure preserving method for teacher-student similarity measurement and experimentally show its effectiveness in GI-based 3D object detection tasks.
Along with this line, Yang et al. \cite{yang2021extract} combine parameterized label propagation and feature-based MLP in the KD's student model, allowing better utilization of the knowledge from both the teacher model and prior basement.
TinyGNN \cite{yan2020tinygnn} focuses the KD principle on accelerating model inference and designs a peer-aware module and a neighbor distillation strategy for preserving model accuracy.
They all focus on the edge inference aspect and are Level-3 EGI systems in the rating taxonomy.

\begin{mybox}
\textbf{Lessons Learned (Sec. \ref{sec:edge_computation_gnn})}

Edge networks as a physical infrastructure can support GI in various architectures, including FGL with edge-cloud synergy, collaborative computing with distributed edge platforms, and in-situ computing on individual edge devices.
The architecture used for EGI services depends on the demand of SLO (e.g., latency and privacy requirements) and the availability of the cloud.
The lower reliance on the cloud the system has, the more stringent the constraint of computing resources, and thus the more attention is paid to resource-efficient optimizations.
\end{mybox}
\section{Graph Intelligence for Edge Networks}
\label{sec:gnn-optimization-edge}

Besides directly using GI models to specific edge applications discussed in Sec. \ref{sec:gnn-application-edge}, GI techniques are also employed for optimizing edge networks.
This can be relevant since many components in edge networks, e.g., network hierarchy as illustrated in Fig. \ref{fig:graph-example}(e), can be naturally recognized in graphs.
Towards that, we discuss GI-based optimization on edge networks with respect to their resource management, graph-based decision-making, security issues, and GI-assisted FL in the following.
Using GI for optimizing edge not only involves GI model inference at the edge but also training tailored GI at targeted edge platforms, thus their EGI systems are at Level 4 in the rating taxonomy.

\subsection{Network Resource Management}
\label{sec:network-resource-management}
Given the graph structure of massively connected edge networks, GI models are deemed to be a useful tool to achieve that and are leveraged for optimizing network resource allocation and network function orchestration.

\subsubsection{Resource Allocation}
Resource allocation is critical for ensuring networking efficiency and robustness, and has been extensively studied using traditional optimization means.
Instead of applying complicated formulation and optimization, GI-based methods model the whole network as a graph and schedule an allocation strategy in a holistic manner \cite{shen2020graph, ma2024flocoff, huang2025hyperjet}.

Power control in wireless networks seeks the optimal power allocation to transmitters to minimize power consumption while maintaining signal quality \cite{chiang2008power,cruz2003optimal}.
Shen et al. \cite{shen2019graph} introduce an interference GCN to solve the power control problem in wireless networks, designed to model K-user interference channels as complete graphs and incorporate wireless channel data as graph features for scalability.
Their approach is complemented by Eisen et al. \cite{eisen2020optimal,eisen2020transferable}, who employed spectral-based GCNs for power allocation in Device-to-Device (D2D) wireless networks, highlighting the effectiveness of GCNs in managing interference.
Sala{\"u}n \cite{salaun2022gnn} study cell-free MIMO networks and formulate a graph of the dominant dependence relationship between access points and user equipment, using GNN for scheduling the power control.
Building on these ideas, Nikoloska et al. \cite{nikoloska2021fast} utilized REGNN to tackle power control challenges in decentralized wireless networks, leveraging the adaptability of GNNs to changing network topologies. Similarly, Yang et al. \cite{yang2024knowledge} propose UWGNN, which enhances GNNs by unrolling modules in the weighted minimum mean-square error algorithm into GI frameworks, demonstrating a method to integrate knowledge more effectively. Together, these studies underscore the versatility of GNNs in addressing power control problems across various wireless network settings, from centralized to decentralized architectures, and from static to dynamic topologies.

Link and channel scheduling is another key issue in communication networks as it drastically impacts the overall performance of the system (throughput, delay, etc.) \cite{cruz2003optimal}.
Some researchers apply GNN or combine GNN with other models for channel tracking and throughput optimization in MIMO networks, where the channel states and links are formulated in graphs \cite{yang2020graph, salaun2022gnn, he2023message, ranasinghe2021graph}.
Yang et al. \cite{yang2020graph} intend to represent the obtained channel data in graphs by associating channel correlations as links' weights, based on which a GNN model is trained for channel tracking.
D2D networks are also studied in a similar vein, e.g., GCN-GAN \cite{lei2019gcn} integrates the strengths of GCN, LSTM, and GAN to capture the dynamics, topology, and evolutionary patterns of weighted dynamic networks for temporal link prediction. 
These methods collectively leverage graph-based models to enhance link scheduling and prediction, demonstrating the versatility and effectiveness of graph neural networks in addressing various aspects of network performance optimization.

Beamforming techniques are regarded as promising solutions in multi-user, multi-antenna communication systems \cite{bjornson2014optimal}, and also play a crucial role in wireless resource management. 
Chen et al. \cite{chen2021graph} present an innovative unsupervised GNN approach for beamforming design in D2D wireless networks. Their method significantly reduces the problem dimension by transforming beamforming into primal power and dual variables, showcasing superior performance with fewer samples. Further extending the application of GNNs, Kim et al. \cite{kim2022bipartite} develop a bipartite graph neural network framework for multi-antenna beamforming optimization in multi-user MISO systems. This approach partitions the beamforming optimization into suboperations for individual antennas and users, allowing for scalable and efficient computations. Both studies leverage the flexibility and efficiency of GNNs to handle complex beamforming tasks, demonstrating the potential of graph-based methods in optimizing wireless communication systems.

\subsubsection{Network Function Orchestration}
Virtual Network Function (VNF) virtualizes the functions that traditionally run on dedicated hardware devices such as routers and firewalls and has become a key enabler in software-defined networks (SDNs) \cite{shen2021holistic}.
Researchers employ GI models on VNF for orchestrating various tasks including adaptive Service Function Chain (SFC) and network slicing.
SFC, crucial for identifying efficient paths in network servers for processing requested VNFs, faces challenges in maintaining high QoS in complex scenarios.

Heo et al. \cite{heo2020graph} introduce a GAE for dynamic network topologies, where the encoder represents the network topology and the decoder evaluates nodes for processing VNFs, adapting to topology changes.
Heo et al.'s subsequent work \cite{heo2020reinforcement} overcomes the limitation of earlier approaches restricted to fixed topologies, by employing RL for GNN-based model training across various unlabeled network topologies. 
Pioneering studies on network slicing \cite{wang2020graphneural, naeem2021digital} employ Digital Twin (DT) technology to digitally replicate slicing-enabled networks, creating detailed virtual models that mirror the physical network's dynamic characteristics.
These models are then used to develop GNNs that learn and optimize network behaviors directly from these non-Euclidean graph representations, thereby addressing challenges in resource allocation and QoS management in highly dynamic network environments.
Besides the above aspects, we refer readers interested more about GNN for communications to related discussions \cite{jiang2022graph, suarez2022graph}.

\subsection{Graph-Based Decision Making}

\begin{figure}[t]
  \centering
  \includegraphics[width=0.95\linewidth]{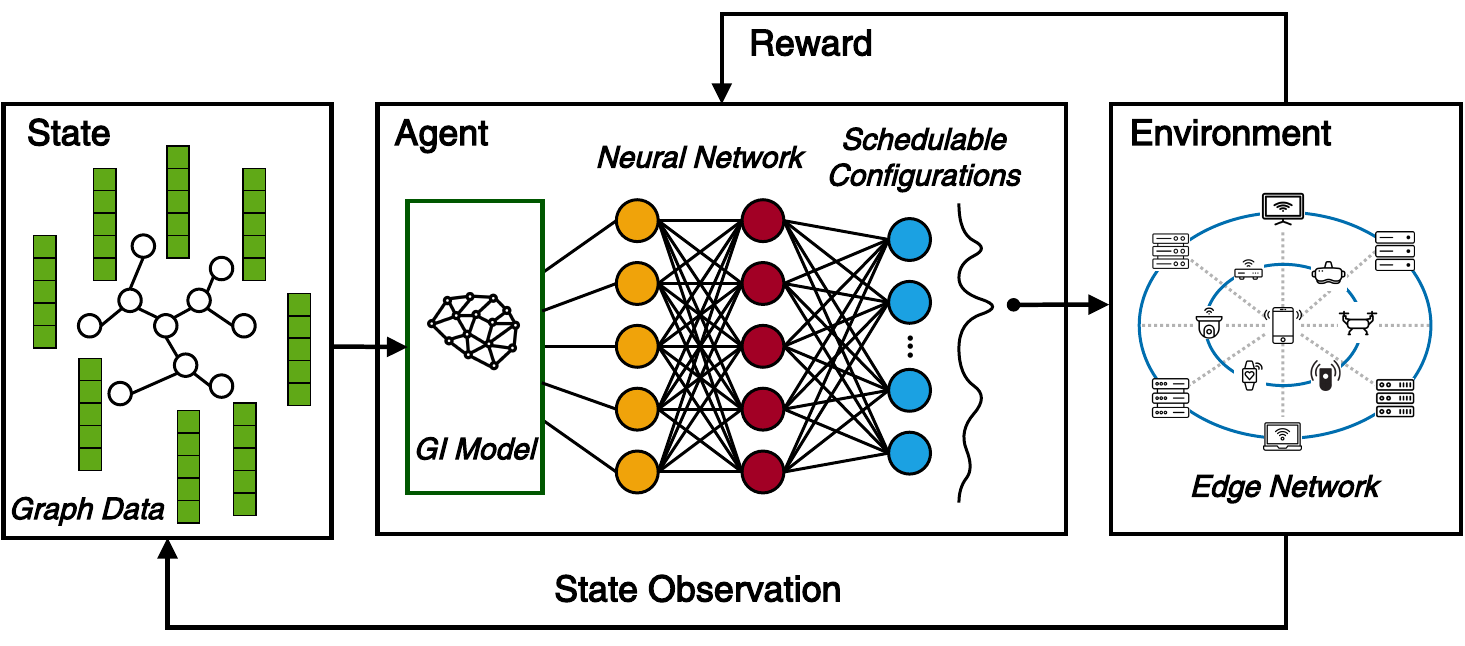}
  \caption{In a general GI-incorporated DRL workflow, an agent observes states in a graph format and, through the GI model, abstracts relational semantics from states. Next, the agent passes the encoded data to a neural network operator to decide a scheduling action and receives a reward from the targeted environment (i.e., edge networks).
  }
  \label{fig:gnn-drl}
\end{figure}

Given the powerful ability to abstract graphs, GI models can be integrated with DRL techniques to enable decision-making on graphs.
Their confluence leads to DGRL, which is extensively utilized for scheduling decisions on edge networks, e.g., offloading, routing, and wireless network control.
In general, the DGRL framework can be illustrated in Fig. \ref{fig:gnn-drl}, where an agent observes the state, schedules configurations, and feeds the agent (neural network) back with a reward after each action.
The GI model is incorporated as a graph information encoder to abstract relational semantics from the state.

\subsubsection{Offloading and Routing}
Offloading and routing are classic techniques in edge networks that allow workload migration across networking entities and some researchers \cite{chen2021multitask, swaminathan2021graphnet, ma2024mogr} use graphs to abstract tasks to be scheduled for offloading decisions.
Specifically, ACED \cite{chen2021multitask} introduces an actor-critic mechanism for DAGs-based computation offloading decisions considering task dependencies and wireless interference.
Swaminathan et al. \cite{swaminathan2021graphnet} enhance a routing algorithm for SDN using a GNN functioning with a deep Q-Network.
Trained to forecast efficient routing paths, this GI model minimizes packet delivery delays through simultaneous training and inference, enabling continuous learning and real-time adaptation.

\subsubsection{Wireless-Enabled Control Systems}
DRL techniques have been frequently used in attaining efficient control of communication resources in wireless networks.
However, trivially applying DRL does not scale well with the network size, and GI models are thus employed to overcome this challenge.
For example, Nakashima et al. \cite{nakashima2020deep} design a DGRL method for scalable, model-free resource allocation in large-scale wireless control systems. 
It capitalizes on the ability of GNNs to apply spatial-temporal convolutions to graph-structured data, harnessing the natural graph form of wireless networks for efficient policy design.
Some of its follow-up works \cite{chowdhury2021unfolding,szott2022wi} extend this approach by incorporating long-term constraints and demonstrate the permutation invariance and transferability of graph-based reinforcement learning policies in wireless control systems.

\subsection{Security and Anomaly Detection}
\begin{figure}[t]
  \centering
  \includegraphics[width=0.9\linewidth]{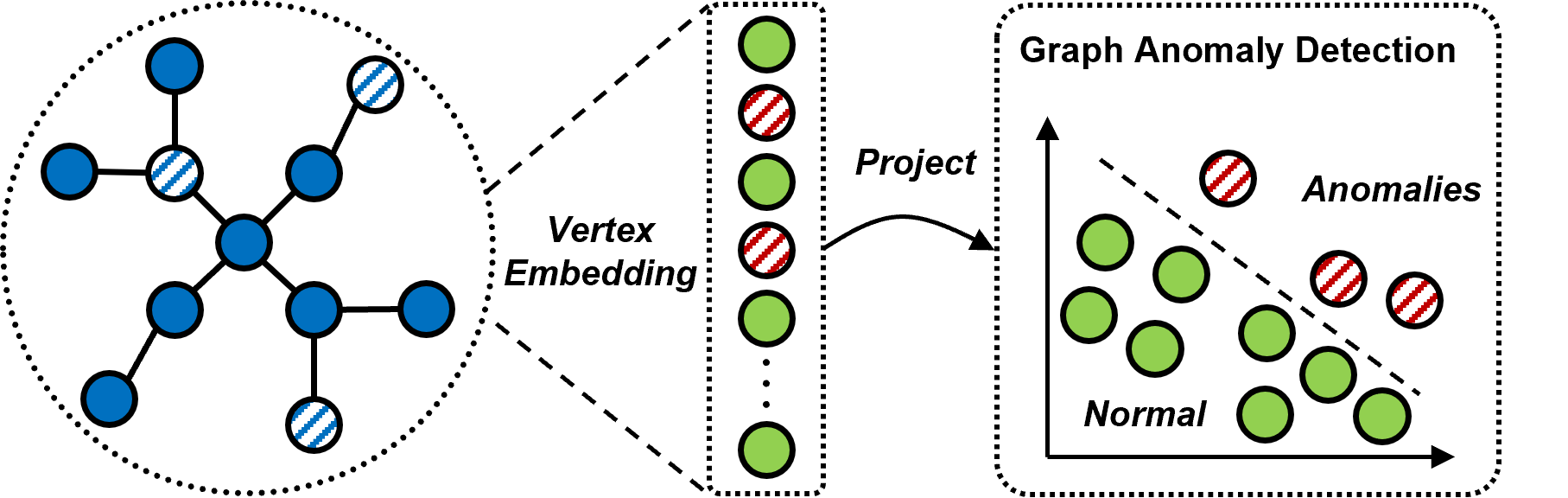}
  \caption{
  GI models can abstract vertices's features into embeddings, which can be used for detecting anomalies by classifying in a latent space.
  }
  \label{fig:anomaly-detection}
\end{figure}

Detecting anomalies in a graph is an important task in the analysis of complex systems, especially for ubiquitous edge networks.
The anomalies, in the context of graphs, are patterns that fail to conform to normal patterns expected of the graph's attributes or structures.
In this respect, the community has developed GI models for anomaly detection tasks and applied them to optimize edge networks.
Fig. \ref{fig:anomaly-detection} illustrates its rationale, where vertices (with features) can be transformed into embeddings, which are used for anomaly identification in a latent space.

\subsubsection{Static Network}
For edge networks that are static within a period, e.g., a smart home network in a short slot, we call it a static network. 
Extensive research has been conducted focusing on the detection of anomalies within static graphs, encompassing various categories including vertex, link, and graph-level anomalies \cite{kim2022graph}.
AnomMAN \cite{chen2023anomman} incorporates a graph auto-encoder module to leverage the low-pass filtering characteristic of graph convolution, typically a disadvantage for anomaly detection, transforming it into a beneficial feature by utilizing reconstruction errors to effectively identify anomalies.
EFraudCom \cite{zhang2022efraudcom} proposes a fraud detection system employing GI to identify anomalous edges in e-commerce platforms.
The system's core innovation lies in modeling the distributions of normal and fraudulent behaviors separately, enhancing robustness against evolving fraud patterns.
GLocalKD \cite{ma2022deep} and OCGTL \cite{qiu2022raising} both tackle graph-level anomaly detection. The former introduces a method using random distillation of graph and node representations for deep anomaly detection, and the latter employs self-supervised and transformation learning to enhance existing deep one-class approaches \cite{ruff2018deep}.

\subsubsection{Dynamic Network}
Dynamic edge networks take temporality into consideration where their topology and attributes may change over time \cite{chen2023graph}.
Recent studies, such as those by Deng et al. \cite{deng2021graph} and Zhang et al. \cite{zhang2022grelen}, have delved into the intricacies of multivariate time series anomaly detection within cyber-physical systems.
Both of them underscore the significance of understanding the complex interplay between sensors over time.
Deng et al. \cite{deng2021graph} propose an attention-based graph deviation network to learn the complex relationships between sensors in multivariate time series, not only accurately detecting anomalies but also providing an explainable model that identifies and clarifies deviations from normal sensor interactions. Similarly, Zhang et al. \cite{zhang2022grelen} focus on sensor dependence relationships in cyber-physical systems, utilizing a VAE for feature extraction and a GNN with stochastic learning to capture inter-sensor dependencies.
Building upon these methodologies, FuSAGNet \cite{han2022learning} combines Sparse Autoencoder with GNN for anomaly detection in high-dimensional time series, which optimizes both reconstruction and forecasting while explicitly modeling the relationships within multivariate time series.

\subsection{GI-Assisted Federated Learning}
\label{sec:GFL}

In certain FL scenarios, clients exhibit relational dependencies, a structural representation of which can be established through a graph.
For instance, traffic sensors in physical proximity often result appear in similar traffic patterns.
In this case, a cross-client graph can be established by identifying them as respective vertices.
The presence of this cross-client graph prompts the utilization of GI algorithms to enhance the FL training process, resulting in the paradigm of GI-assisted FL (GFL), as shown in Fig. \ref{fig:gl-asssited-fl}.
As we have discussed in Sec. \ref{sec:fgl}, GFL and FGL are different mechanisms, where the former applies GI techniques to optimize FL on arbitrary ML models (which do not necessarily be GI models) and the latter applies FL to jointly train GI models.
In other words, GI models are optimization tools in GFL, whereas they are the targeted models to be trained in FGL.
Due to such distinct functionality, we categorize GFL into the GI-based optimization for edge networks and discuss it in this subsection, while FGL will be introduced in the context of edge computation of GI models in Sec. \ref{sec:fgl}.

GFL works upon the premise that clients closely connected in the graph likely share similar data distributions.
Based on that, GNNs can utilize the graph structure inherent in the FL system to tackle the non-IID data heterogeneity among clients.
With respect to the utilization of centralized servers, we discuss GFL in the centralized setting and the decentralized setting, respectively.

\subsubsection{Centralized Federated Learning}
\begin{figure}[t]
  \centering
  \includegraphics[width=0.9\linewidth]{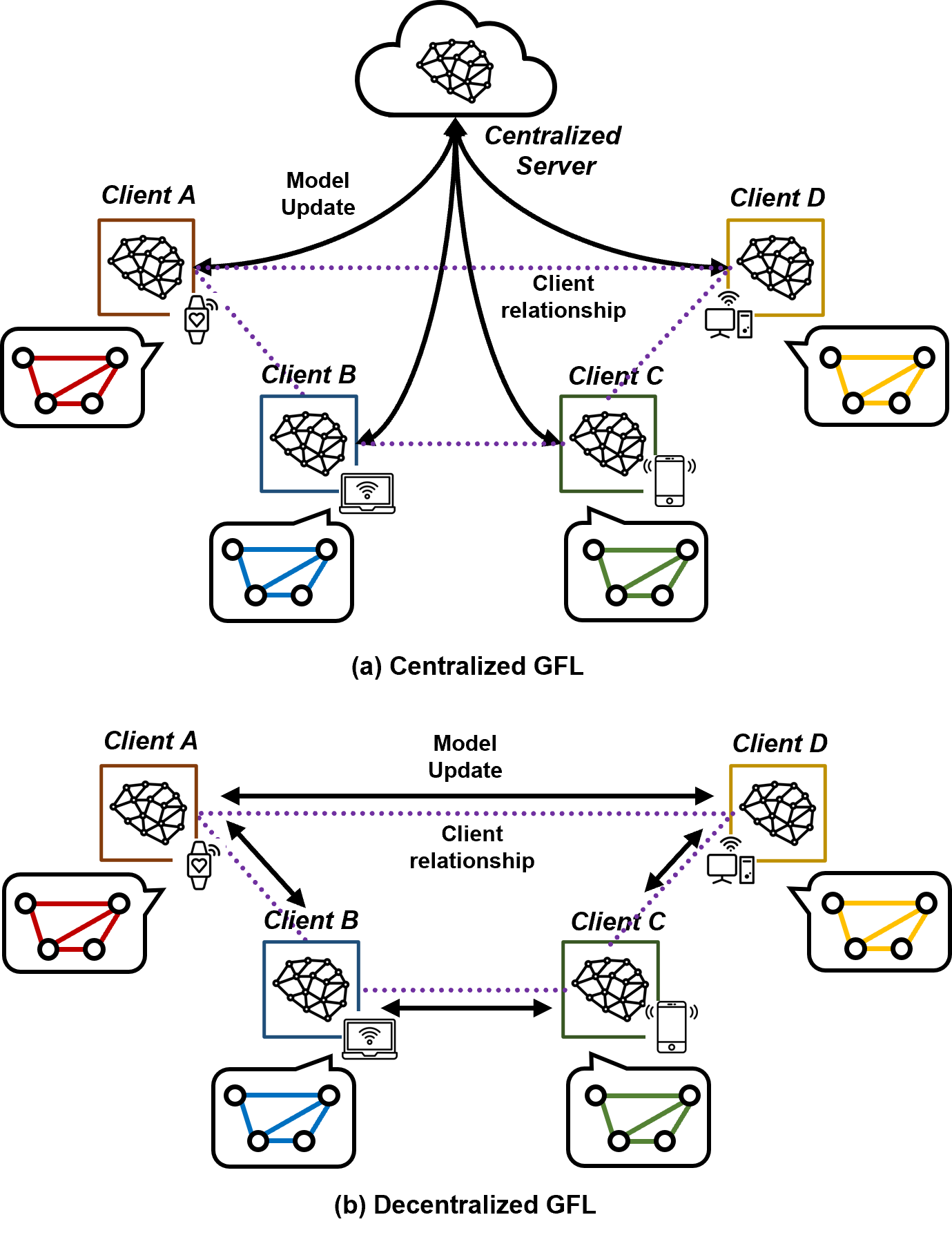}
  \caption{GI-assisted FL employs GI models as a tool to orchestrate FL clients with relationships and can be categorized into (a) centralized settings and (b) decentralized settings.
  }
  \label{fig:gl-asssited-fl}
\end{figure}

Under the centralized setting for GFL, a centralized server is employed to coordinate FL among clients, as depicted in Fig. \ref{fig:gl-asssited-fl}(a).
The GI model, which is applied for optimizing the FL procedure based on cross-client graphs, can be executed either on the server or the clients.

For the case of server-side GI execution, a GI model undergoes training on the server using graph relationships between clients, operating under the assumption that proximate clients exhibit similarities in their local models or feature embeddings.
Initially, the server initiates the collection of parameters from clients, following a standard FL protocol.
These uploaded parameters collected from client models are construed as node features within the context of the cross-client graph, using which the server trains a GI model to facilitate the aggregation process in FL.
Note that the specification of the cross-client graph may either be predefined or dynamically extracted through the application of a self-attention module during the training phase \cite{wu2022multi}.
The resultant updated parameters are then transmitted back to the respective clients.

In this process, researchers have proposed different routines to tackle different issues in training the coordinated GI model.
For instance, some researchers focus on the convergence speed of GI models under non-IID circumstances, designing bi-level optimization to train both the client models and the server-side GI model simultaneously, e.g., using unsupervised contrastive learning \cite{xing2022big} or regularization-aided supervised learning methods \cite{chen2022personalized}.
Others ponder the specific functionality of the GI models and develop training strategies in various aspects \cite{li2022power,lee2022privacy}.

For the case of client-side GI execution, each client is assumed to possess a global cross-client graph and trains not only its local model but a GI model to obtain global knowledge from other clients.
Building different graphs within the clients can be employed to address various problem-solving objectives like data heterogeneity and model heterogeneity.
For instance, FedCG \cite{caldarola2021cluster} establishes a fully connected graph by leveraging the similarity among clients' model weights or pattern features. 
Each client undertakes the training of a GNN using the cross-client graph and derives a global embedding.
Subsequently, the local model embedding and the global embedding are amalgamated by applying a trainable weight, which enhances the capacity of the model to incorporate both local and global information, addressing data heterogeneity.

\subsubsection{Decentralized Federated Learning}

Contrary to the centralized setting, decentralized GFL allows clients to communicate with their neighbors but does not have a central server to coordinate them (Fig. \ref{fig:gl-asssited-fl}(b)), making FL model aggregation a key challenge for global model convergence.
To address that, existing literature usually path two ways, i.e., weight summation across clients and graph regularization.

Weight summation methods update clients' local models by aggregating their neighbors’ local model parameters following the global cross-client graph topology.
Under this principle, DSGT \cite{lu2020decentralized} employs decentralized stochastic gradient tracking to achieve faster convergence, and PSO-GFML \cite{gogineni2022decentralized} selectively exchanging local model parameters with the servers to improve communication efficiency.
The weights of the cross-client graph's links are determined based on the similarity between unlabeled graph embeddings \cite{tao2022semigraphfl} or hidden parameters \cite{yuan2022fedstn}.

Graph regularization integrates graph Laplacian regularization into the objective function to transform model parameters from neighboring clients \cite{ortega2018graph} and is particularly beneficial to multi-task GFL.
In dFedU \cite{dinh2022new}, a pre-defined fully connected cross-client graph is given ahead where each client is assigned a single task.
When receiving local updated models from neighboring clients, each client conducts model updating with graph regularization.
He et al. \cite{he2021spreadgnn} alternatively posit that each client runs multiple tasks and initializes a cross-client task relationship graph with task classifier parameters, where decentralized periodic averaging SGD is employed to optimize the objective function.

\begin{mybox}
\textbf{Lessons Learned (Sec. \ref{sec:gnn-optimization-edge})}

GI can serve as an optimization tool for enhancing different functionalities of edge networks, including network resource management, graph-based decision-making, security and anomaly detection, and GFL.
To well harness GI in these cases, there are dual knobs: on the one hand, the system needs to concisely identify the optimization constraints and objectives from graph perspectives, where one or more graph(s) should be built to represent the structure of the targeted problem; on the other hand, the graph may associate appropriate (encoded) properties that supply sufficient information and convert optimization objectives into learning objectives, such that GI models can use these graph semantics to induce expected output.
\end{mybox}

\section{Edge Graph Intelligence Ecosystems}
\label{sec:edge-infra-gnn}

\begin{figure*}[t]
  \centering
  \includegraphics[width=0.95\linewidth]{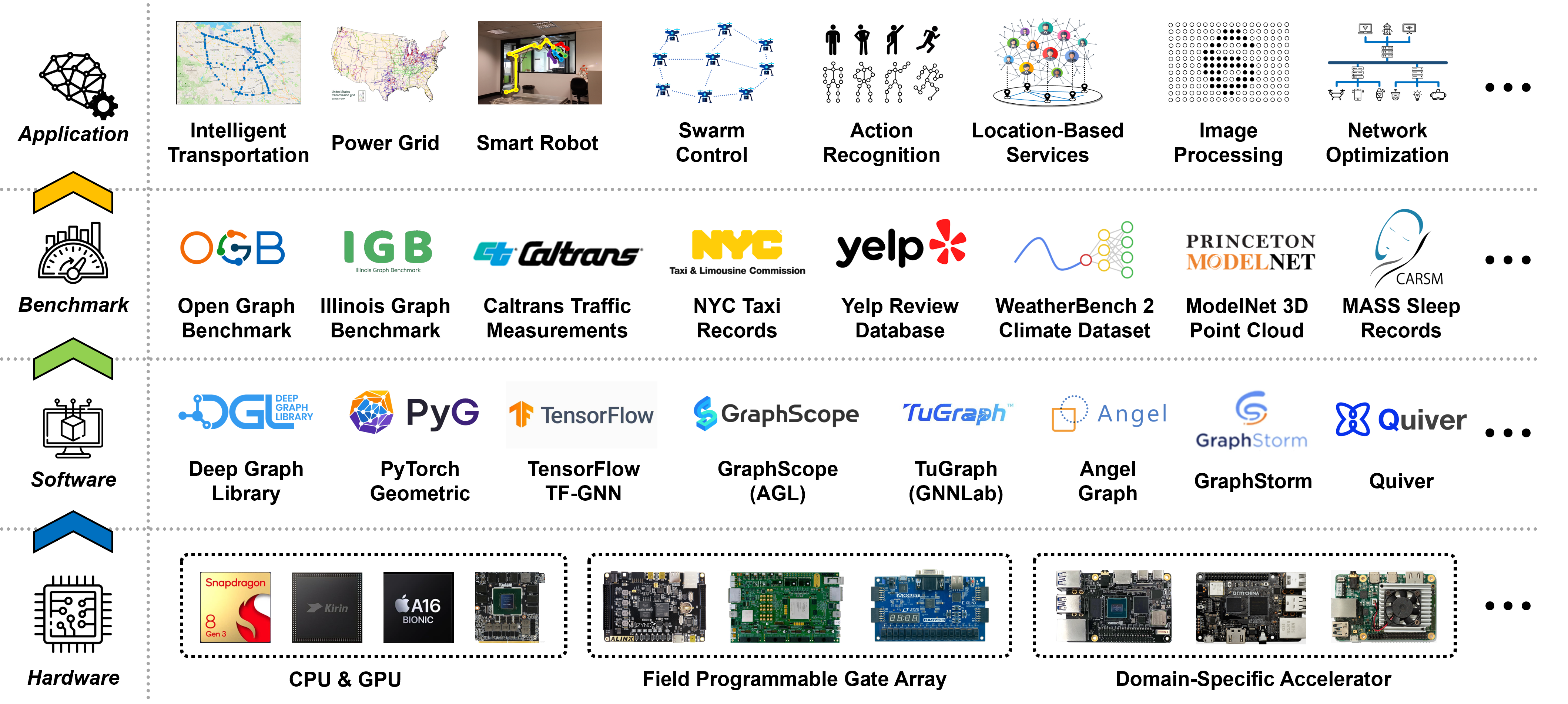}
  \caption{The EGI ecosystem includes all stakeholders related to EGI from hardware, software, benchmark, and applications. The hardware involves edge hardware for sustaining EGI computation, covering CPU, GPU, FPGA, and DSA. The software indicates development frameworks for programming and deploying EGI models such as DGL and PyG. The benchmarks provide datasets and unified toolkits for evaluating dedicated EGI models and services. The applications show a rich set of scenarios where EGI actively works.
  }
  \label{fig:egi-ecosystem}
\end{figure*}

After reviewing the interaction between GI and edge networks in Sec. \ref{sec:edge_computation_gnn} and Sec. \ref{sec:gnn-optimization-edge}, we discuss the EGI ecosystem inflated from this loop.
Fig. \ref{fig:egi-ecosystem} shows the landscape of the EGI ecosystem from levels of hardware, software, benchmark, and applications.

\subsection{Edge Hardware for EGI}

Edge networks cover a wide spectrum of edge platforms from embedded and mobile devices to vehicles and edge servers, as discussed in Sec. \ref{sec:spectrum_edge_network}.
At the core of these edge facilities, different processors deliver different capabilities and require different optimizations for GI model computation.
In general, they can be discussed in three types, mobile CPUs and GPUs, Field Programmable Gate Array (FPGA), and Domain-Specific Accelerators (DSAs).

\subsubsection{CPU and GPU}
A majority of edge facilities, especially wearable and mobile devices, are equipped only with conventional CPUs and GPUs such as Qualcomm SnapDragon series and Huawei Kirin series.
Given the intensive workload of GI-driven applications, efficient execution of them requires meticulously exploiting the capability of low-power processors.
To better understand the characteristics of GI models on conventional platforms, Huang et al. \cite{huang2021understanding} examine in detail representative GI systems and reveal five major gaps in optimizing GI computation performance, i.e, poor data locality, severe workload imbalance, redundant memory access, large memory footprint, and inefficiency on variant feature lengths.
Based on these observations, Huang et al. \cite{huang2021understanding} propose a set of tailored optimizations to bridge the gaps and achieve 1.37$\times$-15.5$\times$ performance improvement on various GI models.
Towards the same objective, Zhang et al. \cite{zhang2022understanding} study the GI execution performance on commercial platforms from the perspective of computational graph and summarize three issues that existing systems suffer, namely redundant neural operator computation, inconsistent thread mapping, and excessive intermediate data.
To tackle these challenges, Zhang et al. \cite{zhang2022understanding} introduce three corresponding designs, i.e., propagation-postponed operator reorganization, unified thread mapping for fusion, and intermediate data recomputation, achieving notable speedup in lower memory IO and footprint.

These empirical studies motivate literature \cite{rahman2021fusedmm,gong2022graphite,md2021distgnn,thorpe2021dorylus,wang2021flexgraph} in efficiently leveraging the potential of conventional CPUs and GPUs.
For instance, Adiletta et al. \cite{adiletta2023characterizing} study a detailed characterization of GI models on Intel's Programmable Integrated Unified Memory Architecture (PIUMA) for scalable training.
Graphite \cite{gong2022graphite} addresses GI model training on CPUs by eliminating the memory limitation through software-hardware co-design including layer fusion, feature compression, and a vertex processing reordering algorithm for improving temporal locality, and implements a high-performance GI computation system on Intel CPUs.
Some works explore optimizing matrix computation in GI tasks from the kernel level.
FeatGraph \cite{hu2020featgraph} provides efficient implementations of Sampled Dense-Dense Matrix Multiplication (SDDMM) and Sparse-dense Matrix Multiplication (SpMM) for both CPUs and GPUs, and FusedMM \cite{rahman2021fusedmm} further develops a general solution for different GI programming backends.
There are also many comprehensive systems developed to tackle the various aspects of GI training and inference for distributed training on CPU or GPU clusters (e.g., Dorylus \cite{thorpe2021dorylus}, FlexGraph \cite{wang2021flexgraph}).
However, most of them rely on the computation infrastructure at datacenter levels and are thus out of the scope of this survey.
Interested readers may refer to recent reviews \cite{shao2022distributed, lin2023comprehensive} on large-scale distributed GI systems.

\subsubsection{Field-Programmable Gate Array}
FPGA is a type of integrated circuit with a differentiated ability to be (re)programmed to desired functionality requirements after manufacturing.
This advanced ability allows FPGA to be used in applications with requirements on flexibility, speed, and parallelism, and thus are embedded and employed in edge platforms for GI workload processing.
HP-GNN \cite{lin2022hp} generates high-throughput GNN training implementations on a given CPU-FPGA platform by designing an automatic algorithm that maps GI training algorithms, GI models, and hardware specifics to the targeted CPU-FPGA platform.
GenGNN \cite{abi2022gengnn} develops a generic GI model acceleration framework using High-Level Synthesis (HLS) techniques, aiming to render GI inference in low latency and support various GI models in flexible extensibility.
It embraces an optimized message-passing structure to accommodate the majority of popular GI models and provides a rich library of model-specific components.
GraphAGILE \cite{zhang2023graphagile} designs a hardware module named adaptive computation kernel for efficient execution of GI matrix operations such as sparse-dense matrix multiplication and sampled dense-dense matrix multiplication, and proposes a set of compiler optimizations to reduce inference latency.
Graph-OPU \cite{chen2023graphopu} proposes an FPGA-based overlay processor for accelerating GI execution, which introduces microarchitecture for fully-pipelined GI processing and customizes the instruction sets for various GI model adoptions.

Besides optimizing FPGA solutions for general GI model execution, some researchers further study how to utilize FGPA-integrated platforms for specific applications efficiently.
For example, Heintz et al. \cite{heintz2020accelerated} develop an FPGA implementation for charged particle tracking based on GI models, incorporating OpenCL and hls4ml in an interaction network architecture.
Zhang et al. \cite{zhang2023accelerating} consider synthetic aperture radar-assisted remote-sensing image recognition tasks and design an FPGA acceleration solution with customized data path and memory organization for various GI model kernels.
To further speed up GI inference, it exploits FPGA's high bandwidth memory for loading data and storing intermediate results, as well as a splitting kernel technique to improve the on-chip routability and frequency.
Huang et al. \cite{huang2023low} specify an FPGA-based parallel particle tracking system with memory-aware data allocation and buffer arrays and collision-aware scheduling on parallel events.

\subsubsection{Domain-Specific Accelerator}
Domain-Specific Accelerators (DSAs) are important computing components and becoming more pervasive in many edge facilities such as dedicated edge servers and some mobile devices \cite{sun2018ultra, google_tpu, qualcomm, apple_ANE}.
By optimizing specifically for a narrow range of computations, DSAs are tailored to meet the needs of targeted algorithms in dedicated domains and enable orders of magnitude improvements in performance or cost compared to general-purpose processors.
Motivated by the achieved success of DSAs in CNNs, the community has made plentiful efforts in developing DSAs for GI workload.
Towards that, EnGN \cite{liang2020engn} presents a unified architecture inspired by CNN accelerators, where GI computations are viewed in a matrix perspective and matrix multiplication optimizations are applied for improved runtime performance.
Auten et al. \cite{auten2020hardware} propose a modular DSA architecture for convolutional GNNs, where each tile as a basic unit is composed of an aggregation module, a DNN accelerator module, a DNN queue, and a graph processing element.

Nonetheless, GNNs are semantically different from traditional DNNs (c.f. Sec. \ref{sec:fundamental-gnn}), bringing new computing characteristics desired for tailored designs.
To better understand such computing characteristics, Yan et al. \cite{yan2020characterizing} and Lin et al. \cite{lin2022characterizing} respectively conduct detailed profiling of general GNN workload and distributed GNN training workload on GPUs, providing useful insight on GNN-oriented DSA design.
Based on that, HyGCN \cite{yan2020hygcn} introduces a hybrid architecture to tame the alternating phases of GI computation, which is composed of separate dedicated engines for the aggregate and update functions as well as a control scheme to pipeline the execution of both functions.
GRIP \cite{kiningham2022grip} leverages the programming abstraction of GReTA \cite{kiningham2020greta} to develop a general DSA for GNNs, which organizes a GI model inference into four steps, i.e., gather, reduce, transform, and activate.
Regarding this abstraction, GRIP specializes in the processing units for links and vertices separately and designs a control module to coordinate data movement between units and buffers.
AWB-GCN \cite{geng2020awb} raises an aggressive adaptation to the structural sparsity of GI models under the motivation that for power-law graphs, GI computation can be polarly dense or sparse and suffers from workload imbalance.
To address this imbalance, AWB-GCN \cite{geng2020awb} develops a custom matrix multiplication engine with autotuning workload balancing techniques, and GNNIE \cite{mondal2022gnnie} advocates a flexible MAC architecture that splits features into blocks with load (re)distribution and graph-specific caching to bypass the high costs of random DRAM.

\subsection{Developmemt Frameworks}

GI models' unique computing workflow introduces different programming patterns from traditional DL models (e.g., CNNs and RNNs), which requires dedicated frameworks for implementation.
Towards that, the community has developed various programming frameworks compatible with edge computing platforms. 
Among them, PyTorch Geometric (PyG) \cite{fey2019fast} and Deep Graph Library (DGL) \cite{wang2019deep} are the two most popular ones and are widely used in various applications.
PyG is a geometric deep learning extension library built upon PyTorch that provides flexible interfaces for programming GI models on both CPUs and GPUs.
It adopts a message-passing programming abstraction for building GI models, which is programmer-friendly to express various GI variants.
DGL is another general framework specialized for deep learning models on graphs.
It also leverages the message-passing primitives in a user-configurable way and introduces a set of parallelization strategies for high-performance GNN execution.
Both PyG and DGL can be used in edge devices and edge servers as long as they enable the PyTorch library in their execution environments.
In addition, DGL also supports TensorFlow and MXNet, and edge platforms that run these two frameworks also allow model execution with DGL.
Besides PyG and DGL, the industry has also developed many frameworks upon their customized demand, such as AGL \cite{zhang2020agl}, BGL \cite{liu2023bgl}, TF-GNN \cite{ferludin2022tf}, and Angel-Graph, etc. 
Nonetheless, most of them are dedicated to cloud-level resources without consideration of the deployment on edge networks and thus require tailored adjustments when building edge GI services.

\begin{table*}[]
\centering
\caption{Selected open datasets for EGI applications.}
\label{tab:dataset}
\begin{tabular}{m{1.5cm}clp{8cm}c}
\hline
\centering \textbf{Domain}                                        & \textbf{Dataset}    & 
\makecell[c]{\textbf{Graph}}                                                                                & \centering \textbf{Description}                                                                                                                                     & \textbf{Link}                                                               \\ \hline
\centering \centering \multirow{15}{*}{\begin{tabular}[c]{@{}c@{}} Smart\\Cities\end{tabular}}                          & PeMS                & \begin{tabular}[c]{@{}l@{}}Vertex: Traffic sensor\\ Link: Road nets\end{tabular}              & \begin{tabular}[c]{@{}p{8cm}@{}}Traffic sensory data collected by CalTrans Performance Measurement System in California. \end{tabular}                                                                & \cite{pems}                                             \\ 
                                                       & METR-LA             & \begin{tabular}[c]{@{}l@{}}Vertex: Traffic sensor\\ Link: Road nets\end{tabular}              & \begin{tabular}[c]{@{}p{8cm}@{}} Traffic sensory data of loop detectors in the highway of Los Angeles.\end{tabular}                                                                                    & \cite{metr-la}                                                      \\
                                                       & NYC Taxi            & \begin{tabular}[c]{@{}l@{}}Vertex: Taxi\\ Link: Road nets\end{tabular}                        & \begin{tabular}[c]{@{}p{8cm}@{}} Origin-Destination demand dataset with taxi trip records capturing pick-up and drop-off locations, trip distances, payment types, etc.\end{tabular}       & \cite{nyc-taxi}              \\
                                                       & Chicago Bike        & \begin{tabular}[c]{@{}l@{}}Vertex: Bike\\ Link: Road nets\end{tabular}                        & \begin{tabular}[c]{@{}p{8cm}@{}} Origin-Destination demand dataset with bike trip records capturing trip start/end time, station, and rider type.\end{tabular}                                         & \cite{chicago-bike}                                      \\
                                                       & Beijing Air Quality & \begin{tabular}[c]{@{}l@{}}Vertex: Region\\ Link: Geographical adjacency\end{tabular}         & \begin{tabular}[c]{@{}p{8cm}@{}} Air quality data of distributed monitoring sites from Beijing Municipal Environmental Monitoring Center.\end{tabular}                                                 & \cite{beijing-air} \\
                                                       & WeatherBench2        & \begin{tabular}[c]{@{}l@{}}Vertex: Region\\ Link: Geographical adjacency\end{tabular}         & \begin{tabular}[c]{@{}p{8cm}@{}} Comprehensive meteorological data for mid-range weather forecasting, including temperature, humidity, wind, and cloud cover, etc.\end{tabular}                        & \cite{weatherbench2}                                 \\
                                                       & PowerGraph          & \begin{tabular}[c]{@{}l@{}}Vertex: Energy station\\ Link: Power Grid\end{tabular}             & \begin{tabular}[c]{@{}p{8cm}@{}} Electrical power grid dataset that models cascading failures in power grids.\end{tabular}                                                                              & \cite{powergraph}                   \\
                                                       & COVID-19         & \begin{tabular}[c]{@{}l@{}}Vertex: Region\\ Link: Geographical adjacency\end{tabular}         & \begin{tabular}[c]{@{}p{8cm}@{}} Data repository of 2019 Novel Coronavirus visual dashboard, which collects global COVID-19 infectious data.\end{tabular}                                              & \cite{covid-19}                                  \\ \hline
\centering \multirow{13}{*}{\begin{tabular}[c]{@{}c@{}}Intelligent\\Robots\\ \& Vehicles\end{tabular}}     
                                                       & US-101              & \begin{tabular}[c]{@{}l@{}}Vertex: Vehicle\\ Link: Vehicle Connection\end{tabular}            & \begin{tabular}[c]{@{}p{8cm}@{}} Vehicle trajectory data on southbound US 101 that provides the precise locations of vehicles every one-tenth of a second.\end{tabular}                                                                                                             & \cite{us-101}                                                     \\
                                                       & Stanford Drone      & \begin{tabular}[c]{@{}l@{}}Vertex: Drone\\ Link: Drone Connection\end{tabular}                & \begin{tabular}[c]{@{}p{8cm}@{}} Flight records of navigated drones that  collects images and videos of various types of agents in a realistic outdoor environment\end{tabular}                                                                                    & \cite{stanford-drone}                               \\
                                                       & Kinetics            & \begin{tabular}[c]{@{}l@{}}Vertex: Entity\\ Link: Entity relationship\end{tabular}            & \begin{tabular}[c]{@{}p{8cm}@{}} Large-scale human action video dataset collected from YouTube, covering human-object interactions and human-human interactions.\end{tabular}            & \cite{kinetics}  \\ 
                                                       & nuScenes            & 
                                                       \begin{tabular}[c]{@{}l@{}}Vertex: Vehicles or pedestrians\\ Link: Entity relationship\end{tabular}            & \begin{tabular}[c]{@{}p{8cm}@{}} A large-scale, multimodal dataset for autonomous driving, containing annotated sensor data from LiDAR, radar, cameras, and GPS across diverse driving scenarios in urban environments.\end{tabular}            & \cite{nuscenes}                           \\
                                                       & CORSMAL            & \begin{tabular}[c]{@{}l@{}}Vertex: Entity\\ Link: Entity relationship\end{tabular}            & \begin{tabular}[c]{@{}p{8cm}@{}}A multimodal dataset focused on robotic perception for manipulating containers, including audio, visual, and depth data to estimate container properties and contents in real-time interactions.\end{tabular}            & \cite{corsmal} \\
                                                       & RoboNet            & \begin{tabular}[c]{@{}l@{}}Vertex: Entity\\ Link: Entity relationship\end{tabular}            & \begin{tabular}[c]{@{}p{8cm}@{}}Collection of robotic manipulation data across diverse objects, environments, and robotypes, designed to enable learning generalizable manipulation policies.\end{tabular}  & \cite{robonet} \\
                                                       \hline
\centering \multirow{9}{*}{\begin{tabular}[c]{@{}c@{}}Human\\Sensing\end{tabular}}                         & NTU RGB+D           & \begin{tabular}[c]{@{}l@{}}Vertex: Body joint\\ Link: Human skeleton\end{tabular}             & \begin{tabular}[c]{@{}p{8cm}@{}} Large-scale dataset for human action recognition that contains RGB videos, depth map sequences, and 3D skeletal data. data.\end{tabular}                                                                                                      & \cite{ntu-rgbd}                         \\
                                                       & MobiAct             & \begin{tabular}[c]{@{}l@{}}Vertex: Body joint\\ Link: Human skeleton\end{tabular}             & \begin{tabular}[c]{@{}p{8cm}@{}} Human action recognition dataset with acclerometer, gyroscope, and orientation sensory data\end{tabular}                                                              & \cite{mobiact}                     \\
                                                       & TST V2              & \begin{tabular}[c]{@{}l@{}}Vertex: Body joint\\ Link: Human skeleton\end{tabular}             & \begin{tabular}[c]{@{}p{8cm}@{}} Fall detection dataset with depth frames, skeleton joints, acceleration sensory data\end{tabular}                                                                     & \cite{tst}           \\
                                                       & MASS-SS3            & \begin{tabular}[c]{@{}l@{}}Vertex: Monitoring channel\\ Link: Channel connection\end{tabular} & \begin{tabular}[c]{@{}p{8cm}@{}} Polysomnography recordings collected in a lab-based environment for sleep quality prediction.\end{tabular}                                                            & \cite{mass}                                                 \\
                                                       & AffectNet           & \begin{tabular}[c]{@{}l@{}}Vertex: Action unit\\ Link: FACS-based link\end{tabular}           & \begin{tabular}[c]{@{}p{8cm}@{}} Large-scale facial expression analysis dataset collected from the Internet.\end{tabular}                                                                              & \cite{affectnet}                                  \\ \hline
\centering \multirow{7}{*}{\begin{tabular}[c]{@{}c@{}}Locations\\-Based\\Social\\Recommendation\end{tabular}} & Gowalla             & \begin{tabular}[c]{@{}l@{}}Vertex: User\\ Link: Friendship\end{tabular}                       & \begin{tabular}[c]{@{}p{8cm}@{}} Location-based social networks where users share their locations by checking-in.\end{tabular}                                                                         & \cite{gowalla}                             \\
                                                       & Foursquare          & \begin{tabular}[c]{@{}l@{}}Vertex: User and Hotel\\ Link: Visiting history\end{tabular}       & \begin{tabular}[c]{@{}p{8cm}@{}} Location-based digital footprints from Foursquare for personalized location recommendation and search.  \end{tabular}                                                  & \cite{foursquare}            \\
                                                       & SIoT                & \begin{tabular}[c]{@{}l@{}}Vertex: IoT device\\ Link: Social connection\end{tabular}          & \begin{tabular}[c]{@{}p{8cm}@{}} Network of IoT devices with social connections with identity information including device type, brand, and ownership.\end{tabular}                                                                                                          & \cite{siot-dataset}                                                              \\
                                                       & Yelp                & \begin{tabular}[c]{@{}l@{}}Vertex: User comment\\ Link: Comment history\end{tabular}          & \begin{tabular}[c]{@{}p{8cm}@{}} Users' review comments of POIs from Yelp with location, purchasing, and social information.\end{tabular}                                                                        & \cite{yelp-dataset}                                                \\ \hline
\centering \multirow{10}{*}{\begin{tabular}[c]{@{}c@{}}Mobile\\Vision\end{tabular}}                         & MNIST               & \begin{tabular}[c]{@{}l@{}}Vertex: Pixel\\ Link: Pixel adjacency\end{tabular}                 & \begin{tabular}[c]{@{}p{8cm}@{}} Handwritten digits dataset, where pixels are viewed in grid-structure graphs.\end{tabular}                                                                            & \cite{mnist}                                          \\
                                                       & DAVIS               & \begin{tabular}[c]{@{}l@{}}Vertex: Entity\\ Link: Entity relationship\end{tabular}            & \begin{tabular}[c]{@{}p{8cm}@{}} Video object segmentation dataset that consists of 50 videos in total with pixel-wise annotations for every frame.\end{tabular}                                       & \cite{davis}                                                 \\
                                                       & YouTubeVIS          & \begin{tabular}[c]{@{}l@{}}Vertex: Entity\\ Link: Entity relationship\end{tabular}            & \begin{tabular}[c]{@{}p{8cm}@{}} Video instance segmentation dataset that contains video clips from YouTube with segmentation masks, instance identity labels, etc.\end{tabular} & \cite{youtube-vos}                                        \\
                                                       & KITTI               & \begin{tabular}[c]{@{}l@{}}Vertex: Point\\ Link: Point adjacency\end{tabular}                 & \begin{tabular}[c]{@{}p{8cm}@{}} Point cloud dataset of various kinds of objects for 3D object detection.\end{tabular}                                                                                 & \cite{kitti}                                      \\
                                                       & ModelNet40          & \begin{tabular}[c]{@{}l@{}}Vertex: Point\\ Link: Point adjacency\end{tabular}                 & \begin{tabular}[c]{@{}p{8cm}@{}} Synthetic object point clouds with CAD-generated meshes in various categories.\end{tabular}                                                                         & \cite{modelnet}                                          \\ \hline
\end{tabular}
\end{table*}

\subsection{Open Datasets}

To assess how well edge GI systems perform and learn their behaviors upon deployment, versatile datasets are collected from diverse edge scenarios, serving as performance benchmarks.
Table \ref{tab:dataset} lists some representative datasets in several application domains discussed in Sec. \ref{sec:gnn-application-edge}.

The first dataset series lies on smart cities, including data collected from intelligent transportation systems, meteorological stations, power grids, and governments.
Their graphs are mostly derived from the geographical network relations in cities, e.g., road nets, power networks, and regional adjacencies.
The second domain is intelligent robotics and vehicles, taking datasets from various robot manipulation scenarios such as UAV swarms and vehicle trajectories
The third category is for human sensing applications, including human action recognition, facial affective analysis, and sleep quality prediction.
Specifically, for the sleep quality dataset, researchers construct sleep stage graphs with electronic channel records and employ GI models for prediction.
The fourth domain is location-based social recommendation, where social networks are naturally graphs and the location information of users puts its analysis into the context of edge networks.
The last group focuses on mobile vision, where the datasets are also widely utilized in many computer vision models.
Readers are encouraged to exact their interested datasets from the links to learn more about their sources and applications.

\begin{mybox}
\textbf{Lessons Learned (Sec. \ref{sec:edge-infra-gnn})}

The promising fusion of EGI facilitates the EGI ecosystem in a full-stack manner. Specifically, the ecosystem provides an abundant set of facilities covering hardware, software, and testing benchmarks, acting as a stage for developing and deploying EGI services.
Developers are suggested to develop their EGI services agilely with well-established open-sourced toolkits. 
\end{mybox}
\section{Open Challenges and Future Research Opportunities}
\label{sec:challenge}

The confluence of GI and edge networks is still in its infancy stage and many of the considered parts of the EGI landscape are yet under exploration.
In this section, we articulate key open challenges and future research directions for EGI.

\subsection{New Promising Applications and Optimizations}
The integration of GI and edge networks opens up new possibilities for low-latency, context-aware analysis, and decision-making, holding great potential for miscellaneous promising applications.
Given the ubiquitous nature of graph data, there are still many areas to be explored for service providers, network operators, and end users with optimization opportunities as well as business revenues.


\subsubsection{New EGI Applications}
Besides applications discussed in Sec. \ref{sec:gnn-application-edge}, there are still many burgeoning edge scenarios well matching EGI's capability.
For instance, in digital twin networks, the mapping between physical objects and digital twins can be regarded as graphs and EGI may be applied for task offloading, power allocation, and network slicing \cite{yao2023cooperative, zhang2023gnn, wang2020graphneural}.
Satellite-aerial-ground intergaetd networks are also one of the emerging edge networks \cite{kawamoto2023haps, kawamoto2023traffic, ariyoshi2024challenges}, where high-altitude platform stations and ground stations constitute vertices with their communication channels as links.
EGI thus can also be used for them.

\subsubsection{B5G and 6G Protocols Design}
Communication in edge networks, particularly for cross-device services, often involves complex interactions that can be naturally represented as graphs. The networking topology, defined by communication nodes, forms the vertices, while relevant data such as channel capacity, queuing delays, and link states can be modeled as graph properties. This representation opens up opportunities to apply GI for developing advanced communication protocols, such as those for B5G and 6G networks, enabling smarter, faster, and more energy-efficient data transmission.
For instance, GNNs can be utilized to dynamically optimize routing strategies and detect intermittent anomalies within edge networks \cite{jiang2022graphcomm, yang2023task}. Furthermore, communication channel states can be incorporated as link attributes in graph models, allowing GI methods to enhance network management and resource allocation \cite{yang2020graph, salaun2022gnn, he2023message, ranasinghe2021graph}.

\subsubsection{Personalized Services}
End users nowadays are immersively surrounded by a vast amount of IoT devices like mobile smartphones, wearable kits, and smart home assets.
For individual users, we can weave a user-centric network \cite{yang20226g} by collecting all user-related data from these surrounding devices and connecting them in a logic graph, on which EGI techniques can be applied to process, understand, and learn.
In other words, EGI can serve as a tool to summarize data points scattered around individual users and be leveraged as a core component to build everyone-centric customized services.

\subsection{Edge-Centric Graph Learning Models and Systems}

Many existing GI models are originally designed to improve accuracy on specific tasks.
In real deployment, however, a set of SLOs beyond accuracy are required by service providers such as responsive latency, system energy, and data privacy, and implementing GI models at edge faces unique challenges.
Note that the graphs in edge networks may not be as huge as social networks on Twitter \cite{kwak2010twitter}.
However, it does not necessarily mean that training and deploying GNN on edge networks is technically trivial.

\subsubsection{Resource Constraints to GI Models}
As we discussed in Sec. \ref{sec:edge_computation_gnn}, edge devices typically have limited but heterogeneous computational capabilities compared to cloud servers, possessing platforms from MCUs to mobile smartphones to edge servers.
Edge-centric GI models need to be lightweight, have acceptable memory footprints, and minimize computational complexity while still delivering accurate results, especially for EGI-based services with real-time or near real-time requirements.
This directs innovations in both model level and system level, e.g., lightweight algorithms like computation deduplication  \cite{wang2022guide} and memory-efficient on-device systems \cite{xue2023sugar}.
Besides, fine-grained characteristics of GI models also expect precious profiling for refined resource utilization.

\subsubsection{System Dynamics}
There are two types of system dynamics, namely resource dynamics and graph dynamics. On the one hand, edge devices frequently operate in dynamic and unpredictable resource conditions, including intermittent connectivity, dynamic computing capabilities, or fluctuated communication bandwidths \cite{gobieski2019intelligence, rodrigues2024smart}.
A stable, satisfactory EGI system is obliged to be robust to these variations and capable of adapting to different network conditions to ensure reliable and timely analysis, especially for those services that process graph data across multiple edge devices.
On the other hand, many real-world graph data can be dynamic in both properties and topologies, requiring both algorithmic and system flexibility to adapt to the input variation.
Some work \cite{zeng2023serving, zeng2022gnn} exploit online incremental adjustment for minor input variation, where drastic changes like graph reconstitution are still under exploration.

\subsubsection{Scaling to Large Graphs}
Many real-world networks represented as graphs can be massive and evolve in size, but how to scale EGI systems to large and heavy-attributed graphs poses significant challenges on both the algorithm and system sides.
From the algorithm perspective, large graph models borrow wisdom from LLM and incorporate pre-training methods to embrace graphs on giant scales, but also bring overload stress to edge networks.
Distributed edge collaboration (cf. Sec. \ref{sec:distributed_edge_collaboration}), which aggregates multiple edge devices as a resource pool for model computation, can be a potential solution but still requires tailored designs with parallel processing, graph partitioning, and incremental learning, etc \cite{ma2025multi}.
From the system perspective, orchestrating a tremendous volume of graph data through edge network channels tackles the communication bottleneck as well as the storage capacity of edge devices.
Data compression like graph pruning and feature quantization can be helpful, but how to balance the tradeoff between model accuracy and execution latency requires careful design.

\subsection{Emerging Learning Paradigms}

Data generated at the network edge are massive but also usually in varying quality.
To improve the applicability and usability of EGI models, emerging learning paradigms should be utilized to tackle different data cases.
While some have been adopted (e.g., Federated Learning, Reinforcement Learning, etc.), there are still many that can be exploited and we discuss three below.

\subsubsection{Large Graph Model}
LGMs borrow the wisdom of LLMs to the graph domain that simultaneously scales both training data and models.
This scaling mechanism envisions LGMs to be capable of capturing complex relationships and patterns within large-scale graphs, which are common representations of data in various domains such as social networks, biological networks, and transportation systems \cite{liu2023towards,xia2024opengraph}.
Nonetheless, applying LGM in edge networks severely suffers from the constrained computing capability and network capacity.
Edge-cloud collaboration, which marries the local data processing ability of edge networks and the powerful computing power of the cloud, can be a potential solution for deployment.
Given the promising ability of large models, LGMs desire more effort in the context of EGI.
On the other hand, EGI may also explore the translation between graphs and natural language to exploit the capability of LLMs, e.g., by semantically converting a graph into a paragraph of sentences for LLM input and generation \cite{zhang2023large}.

\subsubsection{Learning with Small Data}
Edge devices can collect a large volume of data but only with a small portion of labels, which demands EGI to be capable few-shot learners.
Towards that, few-shot learning techniques can be promising, which aims at training GL models for accurate predictions on unseen classes or categories with limited labeled data.
For example, in some EGI-based object detection tasks \cite{chen2019multi}, GI models can learn to recognize novel objects with a few labeled examples by leveraging information from similar objects and their relationships in a graph representation.
Another potential path is transfer learning, which borrows knowledge learned from a source domain or task to improve performance in a target domain or task.
An instance of transfer graph learning is anomaly detection in networked systems \cite{zhang2022efraudcom, ma2022deep}, where GI can learn patterns of fraud by pre-training on a large graph and transfer this knowledge to identify fraudulent activities in a target domain with limited labeled data.
TL can also be used for STGNN streams, e.g., for traffic prediction in smart cities \cite{yao2023transfer}.

\subsubsection{Continual Learning}
Continual Learning (CL) for GI is to incrementally retrain GNN models to adapt to new data or tasks without forgetting previous knowledge. 
It also adjusts the learned latent distribution like Transfer Learning but emphasizes learning patterns from the additional incoming data while preserving previously learned representations.
For EGI services, CL enables incremental service refinement for evolved graphs, e.g., in temporal graph data analysis tasks, without training GI models from scratch.
Examples include predicting traffic conditions in a road network and running the status of power grids, where continual learning of GI models allows the model to adapt over time and incorporate new temporal information while retaining the knowledge learned from past time steps.

\subsection{Native Edge Support for Graph Intelligence}

Edge networks serve as the infrastructural support of GI model computation, for which an EGI-native pipeline is crucial for efficient and effective EGI services provisioning.
Despite its importance, research on EGI-native edge support is yet limited in the context, and it confronts challenges in the full lifespan of GI model computation, i.e., from data collection to data processing to system development.

\subsubsection{Graph Data Collection}
EGI applications can not be realized without pervasive graph data.
For many applications, however, these graph data are generated and distributed in different places, e.g., traffic sensory data scattered in road networks and perception information gathered by vehicles in robotic swarms.
How to collect these (geographically) distributed data over the edge networks efficiently, i.e., in a high-quality, low-latency, and privacy-preserved manner, thereby becomes an unavoidable problem in running GI-based services.
In certain situations, graph generation techniques \cite{guo2022systematic} may be a useful supplement for graph data requirements.

\subsubsection{Edge-Cloud Collaboration}
Edge networks render computing power in versatile forms, e.g., by a single edge device, a pool of collaborated ones, and even a synergetic edge-cloud continuum. 
Provisioning these versatile resources efficiently for EGI applications not only requires flexible resource management but also judicious workload scheduling that well aligns with computation and communication.
In particular, edge-cloud collaboration provides a combined mechanism to embrace both data locality at the edge and resource richness in the cloud, which allows GI to be more efficient and sustainable and attracts growing attention from the community.
Possible means towards this objective include designing GI-oriented communication protocols, GI-specific computation scheduling mechanisms, and a holistic edge network cost optimization methodology.

\subsubsection{System Deployment}
Given the broad use cases of EGI, their development yet lacks a comprehensive framework with full-stack toolkits, particularly for distributed edge collaboration and edge-cloud synergy.
This requires addressing the complexity of distributed systems, including edge device coordination, data synchronization, and fault tolerance, and also tackling long-term interoperability and testing issues.
Existing virtualization techniques such as container and function computing can be further explored for friendly EGI development and deployment.
The communication aspects, such as timely data transmission of (potentially distributed) graph data and graph modeling of network channels, also desire tailored consideration.

\subsubsection{Supporting Large Models}
As an edge-oriented methodology, EGI can support large models in the following angles.
First, EGI provides an effective way to abstract edge data in the form of graphs, acting as a graph data processing module for large model deployment at the edge.
For instance, we may use EGI to transform graph data into prompts acceptable by large language models, allowing language intelligence on edge graphs.
Second, with EGI models, edge data can be abstracted into graph embeddings.
These embeddings can be used as a knowledge base for multi-modal foundation models aiming at specific edge scenarios, i.e., enabling retrieval-augmented generation.
Third, EGI can also be operated as an optimization tool to orchestrate resource allocation for large model computation.
To deploy resource-hungry large models on edge platforms, we may utilize many devices in edge networks, where EGI can be a scheduler, as discussed in Sec. \ref{sec:network-resource-management}, to schedule efficient resource allocation for large model acceleration.

\subsection{Explainability, Security, and Privacy}

The open nature of EGI requires trustworthiness across different entities in the ecosystem to make EGI happen.
Nonetheless, in many scenarios, edge services are practical only if they are reliable.
For EGI, this relates to explainability, security, and privacy.

\subsubsection{Explainability}
Explainability makes GI models predictable and understandable, which is crucial for some action-critical applications such as crime warnings and autonomous vehicle control.
However, taming EGI models' behaviors is challenging, which usually involves specific knowledge in their application scenarios.
Some literature has explored explaining GI models' behaviors in general with mathematical tools like Weisfeiler-Lehman graph isomorphism test \cite{xu2018powerful,maron2019provably,yuan2022explainability}, but those GI models dedicated designed for edge scenarios still lack investigation.

\subsubsection{Security}
Security holds a dual significance in the context of EGI services.
From a model perspective, GI models must be both secure and robust, as they are particularly vulnerable to attacks when malicious vertices are present in the graph input.
The risk is further amplified when input graph data is aggregated from multiple entities, increasing the likelihood of fraudulent activity, and underscoring the importance of developing and implementing secure GI models.
From a system perspective, edge networks are also susceptible to malicious devices, which can disrupt the learning process or compromise data integrity.
This calls for the establishment of verifiable trust mechanisms among edge networks, alongside the development of secure tools to defend against and recover from such attacks.

\subsubsection{Privacy}
Edge networks typically operate in close proximity to individuals, collecting data from edge devices that may include personal and sensitive information, such as geographic location, health and activity records, and electricity consumption patterns \cite{rodrigues2024smart}.
In light of privacy protection regulations, such as the European Union's General Data Protection Regulation (GDPR), the sharing and processing of this data across edge service providers poses significant risks of privacy leakage and unforeseen legal liabilities. 
Consequently, this raises critical research questions regarding not only secure but privacy-preserving computation in EGI models, where cryptographic techniques such as differential privacy and homomorphic encryption can be utilized to mitigate these risks.
\section{Conclusion}
\label{sec:conclusion}

Recent advances in ML have extrapolated representation learning techniques to the graph domain, cultivating GI as a powerful tool to learn from graphs.
Meanwhile, edge computing networks are thriving with the rapid proliferation of edge facilities and becoming a fundamental infrastructure of miscellaneous user-centric intelligent services.
These game-changing trends push GI and edge networks to a close-looped confluence, where edge networks serve as infrastructural support to GI-based tasks and GI conversely serves as a key enabler to build and optimize edge services.
Their complementary interaction raises a promising paradigm, i.e., Edge GI or EGI for brevity, to model, abstract, and analyze the ubiquitous graph data in edge networks, flourishing versatile graph-based applications at the network edge.

In this paper, we advocate EGI as a brand-new principle in the context of edge intelligence by revealing its motivational benefits, conceptual scope, and rating principle.
Based on that, we conduct a comprehensive survey of recent research efforts in this emerging field from multiple dimensions. 
We first provide primers on GI and edge networks and summarize miscellaneous edge applications built with GI models. 
Next, from the perspectives of ``edge for GI" and ``GI for edge", we thoroughly review related concepts, techniques, and systems on computing GI models on edge networks and applying GI for optimizing edge networks, respectively.
We finally present the overview of EGI ecosystems, discuss future directions, and conclude.
We hope that this survey can garner increased attention, foster meaningful discussions, and inspire future research in EGI.

\appendices

\bibliographystyle{IEEEtran}
\bibliography{main.bib}

\end{document}